\documentclass[twocolumn, tighten]{aastex62}
\newcommand\kms{km~s$^{-1}$}
\usepackage{multirow}
\usepackage{float}
\usepackage{booktabs}
\usepackage{txfonts}
\begin{document}

\title{On the magnetic field properties of protostellar envelopes in Orion}

\author[0000-0001-7393-8583]{Bo Huang}
\email{huang@ice.csic.es}
\affiliation{Institut de Ciències de l'Espai (ICE-CSIC), Campus UAB, Can Magrans S/N, E-08193 Cerdanyola del Vallès, Catalonia, Spain}

\author[0000-0002-3829-5591]{Josep M. Girart}
\email{girart@ieec.cat}
\affiliation{Institut de Ciències de l'Espai (ICE-CSIC), Campus UAB, Can Magrans S/N, E-08193 Cerdanyola del Vallès, Catalonia, Spain}
\affiliation{Institut d'Estudis Espacials de Catalunya (IEEC), c/Gran Capita, 2-4, E-08034 Barcelona, Catalonia, Spain}

\author[0000-0003-3017-4418]{Ian W. Stephens}
\affiliation{Department of Earth, Environment, and Physics, Worcester State University, Worcester, MA 01602, USA}

\author[0000-0001-5811-0454]{Manuel Fern\'andez L\'opez}
\affiliation{Instituto Argentino de Radioastronomía (CCT-La Plata, CONICET; CICPBA), C.C. No. 5, 1894, Villa Elisa, Buenos Aires, Argentina}

\author[0000-0001-5653-7817]{Hector G. Arce}
\affiliation{Department of Astronomy, Yale University, New Haven, CT 06511, USA}

\author[0000-0003-2251-0602]{John M. Carpenter}
\affiliation{Joint ALMA Observatory, Av. Alonso de Córdova 3107, Vitacura, Santiago, Chile}

\author[0000-0002-3583-780X]{Paulo Cortes}
\affiliation{National Radio Astronomy Observatory, 520 Edgemont Rd., Charlottesville, VA 22093, USA}
\affiliation{Joint ALMA Observatory, Alonso de Córdova 3107, Vitacura, Santiago, Chile}

\author[0000-0002-5216-8062]{Erin G. Cox}
\affiliation{Center for Interdisciplinary Exploration and Research in Astrophysics (CIERA), 1800 Sherman Avenue, Evanston, IL 60201, USA}

\author[0000-0001-7594-8128]{Rachel Friesen}
\affiliation{Department of Astronomy $\&$ Astrophysics, University of Toronto, 50 St. George St., Toronto, ON, M5S 3H4, Canada}

\author[0000-0002-5714-799X]{Valentin J. M. Le Gouellec}
\affiliation{NASA Ames Research Center, Space Science and Astrobiology Division M.S. 245-6 Moffett Field, CA 94035, USA}
\affiliation{NASA Postdoctoral Program Fellow}

\author[0000-0002-8975-7573]{Charles L. H. Hull}
\affiliation{Joint ALMA Observatory, Av. Alonso de Córdova 3107, Vitacura, Santiago, Chile}
\affiliation{National Astronomical Observatory of Japan, Alonso de Córdova 3788, Office 61B, Vitacura, Santiago, Chile }

\author[0000-0003-3682-854X]{Nicole Karnath}
\affiliation{SOFIA Science Center, Universities Space Research Association, NASA Ames Research Center, Moffett Field, California 94035, USA}
\affiliation{Space Science Institute, 4765 Walnut St, Suite B Boulder, CO 80301, USA}
\affiliation{Center for Astrophysics $\mid$ Harvard $\&$ Smithsonian, 60 Garden Street, Cambridge, MA 02138, USA}

\author[0000-0003-4022-4132]{Woojin Kwon}
\affiliation{Department of Earth Science Education, Seoul National University, 1 Gwanak-ro, Gwanak-gu, Seoul 08826, Republic of Korea}
\affiliation{SNU Astronomy Research Center, Seoul National University, 1 Gwanak-ro, Gwanak-gu, Seoul 08826, Republic of Korea}

\author[0000-0002-7402-6487]{Zhi-Yun Li}
\affiliation{Astronomy Department, University of Virginia, Charlottesville, VA 22904, USA}

\author[0000-0002-4540-6587]{Leslie W. Looney}
\affiliation{Department of Astronomy, University of Illinois, 1002 West Green Street, Urbana, IL 61801, USA}

\author[0000-0001-7629-3573]{Tom Megeath}
\affiliation{Department of Astronomy, University of Toledo, Toledo, OH 43606, USA}

\author[0000-0002-2885-1806]{Philip C. Myers}
\affiliation{Center for Astrophysics $\mid$ Harvard $\&$ Smithsonian, 60 Garden Street, Cambridge, MA 02138, USA}

\author{Nadia M. Murillo}
\affiliation{Instituto de Astronom\'ia, Universidad Nacional Aut\'onoma de M\'exico, AP106, Ensenada CP 22830, B. C., M\'exico}
\affiliation{Star and Planet Formation Laboratory, RIKEN Cluster for Pioneering Research, Wako, Saitama 351-0198, Japan}

\author[0000-0002-3972-1978]{Jaime E. Pineda}
\affiliation{Center for Astrochemical Studies, Max Planck Institute for Extraterrestrial Physics, D-85748 Garching, Germany}

\author[0000-0001-7474-6874]{Sarah Sadavoy}
\affiliation{Department of Physics, Engineering and Astronomy, Queen’s University, 64 Bader Lane, Kingston, ON, K7L 3N6, Canada}

\author[0000-0002-3078-9482]{\'Alvaro S\'anchez-Monge}
\affiliation{Institut de Ciències de l'Espai (ICE-CSIC), Campus UAB, Can Magrans S/N, E-08193 Cerdanyola del Vallès, Catalonia, Spain}
\affiliation{Institut d'Estudis Espacials de Catalunya (IEEC), c/Gran Capita, 2-4, E-08034 Barcelona, Catalonia, Spain}

\author[0000-0002-7125-7685]{Patricio Sanhueza}
\affiliation{National Astronomical Observatory of Japan, 2-21-1 Osawa, Mitaka, Tokyo 181-8588, Japan}
\affiliation{Astronomical Science Program, The Graduate University for Advanced Studies, SOKENDAI, 2-21-1 Osawa, Mitaka, Tokyo 181-8588, Japan}

\author[0000-0002-6195-0152]{John J. Tobin}
\affiliation{National Radio Astronomy Observatory, 520 Edgemont Rd., Charlottesville, VA 22093, USA}

\author[0000-0003-2384-6589]{Qizhou Zhang}
\affiliation{Center for Astrophysics $\mid$ Harvard $\&$ Smithsonian, 60 Garden Street, Cambridge, MA 02138, USA}

\author[0000-0002-3466-6164]{James M. Jackson}
\affiliation{National Radio Astronomy Observatory, 520 Edgemont Rd., Charlottesville, VA 22093, USA}

\author[0000-0003-3172-6763]{Dominique Segura-Cox}
\affiliation{Center for Astrochemical Studies, Max Planck Institute for Extraterrestrial Physics, D-85748 Garching, Germany}
\affiliation{Department of Astronomy, University of Texas, 2515 Speedway, Stop C1400, Austin, TX 78712, USA}

\begin{abstract}

We present 870 $\mu$m polarimetric observations toward 61 protostars in the Orion molecular clouds, with $\sim$400 au (1$\arcsec$) resolution using the Atacama Large Millimeter/submillimeter Array.  
We successfully detect dust polarization and outflow emission in 56 protostars, in 16 of them the polarization is likely produced by self-scattering.
Self-scattering signatures are seen in several Class 0 sources, suggesting that grain growth appears to be significant in disks at earlier protostellar phases.
For the rest of the protostars, the dust polarization traces the magnetic field, whose morphology can be approximately classified into three categories: standard-hourglass, rotated-hourglass (with its axis perpendicular to outflow), and spiral-like morphology. 
40.0\% ($\pm$3.0\%) of the protostars exhibit a mean magnetic field direction approximately perpendicular to the outflow on several 10$^{2}$--10$^{3}$ au scales. However, in the remaining sample, this relative orientation appears to be random, probably due to the complex set of morphologies observed. Furthermore, we classify the protostars into three types based on the C$^{17}$O (3--2) velocity envelope's gradient: 
perpendicular to outflow, non-perpendicular to outflow, and unresolved gradient ($\lesssim$1.0~\kms ~arcsec$^{-1}$).
In protostars with a velocity gradient perpendicular to outflow, the magnetic field lines are preferentially perpendicular to outflow, most of them exhibit a rotated hourglass morphology, suggesting that the magnetic field has been overwhelmed by gravity and angular momentum.
Spiral-like magnetic fields are associated with envelopes having large velocity gradients, indicating that the rotation motions are strong enough to twist the field lines.
All of the protostars with a standard-hourglass field morphology show no significant velocity gradient due to the strong magnetic braking.

\end{abstract}


\section{Introduction} \label{sec:intro}

Magnetic fields (henceforth {\textit B}-fields) are thought to play a crucial role in the star-forming processes \citep[e.g.,][]{maury2022, pattle2023ppvii}.
In ideal MHD models, during the collapse phase of a prestellar core, the {\textit B}-field lines are drawn inward into an hourglass morphology, forming a pseudo-disk. They carry away angular momentum very efficiently, which ultimately leads to catastrophic magnetic braking, preventing the formation of a centrifugally supported disk \citep{ciolek1994ambipolar,allen2003collapse,galli2006gravitational}.
In real astrophysical environments, various non-ideal MHD effects may come into play, allowing the formation of an accretion disk around the central protostar \citep[][]{Dapp2010disk, Hennebelle2016diskform, Hennebelle2020diskform, Zhao2020disk}.
In addition, an initial misalignment between the \textit{B}-field and  the rotation axis leads to weaker magnetic braking of the collapsing core, enabling early disk formation \citep[e.g.,][]{joos2012protostellar, li2013misalignment, hirano2019Bmisalign, machida2020misalignment}. In any case, {\textit B}-fields are also thought to be important in the formation, collimation, acceleration, and regulation of outflows associated with protostellar systems \cite[e.g.,][]{pudritz2019role}.

Polarized dust continuum emission allows us to examine the {\textit B}-field morphology in star-forming regions.
In the presence of {\textit B}-fields, spinning and elongated dust grains with paramagnetic properties are expected to align their long axes perpendicular to the field direction \citep[e.g.,][]{hoang2009grain, andersson2015interstellar}.
In recent decades, dust polarization observations carried out with (sub)millimeter interferometers have increasingly demonstrated their effectiveness in mapping {\textit B}-fields at the scales of cores ($\sim 10^{4}$ au) and envelopes ($\sim10^{2}$ to $10^{4}$ au) \citep[e.g.,][]{girart1999detection,girart2013dr, cox2018alma,  galametz2018sma, hull2019interferometric, le2020IMS, cortes2021magnetic}.
Hourglass-shaped {\textit B}-fields have been observed around protostellar envelopes and trace gravitationally infalling material \citep[e.g.,][]{girart2006magnetic, girart2009magnetic, qiu2014submillimeter, beltran2019alma, le2019characterizing, hull2020understanding}.
Building upon these observations, some studies developed 3D analytical models of hourglass morphology \citep[e.g.,][]{myers2018magnetic,myers2020magnetic}, further reinforcing the prevalence of this structure.
However, interferometric polarization surveys toward low and high-mass protostellar at core and envelope scales found that the {\textit B}-field does not correlate with the axis of outflows.
This suggests that the {\textit B}-field may be less dynamically important than angular momentum and gravity \citep{hull2013misalignment, zhang2014magnetic}.

On the other hand, millimeter polarized emission in planet-forming disks is dominated by self-scattering of large dust grains \citep[$\sim$ several hundreds of microns,][]{kataoka2015millimeter, kataoka2016grain, yang2016inclination, yang2016disc}. Recent high-resolution observations of dust polarization have revealed polarization patterns that more closely align with self-scattering, rather than the signature expected from {\em B}-fields within protostellar and protoplanetary disks \citep{kataoka2017evidence, stephens2017alma, bacciotti2018alma, sadavoy2018dusta, cox2018alma, girart2018resolving, hull2018alma, yang2019does}.
Only a few observations show cases where the polarization is consistent with magnetically aligned dust grains on disk scales \citep{lee2018alma, sadavoy2018dustb, alves2018magnetic, ohashi2018two}.

The Orion molecular clouds (OMCs) are one of the best regions to study the role of {\textit B}-fields in the star formation process.
They are the closest \citep[$\sim$ 400 pc,][]{Kounkel2017dis} high-mass/intermediate-mass star-forming regions, and have been extensively studied across multiple wavelengths, providing a wealth of ancillary data and information.
Moreover, most solar-type stars, including the Sun, formed in massive, clustered star-forming regions \citep{lada2003embedded}, making the OMCs a more typical representation of star-forming conditions within the Galaxy.
Finally, the OMCs have the largest population of Class~0 protostars within 500~pc \citep{stutz2013discovery, furlan2016herschel}.
Class~0 protostars are exceptionally well-suited for studying the role of {\textit B}-fields in star formation due to their early evolutionary stage that preserves information about their initial collapse.

The {\textit B}-field Orion Protostellar Survey (BOPS) used ALMA to observe 870~$\mu$m dust polarization toward 61 young low-mass protostars in the OMCs. 
Its main objective is to investigate the role of {\textit B}-fields on spatial scales ranging from 400 to thousands of au, fully encompassing the protostellar envelope surrounding the youngest protostars.
The limited sensitivity of previous surveys \citep[e.g.,][]{hull2014tadpol, zhang2014magnetic} probed sources at varying evolutionary stages and in different star-forming regions, thus resulting in samples that were biased and non-uniform. To mitigate this, the BOPS observations uniformly probe {\textit B}-field structures within the envelopes surrounding protostars in one star-forming region.
In this paper, we present the first results of the BOPS project, organized as follows:
Section \ref{sec:Obs} introduces the observations and the processes of data reduction. The main results are presented in Section \ref{sec:Res}, followed by a detailed discussion of these results in Section \ref{sec:Dis}.
Finally, the summary is given in Section \ref{sec:Sum}.

\section{Observations and Data Reduction} \label{sec:Obs}

The BOPS (PI: Ian Stephens, 2019.1.00086) survey used ALMA Band 7 (870\,$\mu$m) to observe 57 fields, each centered on a different protostar as identified by the VLA/ALMA Nascent Disk and Multiplicity (VANDAM) Survey of Orion Protostars \citep{tobin2020vla}. The vast majority of these protostars are Class 0, though a few bright Class I protostars were also included in the sample. 
The names and coordinates of these protostars are listed in Table \ref{Tab:parameters} of Appendix \ref{App:Table}.
We targeted the brightest Class 0 sources using their VLA, C-array, 9 mm fluxes ($\sim 1\arcsec$ resolution). The sample size was selected to be approximately twice that of the TADPOL survey \citep{hull2014tadpol}, which included 30 sources throughout different star forming regions. Observations were made from November 29, 2019, to December 20, 2019, using the ALMA compact configurations C43-1 and C43-2, and an intermediate configuration of both, which provided baselines between about 14\,m and 312\,m. The observations were taken in Frequency Division Mode (FDM), providing modest spectral resolutions.
We used four spectral windows, two in each sideband, with the upper sideband targeting $^{12}$CO (3--2) and continuum, and the lower sideband targeting the continuum only. The maximum bandwidth (1.875~GHz per spectral window) was selected for the continuum, while a more modest bandwidth but higher spectral resolution was used for $^{12}$CO (3--2).
Notably, C$^{17}$O (3--2) in spectral window 4 was detected toward each protostar, which we used to trace the envelope kinematics.
Other molecular transition lines were detected because of the resolution provided by the ALMA FDM mode, but are not considered here.
The rest frequency, bandwidth, spectral resolution, and velocity resolution of each spectral window are listed in Table~\ref{Tab:spectralsetup} of Appendix~\ref{App:Table}.

The dust continuum images were produced using the \texttt{tclean} task in CASA \citep{casateam2022casa}, with a {\em Briggs} weighting {\em robust} parameter set to 0.5, which is a good compromise between resolution and sensitivity \citep{Briggs95}.
The extremely bright line $^{12}$CO (3--2) is flagged before self-calibration.
Then we performed three successive rounds of phase-only self-calibration for each source to improve the image quality.
The Stokes {\textit I} image was used as a model for self-calibration, with solution intervals set to 600, 30, and 10 seconds for the first, second, and third iterations, respectively.
To avoid the effects of bright lines on the dust continuum emission, channels exceeding 1.5 times the continuum baseline were flagged after the third round of self-calibration.
The final Stokes {\textit I}, {\textit Q}, {\textit U} continuum maps were produced independently using line-free and self-calibrated data of all spectral windows.

The self-calibrated continuum emission (Stokes {\textit I}) is strongly detected with a signal-to-noise ratio (S/N) ranging from 750 to 4200 for all targets.
The average noise level in the final Stokes {\textit I} image is $\sim0.1$~mJy~beam$^{-1}$, which is higher than the noise level of $\sigma_{QU} \sim0.07$~mJy/beam in both the Stokes {\textit Q} and {\textit U} maps.
This difference arises from the total-intensity image being more dynamic-range-limited than the polarized intensity images.
The debiased polarized intensity is defined as  $P=(Q^{2}+U^{2}-\sigma_{QU}^{2})^{1/2}$ \citep{vaillancourt2006placing, hull2015stokes}, using a 3$\sigma_{QU}$ cutoff value. The fractional polarization is derived as $P_{\text{frac}} = P/I$. The polarization angle is calculated as $\theta=0.5\cdot \arctan(U/Q)$, using a 3$\sigma$ value of the polarization intensity map as a threshold. 

For the line data, we first applied the self-calibration solutions to the entire dataset.
Then, the full, non-channel-averaged dirty image cubes were produced to identify the channels of continuum.
The \texttt{uvcontsub} task was used to perform continuum fitting and subtraction in the UV plane.
Finally, we performed the \texttt{tclean} task to produce Stokes {\textit I}, {\textit Q} and {\textit U} image cubes using self-calibrated, continuum-subtracted data for each spectral window. 

The $^{12}$CO (3--2) and C$^{17}$O (3--2) channel maps were used to identify the molecular outflows and the molecular emission from the envelope, respectively.
The envelope's centroid velocity ($V_{\rm LSR}$) was estimated by fitting a Gaussian to the spectrum obtained from averaging the C$^{17}$O (3--2) emission within a scale of 1000 au around the position of the protostar.  
The \texttt{immoments} task was used to generate the outflow maps. 
The blue- and red-shifted $^{12}$CO (3--2) outflow images were obtained using a velocity range starting from +/- 5 \kms with respect to $V_{\rm LSR}$ and included all channels where the emission was at least 5$\sigma$. 
However, in some sources (HOPS-78, HOPS-88, HOPS-124, HOPS-182, HOPS-310, HOPS-340, HOPS-341, HOPS-354, HOPS-361N, HOPS-361S, HOPS-370, HOPS-384, HOPS-399, and OMC1N-6-7), the channels near the V$_{\rm LSR}$ are significantly affected by the spatial filtering of the large scale emission or by optical depth effects.
In these cases we manually selected the channels.
The velocity field of the envelope was traced using the C$^{17}$O (3--2) line, specifically the moment 1 map.
This map covers all channels with emission exceeding 3$\sigma$.
Our sample included 9 protostars that had companions in one field (HH270IRS, HOPS-310, HOPS-317S, HOPS-354, HOPS-399, HOPS-400, HOPS-402, HOPS-403, and HOPS-404). 
Since C$^{17}$O emission is optically thin in the envelope region (see Appendix \ref{App:VG}), we select the stronger component and carefully choose which channels to include to minimize any effects from the weaker component.

\begin{figure*}
\centering
\text{\textbf{Perp-Type: velocity gradient direction $\perp$ outflow direction ($67.5^{\circ}\le \vert \theta_{\rm Out}-\theta_{v} \vert \le 90^{\circ}$)}}
\includegraphics[clip=true,trim=0cm 1.5cm 0cm 2.5cm,width=0.72 \textwidth]{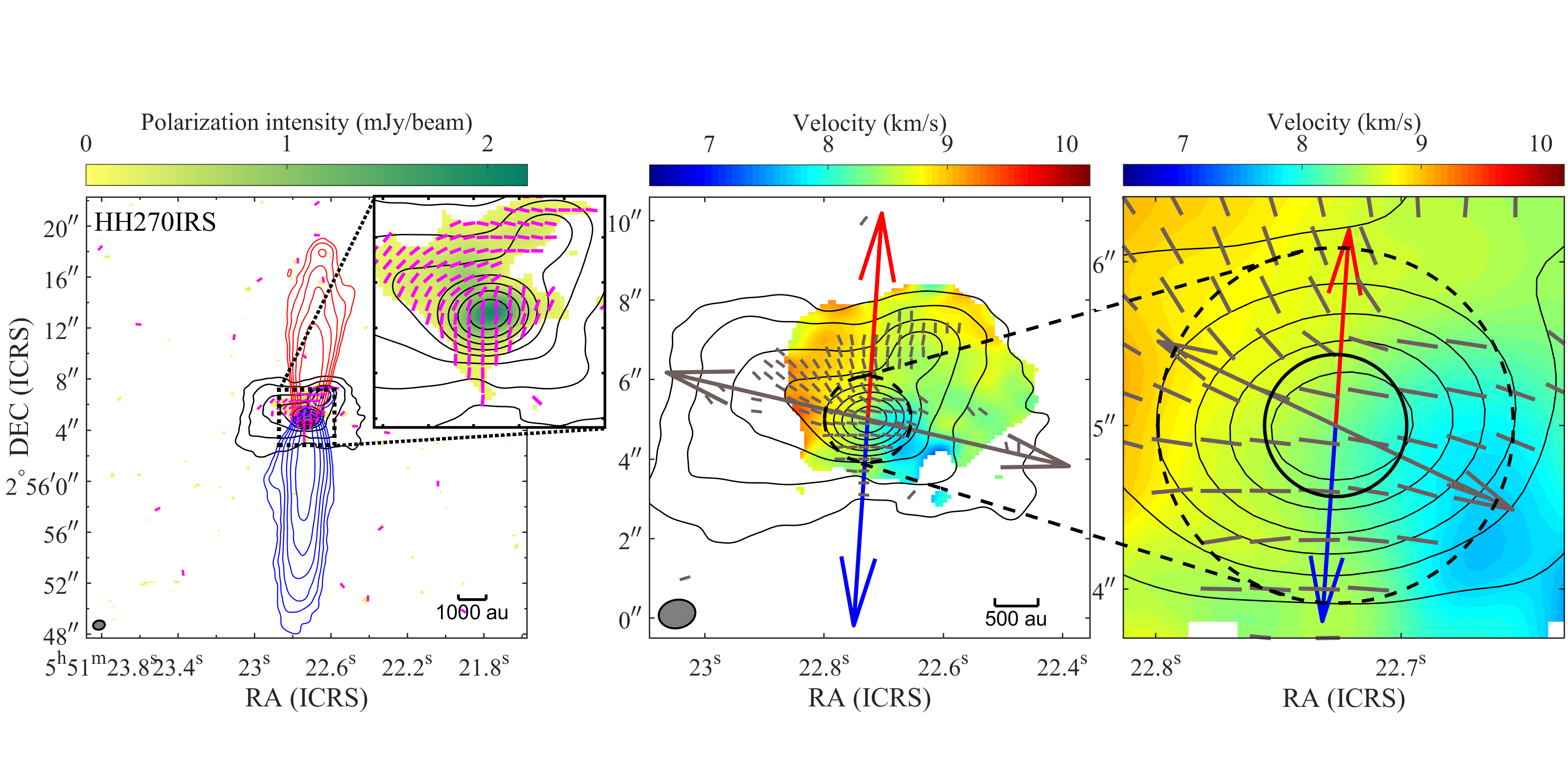}
\includegraphics[clip=true,trim=11.8cm -0.3cm 11.5cm 0cm,width=0.24 \textwidth]{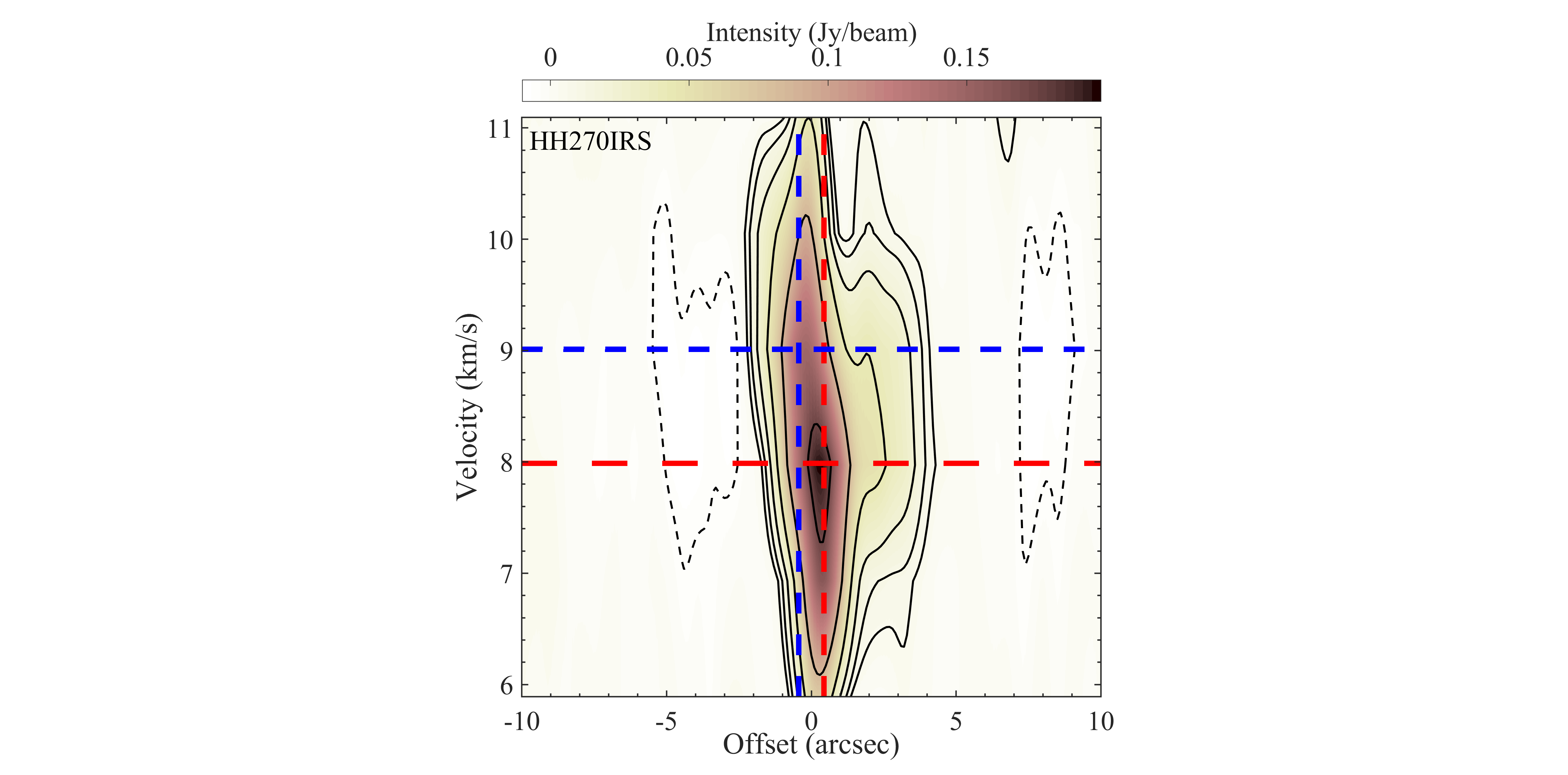}
\text{\textbf{Nonperp-Type: velocity gradient direction is not perpendicular to outflow direction ($\vert \theta_{\rm Out}-\theta_{v} \vert < 67.5^{\circ}$)}}
\includegraphics[clip=true,trim=0cm 1.5cm 0cm 2.5cm,width=0.72 \textwidth]{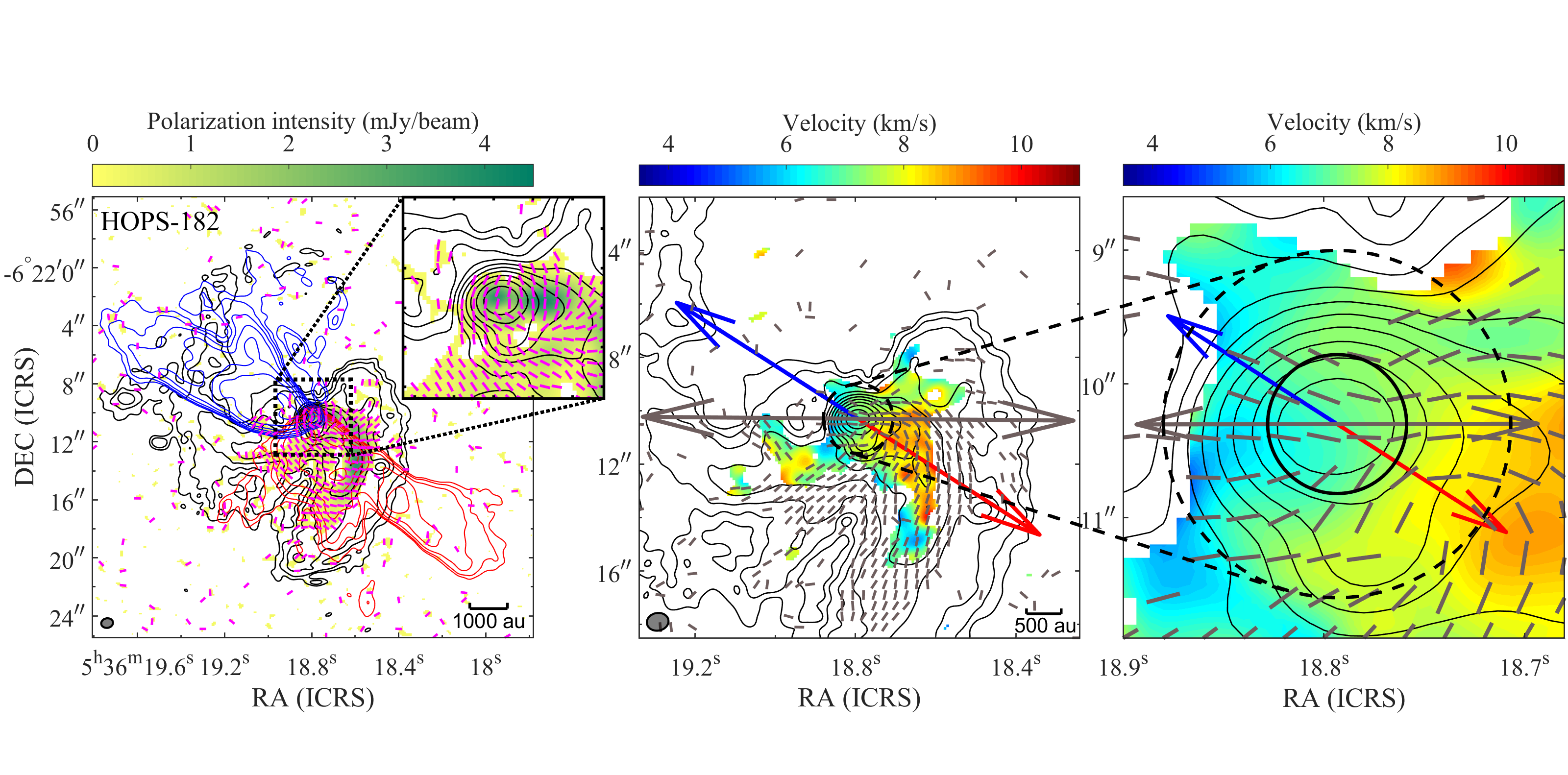}
\includegraphics[clip=true,trim=11.8cm -0.3cm 11.5cm 0cm,width=0.24 \textwidth]{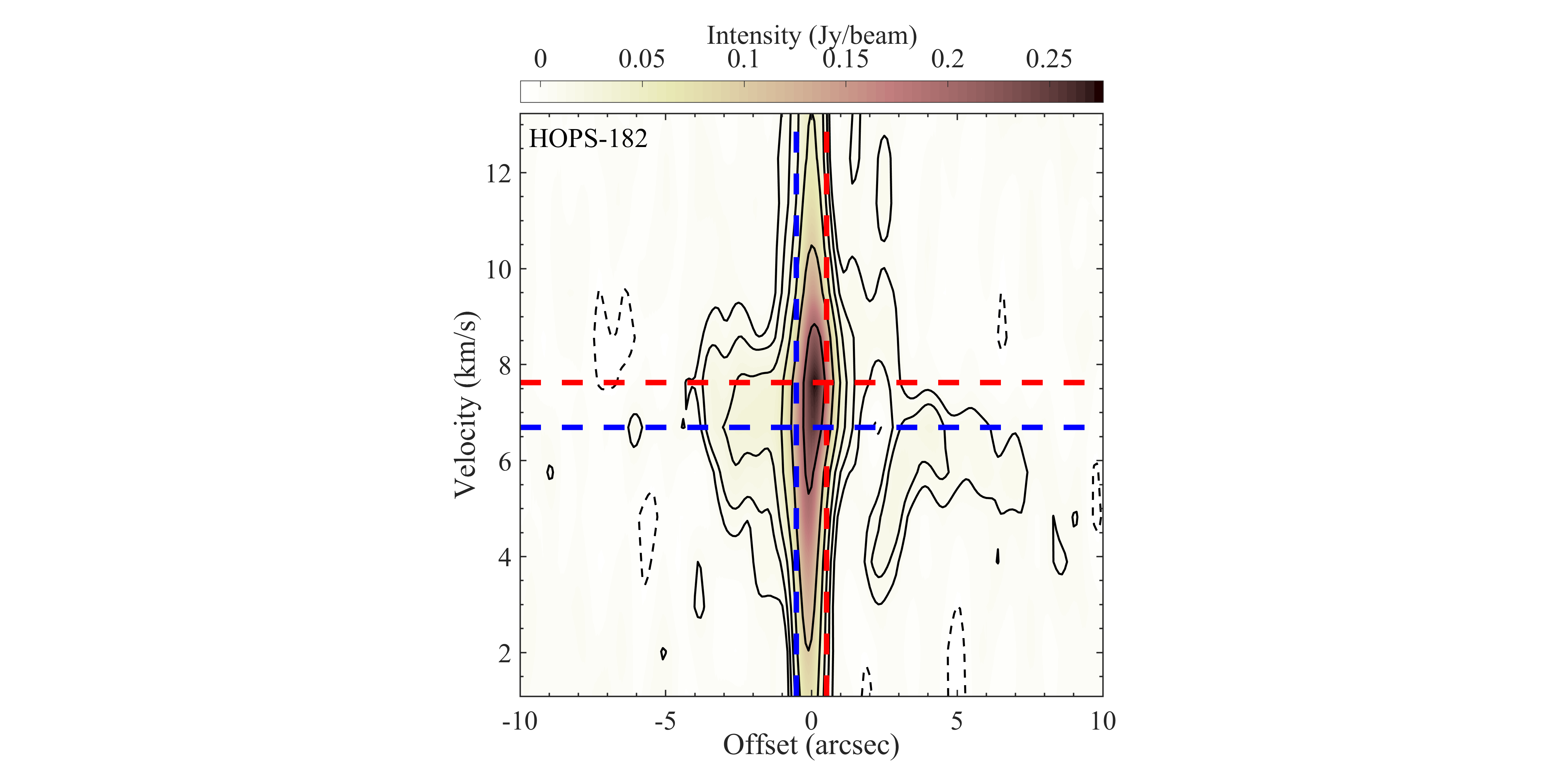}
\text{\textbf{Unres-Type: velocity gradient is unresolved}}
\includegraphics[clip=true,trim=0cm 1.5cm 0cm 2.5cm,width=0.72 \textwidth]{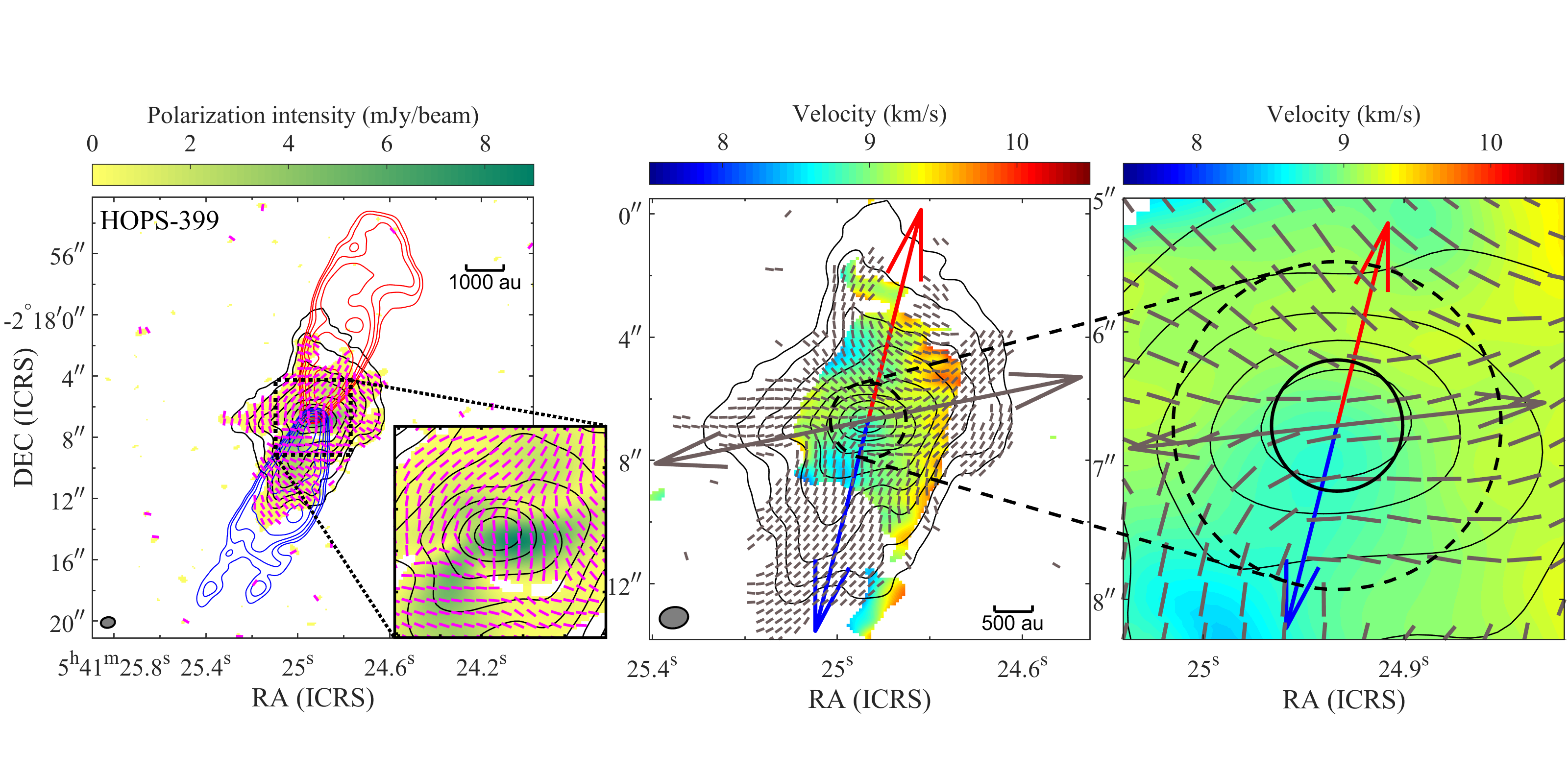}
\includegraphics[clip=true,trim=11.5cm -0.3cm 11.5cm 0cm,width=0.24 \textwidth]{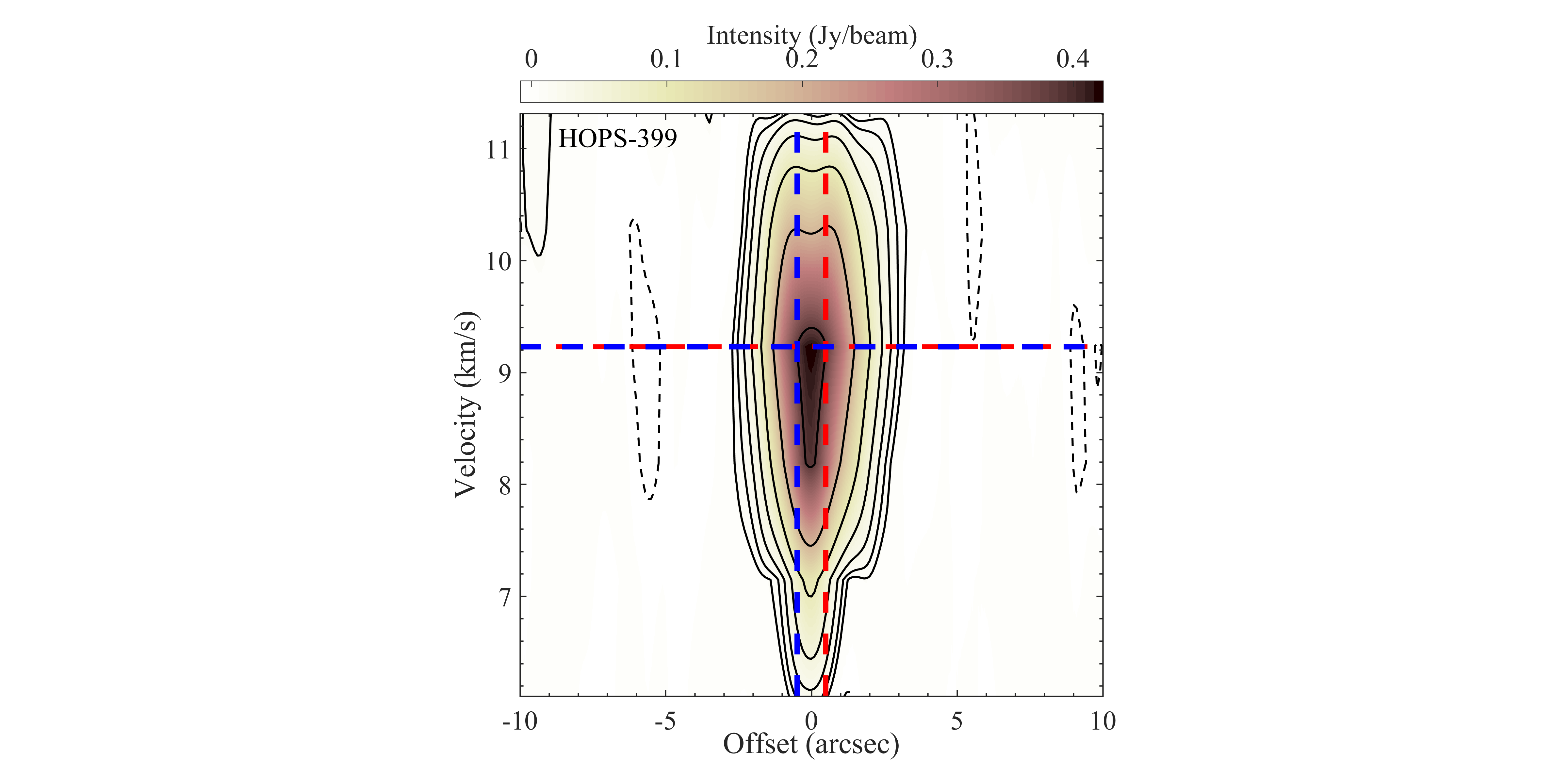}
\caption{Example maps in BOPS survey.
First column: 870~$\mu$m dust polarization intensity in color scale overlaid with the redshifted and blueshifted outflow lobes (obtained from the $^{12}$CO (3--2) line), polarization segments, and dust continuum emission (Stokes {\em I}) contours.
Blue contours indicate the blueshifted outflow, while red contours are redshifted outflow, with counter levels set at 5 times the outflow {\em rms} × (1, 2, 4, 8, 16, 32).
The magenta segments represent the polarization.
The regions of polarization intensity less than 3$\sigma$ have been masked.
Second column: the velocity field in color scale (obtained from the C$^{17}$O (3--2) line) overlaid with the {\textit B}-field segments (i.e., polarization rotated by 90\arcdeg) and Stokes {\em I} contours.
Third column: an enlarged perspective of 1000 au of the second column.
In the second and third panels,
the black segments represent the {\textit B}-fields.
The red and blue arrows indicate the mean direction of the red-shifted and blue-shifted outflows.
For the velocity field, regions with an S/N less than 4 have been flagged.
The grey arrows in the second and third columns indicate the mean \textit{B}-field directions weighted by the intensity including all the polarization segments, and weighted by the uncertainty within annular region of 400--1000 au, respectively.
In the first, second, and third columns, the black contour levels for the Stokes {\textit I} image are 10 times the {\em rms} × (1, 2, 4, 8, 16, 32, 64, 128, 256, 512).
The black dotted square in the first column corresponds to 2000 au scale, while the black dashed, and solid circles correspond to scales of 1000 au, and 400 au, respectively.
Fourth column: C$^{17}$O position-velocity (PV) diagram perpendicular to the outflow direction.
The color scale indicates the total intensity of C$^{17}$O line.
Black solid contour levels are 3 times of the intensity {\em rms} × (1, 2, 4, 8, 16, 32, 64), while dashed contours are set at 3 times of the intensity {\em rms} × (-1, -2, -4).
Vertical dashed lines represent a scale of 400 au, with a blue line indicating a blueshifted offset of 200 au and a red line indicating a redshifted offset of 200 au.
The red and blue horizontal lines indicate the velocity values corresponding to the positions with the highest intensity on the 200 au redshifted blueshifted scales, respectively.}
\label{Fig:SampleMaps}
\end{figure*}

\begin{figure*}
\centering
\includegraphics[clip=true,trim=0cm 0.2cm 0cm 0.5cm,width=0.48 \textwidth]{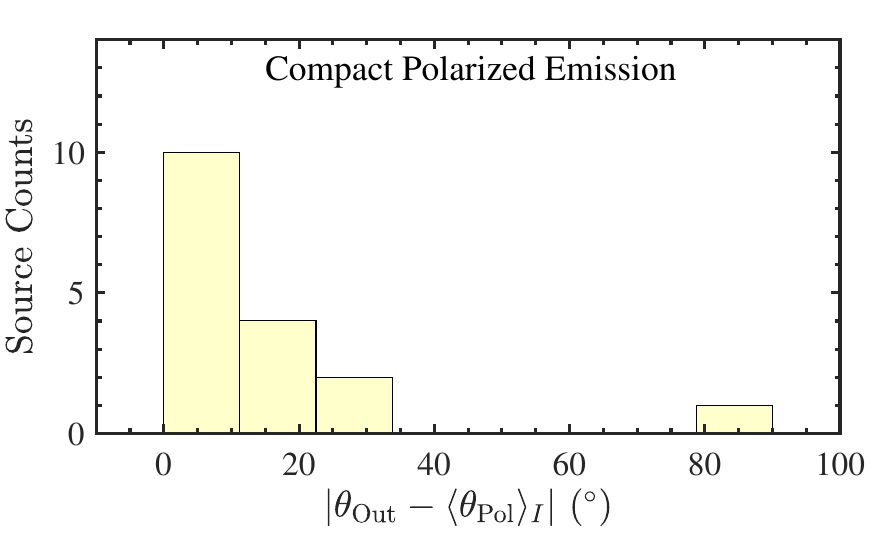}
\includegraphics[clip=true,trim=0cm 0.2cm 0cm 0.5cm,width=0.48 \textwidth]{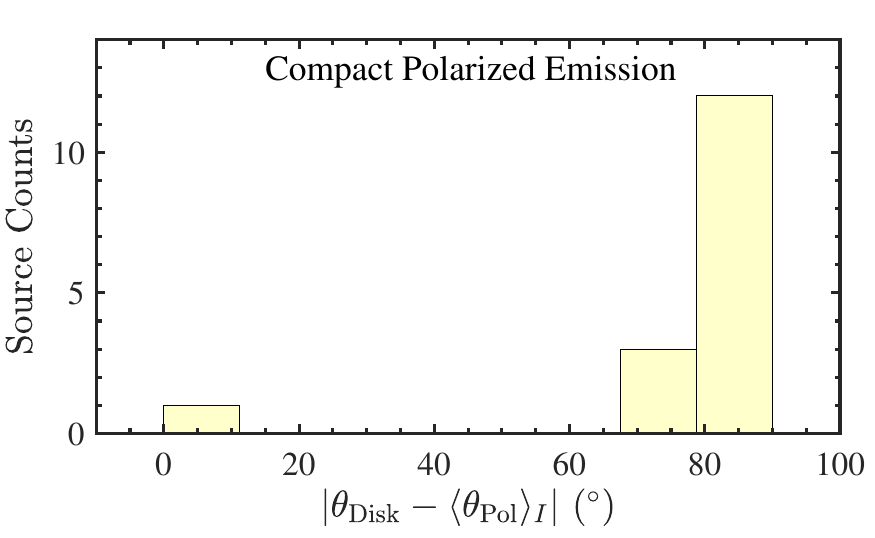}
~\\
~\\
\includegraphics[clip=true,trim=0cm 0.2cm 0cm 0.5cm,width=0.48 \textwidth]{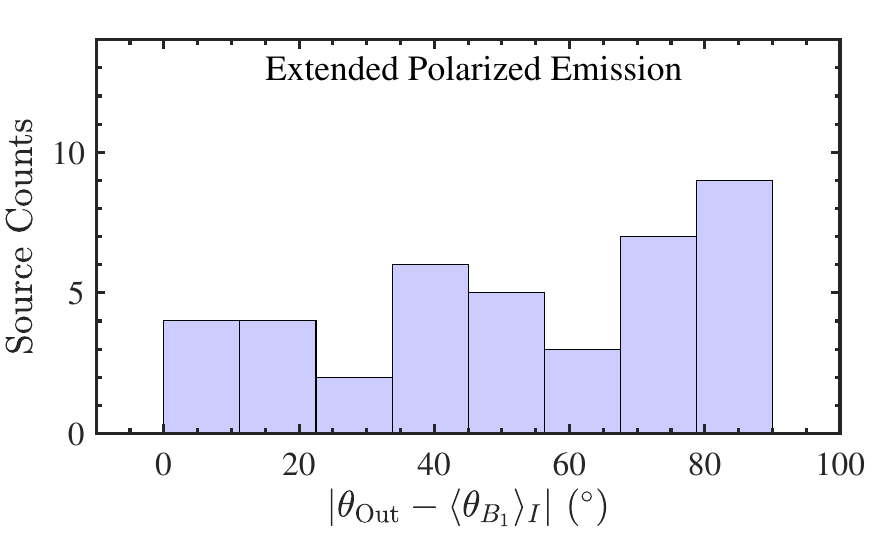}
\includegraphics[clip=true,trim=0cm 0.2cm 0cm 0.5cm,width=0.48 \textwidth]{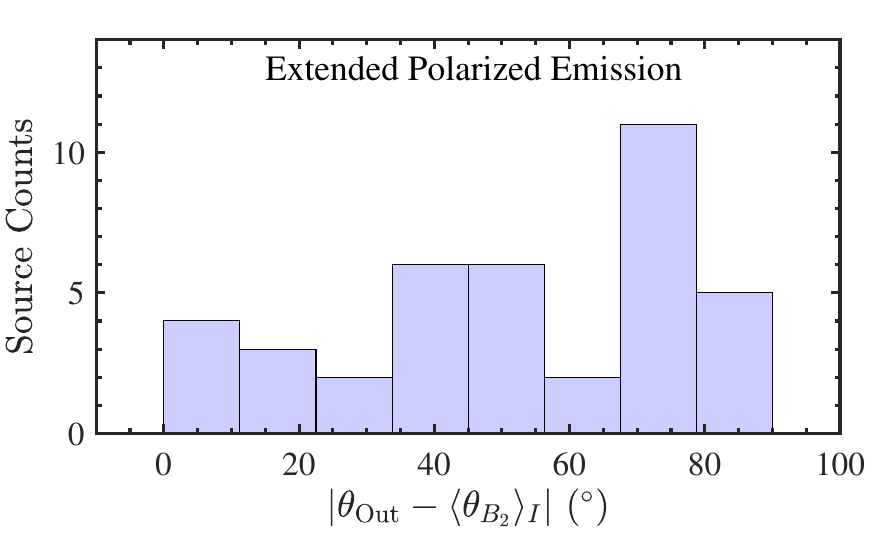}
\caption{
Histograms of different angle's differences for the 17 protostars with compact, barely resolved polarized emission thought to be related to dust polarization caused by dust self-scattering (upper panels), and for the rest of the protostars with extended polarized emission (lower panels).
Upper panels: histogram of angle's difference between the outflow 
direction $\theta_{\rm Out}$ and the mean polarization direction weighted by the intensity $\langle \theta_{\rm Pol} \rangle_{I}$ (left), and between the disk orientation along the major axis $\theta_{\rm Disk}$ and the mean polarization direction $\langle \theta_{\rm Pol} \rangle_{I}$ (right). In this right histogram, one disk, from HOPS-354, is not included because the disk is probably nearly face-on. 
Bottom panels: histogram of angle's difference between the outflow direction $\theta_{\rm Out}$ and the intensity-weighted average {\textit B}-field direction using all the detected polarized emission $\langle \theta_{B_{1}} \rangle_{I}$ (left), and within an annular region with an inner and outer radius of 400 and 1000 au $\langle \theta_{B_{2}} \rangle_{I}$ (right), respectively.
HOPS-250 is not included in the right panel because there is no 4$\sigma$ polarization detection within the annular region.
} 
\label{Fig:Histogram}
\end{figure*}

\section{Results} \label{sec:Res}

Figure \ref{Fig:SampleMaps} shows the results for typical protostars in the sample.
Plots for the entire sample are available in Appendix \ref{App:Figure}.
One field, OMC1N-4-5-W, does not show polarization emission nor CO (3-2) outflow emission, and the dust peak intensity is weak ($\sim$3~mJy~beam$^{-1}$), signifying it could be a starless core.
In the remaining fields, we successfully detected both dust polarization and molecular outflows in 56 sources, and the dust polarization but no clear outflow in 3 sources (HOPS-373E, HOPS-398, HOPS-402).
Additionally, we identified 2 sources (HOPS-87S and OMC1N-8-S) with outflow signature but no polarization detection, which will not be included in this paper.
Among these 56 protostars, 47 clearly exhibit the usual red and blue bipolar patterns, the remaining 9 protostars only show one distinct outflow component, with the other being very weak or overlapping with other outflow lobe.

The properties for the entire sample can be found in Table \ref{Tab:parameters} in Appendix \ref{App:Table}.
The position angles of polarization are determined using two different methods: the total-intensity-weighted average and the uncertainty-weighted average, respectively.
The position angles of polarization weighted by intensity $\langle \theta_{\rm Pol} \rangle_{I}$ and uncertainty $\langle \theta_{\rm Pol} \rangle_{\sigma}$ can be expressed as, respectively: 
\begin{eqnarray}\label{eq}
\langle \theta_{\rm Pol} \rangle_{I}=0.5\cdot {\rm arctan}  \biggl(\frac{D}{N}\biggr), \ \ \ \ \langle \theta_{\rm Pol} \rangle_{\sigma}=0.5\cdot {\rm arctan}  \biggl(\frac{K}{M}\biggr)
\end{eqnarray}
where $N$ and $D$ are intensity-weighted values of Stokes {\em Q} and {\em U}, $M$ and $K$ are uncertainty-weighted values of Stokes {\em Q} and {\em U}, respectively (see Appendix \ref{App:Bfield} for details). 
Note that the weighted polarization angle $\langle \theta_{\rm Pol} \rangle$ should be rotated by 90\arcdeg to infer the mean direction of {\textit{B}-field} $\langle \theta_{B} \rangle$ if the polarization arises from {\textit{B}-field} aligned grains.
The outflow direction is estimated by averaging the position angles of the redshifted and blueshifted emission, as discussed in Appendix \ref{App:Outflow}.
The disk's major axis angles are derived from high resolution ($0\farcs1$, $\sim$40 au) observations as reported in \cite{tobin2020vla}.
We excluded the disks with a low S/N ratio ($<5\sigma$ of integrated intensity)
and the almost face-on disks (with an inclination angle of $<$30\arcdeg).
These position angle of polarization $\langle \theta_{\rm Pol} \rangle$, {\textit{B}-field} $\langle \theta_{B} \rangle$, outflow $\theta_{\rm Out}$, and disk $\theta_{\rm Disk}$ are listed in Table \ref{Tab:PA} and Table \ref{Tab:SelfScattering} in Appendix \ref{App:Table}.

\subsection{Self-scattering} \label{Sec:selfscat}

Self-scattering polarization is observed parallel to the minor axis of an inclined disk \citep[e.g.][]{kataoka2015millimeter, kataoka2016grain, kataoka2017evidence, yang2016inclination, yang2016disc, stephens2017alignment}, thus it is expected to be aligned with the outflow direction.  
Within the BOPS sample with clear outflow detections, there are 17 protostars with compact polarized emission. 
We performed Gaussian fits to the polarized intensity and found that in each cases, the upper limits of the size of polarized emission are comparable to disk sizes, suggesting that the polarization arises from disk scales. 
The deconvolved major and minor axes from the Gaussian fits are consistent with the resolved disk sizes obtained from high resolution data by \cite{tobin2020vla}.
The upper panels in Figure \ref{Fig:Histogram} show the distribution of the difference between the polarization direction and the outflow direction (left-upper panel), and the disk orientation along the major axis (right-upper panel) in compact polarized sources.
In these two plots, almost all of the sources the polarization angle appears to be perpendicular to the disk major axis and parallel to the outflow. Only HOPS-250 shows a polarization angle perpendicular to the outflow.
Excluding this source, we find that the median difference between polarization direction and the the outflow direction is 7\arcdeg, and the median difference between polarization direction and the the disk major axis is 86\arcdeg. 
The fractional polarization of these 16 ptotostars is between 0.5\% to 2.0\%, suggesting that the compact polarization detected arises from self-scattering \citep[e.g.,][]{kataoka2016grain, yang2017scattering, girart2018resolving}.
The outflow position angle, polarization direction, and disk orientation of these sources are listed in Table \ref{Tab:SelfScattering}.

\subsection{{\textit B}-field at envelope scales} \label{Sec:bfielda}

\begin{figure*}
\centering
\includegraphics[clip=true,trim=4cm 2cm 4cm 1.2cm,width=0.98 \textwidth]{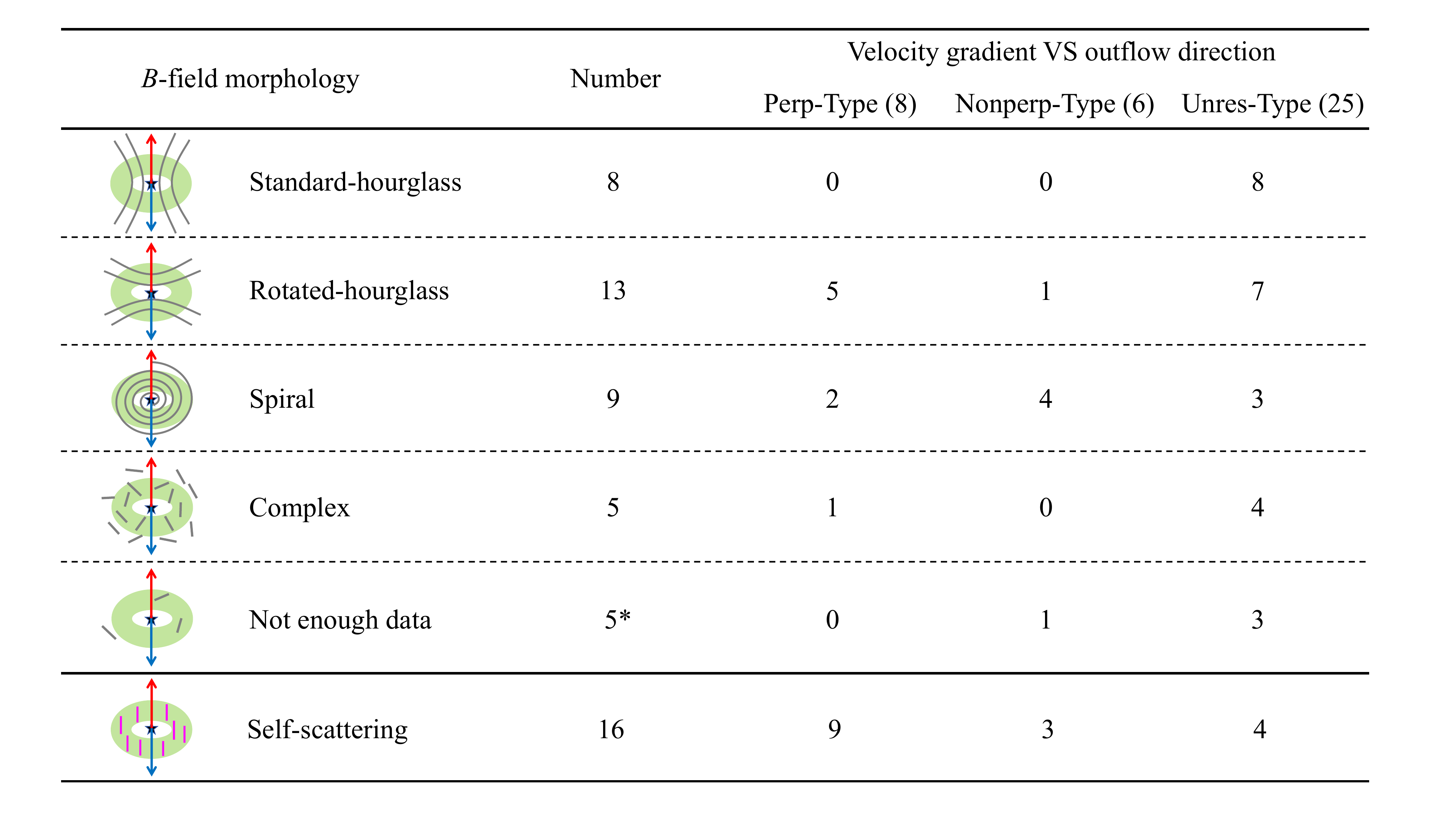}
\caption{Number of protostars for different types of \textit{B}-field morphology.
In the first column, green rings indicate protoplanetary envelope (disk for self-scattering), star symbols located at the center indicate protostars. Red and blue arrows indicate the outflow directions. Black curves and segments indicate the \textit{B}-field orientation, while magenta segments indicate polarization direction.
HOPS-250, does not have 4$\sigma$ polarization detection within annular region from 400 to 1000 au scale, is included in the type of Not enough data, which is marked by the asterisk symbol.} 
\label{Fig:BOPS_summary}
\end{figure*}

We are left with 40 protostars associated with a molecular and envelope polarization emission.
We do not expect self-scattering to be significant in the envelope, because the emission is more isotropic, and the grain size is smaller comparing to disk scale \citep{kataoka2016grain}. Thus, for these cases (including HOPS-250 with compact polarized emission), we assume that the polarization is produced by magnetically aligned grains.
Based on this, we classify each protostar based on its \textit{B}-field pattern, as seen in Figure \ref{Fig:BOPS_summary}.
We find that most of these targets can be classified into three main \textit{B}-field morphologies: standard-hourglass, rotated-hourglass and spiral (note that in many sources there are significant deviations from an ideal spiral shape).
In the standard-hourglass category, 8 protostars show the expected morphology \citep[e.g., ][]{girart2006magnetic}, in which 
their outflow is roughly parallel to the direction of the hourglass axis.
There are 13 protostars that exhibit a similar hourglass structure, but flipped by 90\arcdeg, such that its axis is perpendicular to the outflow, these are classified as rotated-hourglass.  
The spiral category encompasses 9 protostars with well-organized or partial spiral patterns in their {\textit B}-field structure \citep[as in, e.g., ][]{sanhueza2021gravity}.
In addition to these three categories, 5 protostars exhibit a {\textit B}-field pattern that is complex, and 5 do not have enough data (see Table \ref{Tab:PA}). 
In the case of a rotated-hourglass {\textit B}-field morphology, all protostars with this shape in our sample exhibit extended polarized emission, thus the potential ambiguity with self-scattering is not important.
Moveover, there are 6 protostars in our sample that have polarized emission that appears to be along streamer-like dust structures (HOPS-168, HOPS-182, HOPS-361N, HOPS-361S,  HOPS-370 and  OMC1N-8-N, as shown in Appendix \ref{App:Figure}). 
The \textit{B}-field in each of these protostars appears to follow the direction of this streamer-like structures.

As shown by the peripheral vectors in all panels of Figure \ref{Fig:SampleMaps} (particularly in the HOPS-182 field), 3$\sigma$ polarized emission in some regions may be generated by noise in the data, we therefore performed a total-intensity-weighted average {\textit B}-field using a 4$\sigma_{QU}$ threshold in our polarization emission data (see Appendix \ref{App:Bfield}).  The bottom-left panel of Figure~\ref{Fig:Histogram} shows the histogram of the angle difference between the outflow and these average {\textit B}-field directions.
The distribution appears almost random. However, there is a slight preference to the cases where the outflow is perpendicular to the {\textit B}-field direction: two-fifths of the protostars (16 out of 40) are located in the last quartile (angle difference in the $67\fdg5$--$90\arcdeg$ range).

\subsection{{\textit B}-field at scales of 400-1000 au} \label{Sec:bfieldb}

The use of the intensity-weighted average {\textit B}-field direction, as used by \citet{hull2014tadpol}, favors the polarization signal around the dust peak intensity, which may have significant contributions from both total and polarized emission in the circumstellar disk.
In the BOPS sample, the disks have radii between 30 and 250 au (0$\farcs$08 and 0$\farcs$63) \citep{tobin2020vla}. 
To avoid possible contamination from disk self-scattering polarization (see Section \ref{Sec:selfscat}), we select an annular region from 1$\arcsec$ ($\sim$400 au) to 2$\farcs$5 ($\sim$1000 au).
Within this region the difference between the outflow and the intensity-weighted average {\textit B}-field direction  appears to be very similar to the one obtained using all polarization data (as shown in bottom panels of Figure~\ref{Fig:Histogram}).

The envelope kinematics are traced by optically thin C$^{17}$O emission (see Appendix \ref{App:VG}).
The envelope's angular momentum is expected to be parallel to the outflow \citep{pudritz2019role}.
We generated position-velocity (P--V) cuts from the C$^{17}$O (3--2) channel maps centered at the dust peak intensity and perpendicular to the outflow axis. We calculated the absolute velocity gradient ($\vert \nabla v_{\rm {abs}} \vert$) in the P--V image by selecting the most red-/blue-shifted emission at 5$\sigma$ at a distance of 400~au from the protostar position (as shown in column 8 of Table \ref{Tab:PA}).
We found significant velocity gradients (i.e., $\vert \nabla v_{\rm {abs}} \vert \gtrsim$ 1.0~\kms~arcsec$^{-1}$) toward 14 protostars, which is probably indicative of rotation.  However, most of the sources do not show a clear gradient, which may due to the limited spectral resolution of the observations ($\sim$0.9 \kms).
In addition, we use the moment 1 (intensity-weighted velocity maps) of the  C$^{17}$O (3--2) emission to derive the direction of the velocity gradient. To do so, we fit a 2D linear regression fit to the moment 1 map following \cite{goodman1993dense} and \cite{tobin2011ism}. Further details regarding envelope kinematics are discussed in Appendix \ref{App:VG}.
Column 7 in Table \ref{Tab:PA} lists the velocity gradient position angle.
In cases without a significant velocity gradient, we use the spectral resolution divided by the angular resolution as an upper limit. 
We use the velocity gradient position angle $\theta_{v}$ and outflow direction $\theta_{\rm Out}$ (as depicted in Figure \ref{Fig:SampleMaps}) to classify the protostars into three types. This is done for 39 protostars where significant polarization data is detected within the annular region of 400 to 1000 au:
\begin{enumerate}
\item Perp-Type: velocity gradient is perpendicular to the outflow ($67\fdg5\le \vert \theta_{\rm Out}-\theta_{v} \vert \le 90\arcdeg$, 8 out of 39);
\item Nonperp-Type: velocity gradient is not perpendicular to the outflow ($\vert \theta_{\rm Out}-\theta_{v} \vert < 67\fdg5$, 6 out of 39);
\item Unres-Type: velocity gradient is unresolved ($\vert \nabla v_{\rm {abs}} \vert \lesssim 1.0$~km~s$^{-1}$~arcsec$^{-1}$, 25 out of 39).
\end{enumerate}

\begin{figure*}
\centering
\textbf{Perp-Type: velocity gradient direction $\perp$ outflow direction ($67.5^{\circ}\le \vert \theta_{\rm Out}-\theta_{v} \vert \le 90^{\circ}$)}
\includegraphics[clip=true,trim=0cm 0.2cm 1cm 0.5cm,width=0.48 \textwidth]{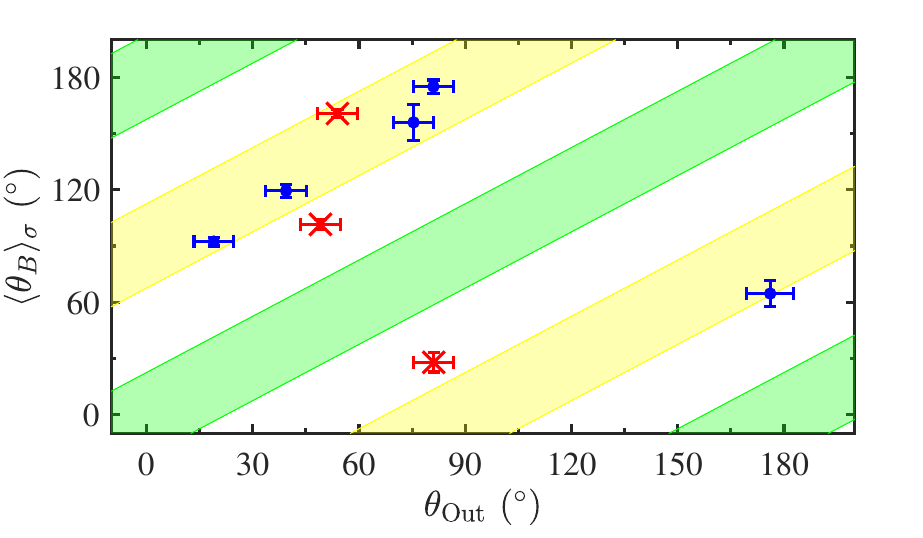}
\includegraphics[clip=true,trim=0cm 0.2cm 1cm 0.5cm,width=0.48 \textwidth] {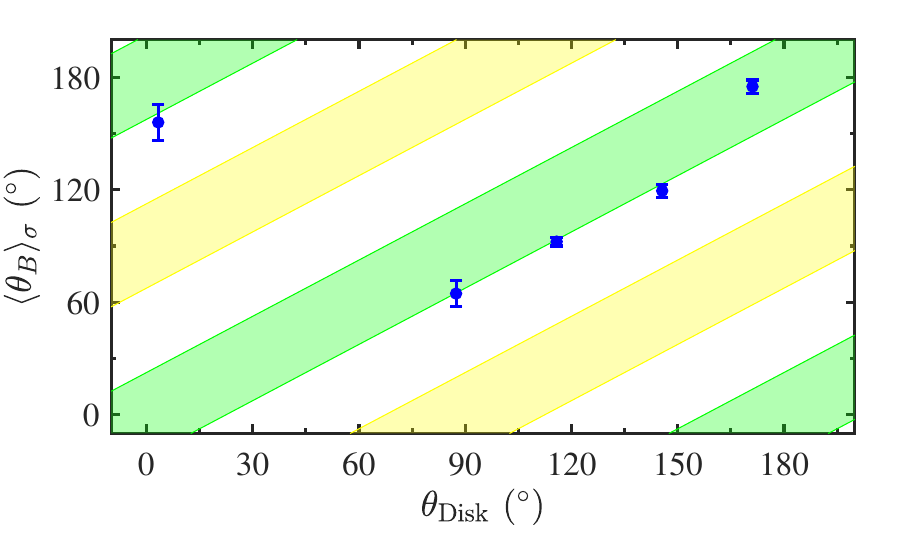}
\\~
\\~
\textbf{Nonperp-Type: velocity gradient direction is not perpendicular to outflow direction ($\vert \theta_{\rm Out}-\theta_{v} \vert < 67.5^{\circ}$)}
\includegraphics[clip=true,trim=0cm 0.2cm 1cm 0.5cm,width=0.48 \textwidth]{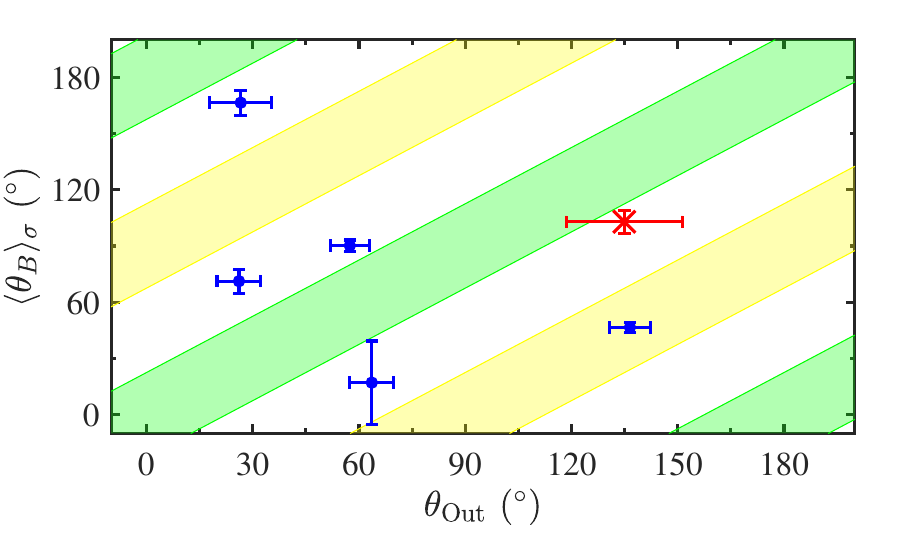}
\includegraphics[clip=true,trim=0cm 0.2cm 1cm 0.5cm,width=0.48 \textwidth]{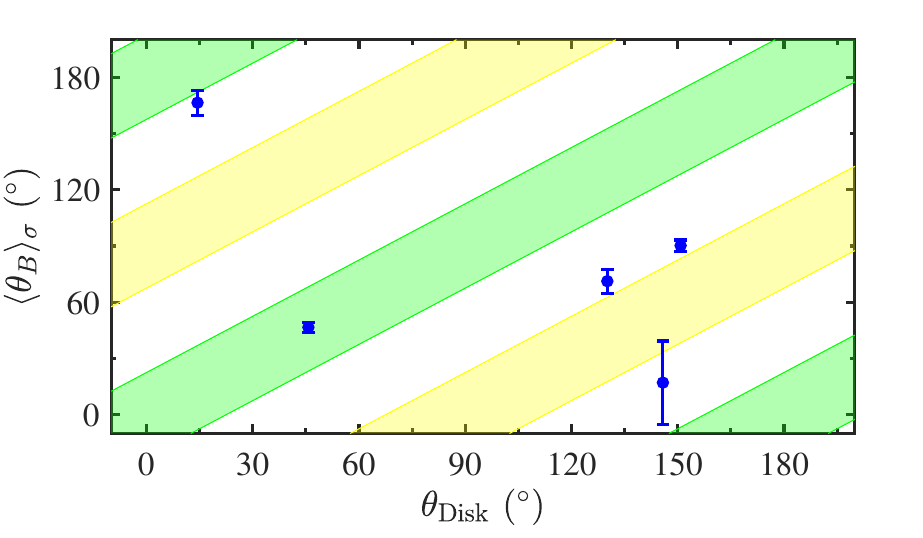}
\\~
\\~
\textbf{Unres-Type: velocity gradient is unresolved}\\
\includegraphics[clip=true,trim=0cm 0.25cm 1cm 0.5cm,width=0.48 \textwidth]{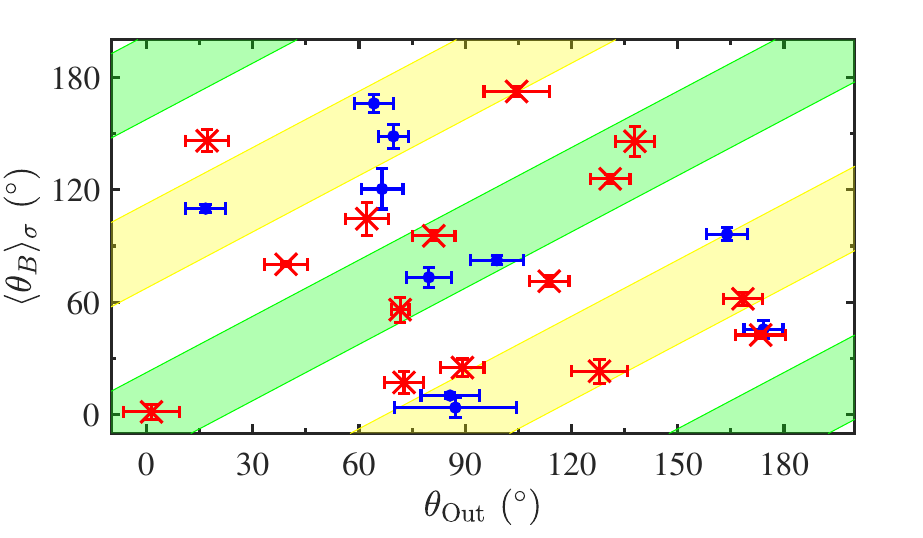}
\includegraphics[clip=true,trim=0cm 0.25cm 1cm 0.5cm,width=0.48 \textwidth]{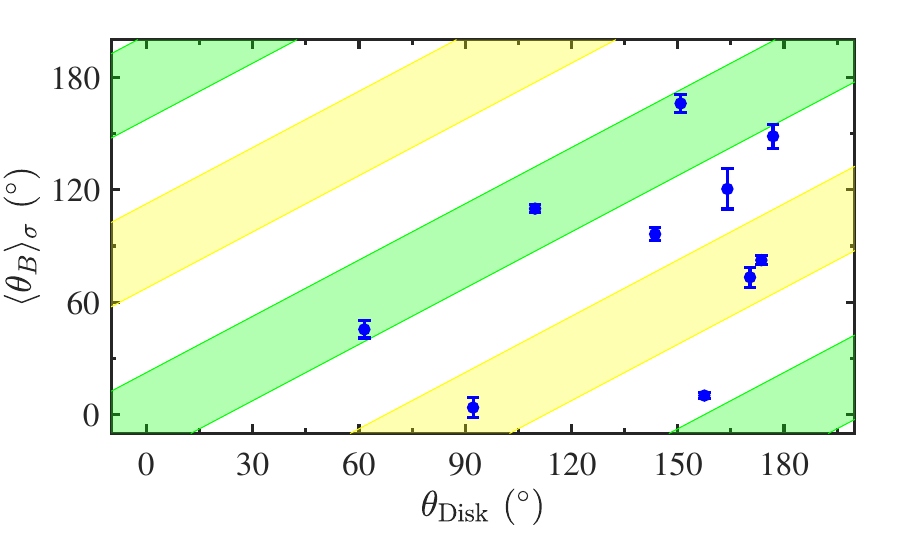}
\caption{Distribution between the uncertainty-weighted mean {\em B}-field direction $\langle \theta_{B} \rangle_{\sigma}$ and outflow direction $\theta_{\rm Out}$ (left panel), and between the mean {\em B}-field $\langle \theta_{B} \rangle_{\sigma}$ and disk orientation $\theta_{\rm Disk}$ (right panel). 
In both panels, yellow stripes indicate cases where the two vectors are perpendicular to each other ($67\fdg5\le \vert \theta_{\rm Diff}\vert \le 90\arcdeg$), while green stripes represent cases where the two vectors are aligned ($\vert \theta_{\rm Diff} \vert \le 22\fdg5$).
Blue dots and red crosses indicate protostars with and without clear disk orientation, respectively.
}
\label{Fig:Distribution}
\end{figure*}

We use the uncertainty-weighted position angles of the {\em B}-field on scales of 400--1000 au (represented as $\langle \theta_{B} \rangle_{\sigma}$) for the following analysis, which are listed in the sixth column of Table \ref{Tab:PA}.
The left panel in Figure \ref{Fig:Distribution} shows the distribution of the {\em B}-field and outflow direction for the three source types.
We find no significant relation between the average {\em B}-field and the outflow directions in Nonperp-Type and Unres-Type.
However, in Perp-Type sources, a correlation between the {\em B}-field and outflow is evident, with a correlation coefficient of 0.83 between $\vert \langle \theta_{B} \rangle_{\sigma}-90\arcdeg \vert$ and outflow position angle $\theta_{\rm Out}$. Specifically, 75.0\% ($\pm$12.5\%) of the sources show a trend that the {\em B}-field is perpendicular to the outflow.

We obtained the disk orientation of 20 sources from \cite{tobin2020vla}, 5 in Perp-Type, 5 in Nonperp-Type, and 10 in Unres-Type.
As shown in the right panel of Figure \ref{Fig:Distribution}, due to the small sample size in each group, it is difficult to characterize the correlation between the {\em B}-field and and the disk orientation.
Globally, there seems to be random alignment between the \textit{B}-field and the disk orientation within the 20 sources. However, we find some trends for the Perp-Type sources with almost all showing parallel alignment between the mean \textit{B}-field and disk orientation, with a correlation coefficient of 0.96.

\subsection{Geometric projection effect}

\begin{figure*}
\centering
\includegraphics[clip=true,trim=0cm 0cm 0cm 0cm,width=0.55 \textwidth]{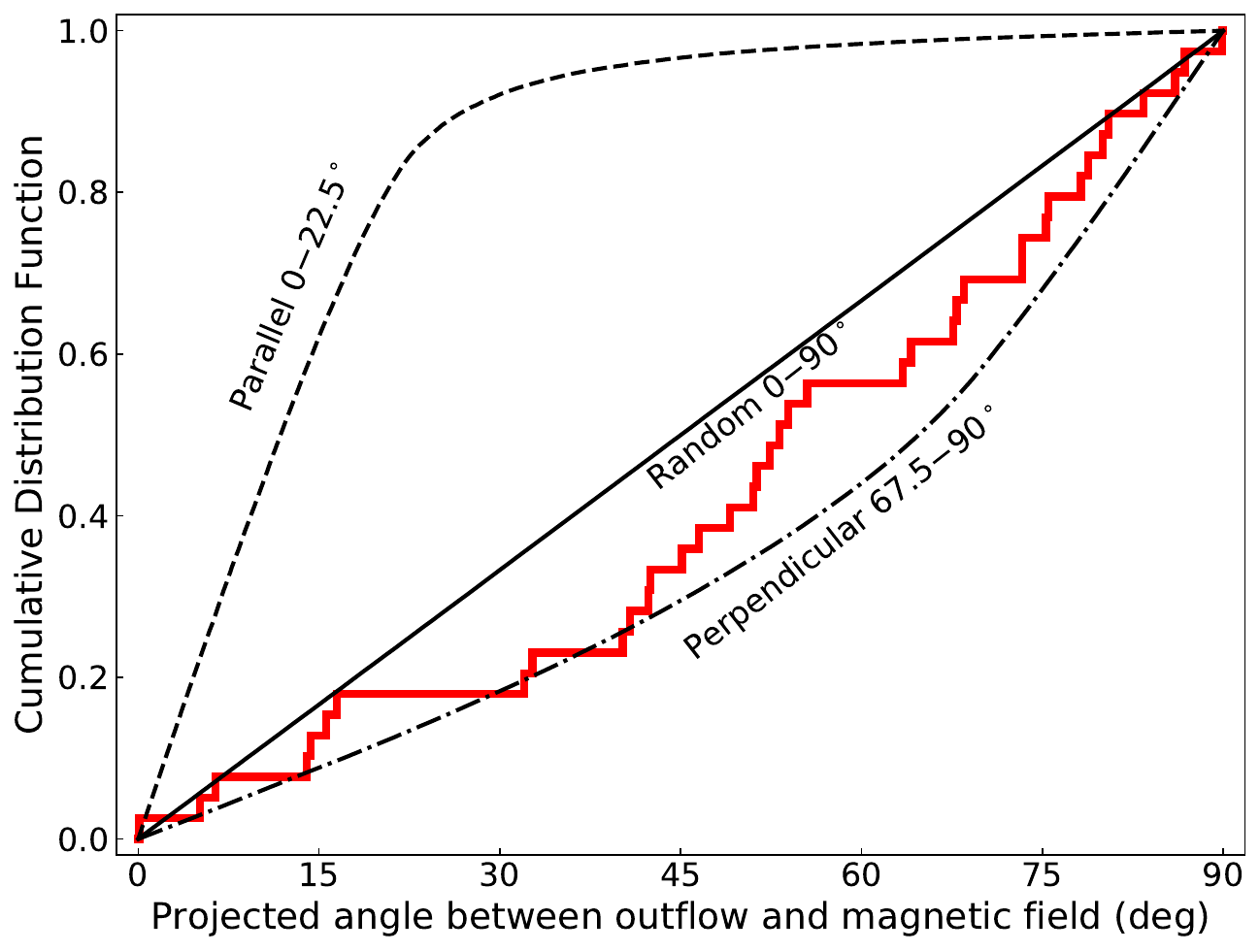}
\caption{Cumulative distribution function of the projected angles.
The red polyline is the observed angle difference between the outflow direction $\theta_{\rm Out}$ and uncertainty-weighted mean \textit{B}-field direction $\langle \theta_{B}\rangle_{\sigma}$.
The black solid line, dashed curve, and dotted-dashed curve indicate Monte Carlo simulations of the expected projected angle for two vectors that are 3D random (difference angle of two vector is between 0\arcdeg and 90\arcdeg), parallel (0\arcdeg-22\fdg5), and perpendicular (67\fdg5-90\arcdeg), respectively.} 
\label{Fig:CDF}
\end{figure*}

The position angles of the outflow and {\em B}-field we measured are projected onto the plane of the sky.
To investigate the influence of projection effects on our results, we plot the cumulative distribution function of the observed angle difference \citep[CDF,][]{hull2013misalignment, hull2014tadpol, stephens2017alignment} and compare it with 2D simulated intersecting angles uniformly projected from a three-dimensional space.
The parallel case uses the angles randomly selected within 0 and 22\fdg5 in 3D, and then project onto the plane of the sky in the simulation, while the range of 3D angles used in random and perpendicular cases are 0\arcdeg-90\arcdeg and 67\fdg5-90\arcdeg, respectively.
For simplicity, it is prudent to exclusively employ the 3D projection analysis on the entire sample only (i.e., not in any of the other subsamples analyzed in this work), represented in Figure \ref{Fig:CDF}.

We run the Kolmogorov–Smirnov (K-S) test on the angle difference distribution of our sample.
When considering the difference between the {\em B}-fields and outflow directions in the entire 39 sources, we compared the observed distribution with the simulated models.
The probability of our results being drawn from parallel models is $<$0.001, ruling out this scenario.
The probability that the distribution is drawn from a random model is 0.325, and from a perpendicular model is 0.615. Though the probability is higher for the perpendicular case, each model could be possible for the distribution.

\section{Discussion} \label{sec:Dis}

The polarization properties in approximately one-third of the sample, 16, strongly suggest that they are tracing self-scattering in disks.
Most of these targets are Class 0 (except for the Class I source of HOPS-84), and their bolometric temperature is similar to protostars without self-scattering (see Table \ref{Tab:parameters}), indicating that self-scattering is independent of evolutionary stages.  This result suggests that grain growth has already occurred in the disks in the earliest stages of the protostellar phase. 
It is possible that other sources in the sample also have polarized emission from self-scattering, but this would only be apparent around the intensity peak. Higher angular resolution observations are needed to properly quantify disk self-scattering in the full sample \citep[see ][]{liu2023omc3}.

Past interferometric polarization surveys using CARMA probed 16 protostars \citep{hull2013misalignment} and 30 star-forming cores \citep{hull2014tadpol} and showed that the {\em B}-fields are randomly oriented with outflows on a scale of a few thousands au.
Subsequently, \cite{zhang2014magnetic} presented SMA observations of 14 massive clumps.
They also found that the {\em B}-fields, on core scale of 0.01--0.1 pc, do not correlate with the outflow direction.
\cite{arce2020outflows} found the same results using 1 mm ALMA observations toward 29 protostellar dense cores in high-mass star-forming regions at a scale of 2700 au.
In our study, both, the histogram of the difference between the outflow and the averaged \textit{B}-field direction (bottom panels of Figure~\ref{Fig:Histogram}) and the CDF of these angle differences (Figure~\ref{Fig:CDF}) show that the \textit{B}-field is almost perpendicular to the outflow axis in a significant number of protostars. 
However, there is also a large fraction of protostars where the distribution of the angle difference appears random. 
These results of relative orientation probably depend on the sample of observed \textit{B}-field morphologies.
Other causes may also explain the differences with previous results in the comparison between the mean \textit{B}-field and the outflow axis. The objects in \cite{zhang2014magnetic} are at much greater distances, so they trace much larger linear scales of the \textit{B}-fields. \cite{hull2013misalignment} had a smaller sample of protostars that spanned multiple cores and lacked sensitivity, thus mapping fewer \textit{B}-field vectors.
Nevertheless, our results are in agreement with the recent numerical study done by \cite{machida2020misalignment}. 
They found that angular momentum, outflow and local \textit{B}-field axes depend on the initial angle difference between the angular momentum and the \textit{B}-field axes, as well at which scales these  directions are measured.
\cite{galametz2020observational} found, in a sample of 20 Class 0 protostars, that the misalignment angle of the \textit{B}-field orientation with the outflow is strongly correlated with the amount of angular momentum on 1000~au scale. We find similar results  for the sources with strong velocity gradients, 7 of 14 (the Perp-Type and the Nonperp-Type) the \textit{B}-field direction is perpendicular to the outflow axis.

Several BOPS protostars shows the expected hourglass \textit{B}-field along the outflow axis. 
This morphology is predicted in the classical models of star formation where the core is initially threaded with a uniform \textit{B}-field, and turbulence and angular momentum are not dynamically important \citep[e.g.,][]{galli2006gravitational}.
All the protostars with the standard-hourglass shape in our sample are observed in the Unres-Type, possibly because the \textit{B}-field is strong enough to slow down the rotation in this case, or because the core's initial angular velocity was energetically less important than the \textit{B}-field.
Most of the protostars with strong envelope velocity gradients (Perp-Type and Nonperp-Type) appear to have a rotated-hourglass or a spiral \textit{B}-field morphology. 
The rotated-hourglass shape tends to have field lines parallel to the velocity gradients and perpendicular to the outflows, however, in most cases of the spiral shape, the envelope \textit{B}-field does not align with either the velocity gradient or the outflow. 
In the case of the spiral morphology, this shape could be due to an initial misalignment between \textit{B}-fields and rotation. 
MHD simulations show that this shape generates magnetic torques, which creates two inflowing spirals aligned with the \textit{B}-field \citep{wang2022magnetic}.
This is what is observed in HOPS-182, HOPS-361S, and possibly in HOPS-361N \citep[and previously in IRAS~18089–1732: ][]{sanhueza2021gravity}. 
The rotated-hourglass could be the extreme case of the standard hourglass, where gravity is so strong \citep[i.e., with a relatively high initial mass-to-flux ratio, see][]{maury2018magnetically} that the bending of the lines appear perpendicular to the outflow axis, but this may happen only in the innermost part of the envelope as 
seen in B335 \citep{maury2018magnetically} and L1448 IRS 2 \citep{kwon2019highly}. 
Alternatively, this could trace the transition from the standard poloidal hourglass to a toroidal \textit{B}-field due the effect of an initially large angular momentum \citep[e.g.,][]{machida2007magnetic}.

Finally, there are few protostars that clearly show filament- or streamers-like dust structures
with the \textit{B}-field along the filament (HOPS-168, HOPS-182, HOPS-361N, HOPS-361S,  HOPS-370 and  OMC1N-8-N, as shown in Appendix \ref{App:Figure}). 
We speculate that these could be accretion streamers \cite[e.g.,][]{alves2020case, pineda2020protostellar, fernandez2023},
however, higher spectral resolution observations are needed to confirm this.

\section{Summary} \label{sec:Sum}

We have presented the first results of the BOPS survey, which targets 61 protostars in the OMCs star-forming complex. 
The outflow structure was traced using $^{12}$CO (3--2), while the velocity field in the dense region was mapped using C$^{17}$O (3--2).
We detect 870~$\mu$m dust polarization emission and outflows around 56 protostars. The main results are:
\begin{enumerate}
\item Self-scattering is observed in 16 sources, most of them are Class 0, indicating that grain growth in disks occurs in the very early stages of the disk evolution.
\item Dust polarization traces the {\textit B}-field in 40 protostars at envelope scales (up to $\sim$ 3000 au). Most of these targets can be classified into three major {\textit B}-field morphologies:  
the standard-hourglass, rotated-hourglass (which may be a highly pinched, standard hourglass), and a spiral configuration.
These morphologies are the result of the complex interplay between gravitational collapse, {\textit B}-fields and rotational motions during the star formation process.
The {\textit B}-fields aligned with the filament- or streamer-like structure could be related with accretion streamers, but high spectral resolution observations are needed to confirm this scenario.
\item Two-fifths of our sample exhibit an averaged {\textit B}-field that is perpendicular to the outflow, however the remaining three-fifths of our sample have a random relative orientation. 
On scales of 400 to 1000~au, the sources with a strong velocity gradient perpendicular to the outflow axis (Perp-Type) have a {\textit B}-field that is also perpendicular to the outflow axis.
\item Most of the protostars with strong velocity gradients $\vert \nabla v_{\rm{abs}}\vert \geq 1.0$ \kms arcsec$^{-1}$ (Perp-Type and Nonperp-Type) tend to have a rotated-hourglass or spiral \textit{B}-field morphology.
In rotated-hourglass \textit{B}-field morphologies, the {\textit B}-field strength is probably less significant with regard to gravity and angular momentum.
While in spiral structures, the rotation motions seem to be strong enough to twist the field lines, contributing to a helical {\textit B}-field morphology.
Notably, all of the sources with \textit{B}-field patterns showing a standard-hourglass structure of {\textit B}-field are in the Unres-Type, i.e., sources without strong velocity gradients probably due to the magnetic braking. 
\end{enumerate}

In summary, three main \textit{B}-field configurations are observed in our study, the rotated-hourglass field shape is more prevalent when the {\em B}-field strength is less significant compared to gravity and angular momentum, while strong rotation tends to give rise to spiral field structures.
In contrast, standard-hourglass field patterns are more commonly observed in sources lacking strong velocity gradients.

\begin{acknowledgments}
We thank Jacob Labonte for early analysis of the BOPS data.
B.H., J.M.G. and A.S.-M.\ acknowledge support by the grant PID2020-117710GB-I00 (MCI-AEI-FEDER, UE). 
B.H. also acknowledges financial support from the China Scholarship Council (CSC) under grant No. 202006660008.
This work is also partially supported by the program Unidad de Excelencia María de Maeztu CEX2020-001058-M.
A.S.-M.\ acknowledges support from the RyC2021-032892-I grant funded by MCIN/AEI/10.13039/501100011033 and by the European Union `Next GenerationEU’/PRTR.
E.G.C. acknowledges support from the National Science Foundation through the NSF MPS Ascend Fellowship Grant
number 2213275.
L.W.L. acknowledges support by NSF AST-1910364 and NSF AST-2307844.
P.S. was partially supported by a Grant-in-Aid for Scientific Research (KAKENHI Number JP22H01271 and JP23H01221) of JSPS.
W.K. was supported by the National Research Foundation of Korea (NRF) grant funded by the Korea government (MSIT) (NRF-2021R1F1A1061794).
Z.Y.L. acknowledges support in part by NASA 80NSSC20K0533 and NSF AST-2307199.
This paper makes use of the following ALMA data: ADS/JAO.ALMA\#2019.1.00086.
ALMA is a partnership of ESO (representing its member states), NSF (USA) and NINS (Japan), together with NRC (Canada), MOST and ASIAA (Taiwan), and KASI (Republic of Korea), in cooperation with the Republic of Chile. The Joint ALMA Observatory is operated by ESO, AUI/NRAO and NAOJ.
\end{acknowledgments}

\bibliographystyle{aasjournal}

\begin{thebibliography}{}
\expandafter\ifx\csname natexlab\endcsname\relax\def\natexlab#1{#1}\fi
\providecommand{\url}[1]{\href{#1}{#1}}
\providecommand{\dodoi}[1]{doi:~\href{http://doi.org/#1}{\nolinkurl{#1}}}
\providecommand{\doeprint}[1]{\href{http://ascl.net/#1}{\nolinkurl{http://ascl.net/#1}}}
\providecommand{\doarXiv}[1]{\href{https://arxiv.org/abs/#1}{\nolinkurl{https://arxiv.org/abs/#1}}}

\bibitem[{Allen {et~al.}(2003)Allen, Li, \& Shu}]{allen2003collapse}
Allen, A., Li, Z.-Y., \& Shu, F.~H. 2003, \apj, 599, 363

\bibitem[{Alves {et~al.}(2020)Alves, Cleeves, Girart, {et~al.}}]{alves2020case}
Alves, F.~O., Cleeves, L.~I., Girart, J.~M., {et~al.} 2020, \apjl, 904, L6

\bibitem[{Alves {et~al.}(2018)Alves, Girart, Padovani, {et~al.}}]{alves2018magnetic}
Alves, F.~O., Girart, J.~M., Padovani, M., {et~al.} 2018, \aap, 616, A56

\bibitem[{Andersson {et~al.}(2015)Andersson, Lazarian, \& Vaillancourt}]{andersson2015interstellar}
Andersson, B.~G., Lazarian, A., \& Vaillancourt, J.~E. 2015, \araa, 53, 501

\bibitem[{Arce-Tord {et~al.}(2020)Arce-Tord, Louvet, Cortes, {et~al.}}]{arce2020outflows}
Arce-Tord, C., Louvet, F., Cortes, P.~C., {et~al.} 2020, \aap, 640, A111

\bibitem[{Bacciotti {et~al.}(2018)Bacciotti, Girart, Padovani, {et~al.}}]{bacciotti2018alma}
Bacciotti, F., Girart, J.~M., Padovani, M., {et~al.} 2018, \apjl, 865, L12

\bibitem[{Beltr{\'a}n {et~al.}(2019)Beltr{\'a}n, Padovani, Girart, {et~al.}}]{beltran2019alma}
Beltr{\'a}n, M.~T., Padovani, M., Girart, J.~M., {et~al.} 2019, \aap, 630, A54

\bibitem[{Briggs(1995)Briggs}]{Briggs95}
Briggs, D.~S. 1995, PhD thesis, New Mexico Institute of Mining and  Technology

\bibitem[{CASA Team {et~al.}(2022)CASA Team, Bean, Bhatnagar, {et~al.}}]{casateam2022casa}
CASA Team, Bean, B., Bhatnagar, S., {et~al.} 2022, \pasp, 134, 114501

\bibitem[{Ciolek \& Mouschovias(1994)Ciolek, \& Mouschovias}]{ciolek1994ambipolar}
Ciolek, G.~E., \& Mouschovias, T.~C. 1994, \apj, 425, 142

\bibitem[{Cort{\'e}s {et~al.}(2021)Cort{\'e}s, Sanhueza, Houde, {et~al.}}]{cortes2021magnetic}
Cort{\'e}s, P.~C., Sanhueza, P., Houde, M., {et~al.} 2021, \apj, 923, 204

\bibitem[{Cox {et~al.}(2018)Cox, Harris, Looney, {et~al.}}]{cox2018alma}
Cox, E.~G., Harris, R.~J., Looney, L.~W., {et~al.} 2018, \apj, 855, 92

\bibitem[{Dapp \& Basu (2010)Dapp, \& Basu}]{Dapp2010disk}
Dapp, W.~B., \& Basu, S. 2010, \aap, 521, L56

\bibitem[{Eisner {et~al.}(2016)Eisner, Bally, Ginsburg, {et~al.}}]{eisner2016proto}
Eisner, J.~A., Bally, J.~M., Ginsburg, A., {et~al.} 2016, \apj, 826, 16

\bibitem[{Federman {et~al.}(2023)Federman, Megeath, Tobin, {et~al.}}]{federman2023300}
Federman, S., Megeath, S.~T., Tobin, J.~J., {et~al.} 2023, \apj, 944, 49

\bibitem[{Fern{\'a}ndez-L{\'opez} {et~al.}(2023)Fern{\'a}ndez-L{\'opez}, Girart, L{\'o}pez-V{\'a}zquez, {et~al.}}]{fernandez2023}
Fern{\'a}ndez-L{\'opez}, M., Girart, J.~M., L{\'o}pez-V{\'a}zquez, J.~A., {et~al.} 2023, \apj, 956, 82

\bibitem[{Furlan {et~al.}(2016)Furlan, Fischer, Ali, {et~al.}}]{furlan2016herschel}
Furlan, E., Fischer, W., Ali, B., {et~al.} 2016, \apjs, 224, 5

\bibitem[{Galametz {et~al.}(2018)Galametz, Maury, Girart, {et~al.}}]{galametz2018sma}
Galametz, M., Maury, A., Girart, J.~M., {et~al.} 2018, \aap, 616, A139

\bibitem[{Galametz {et~al.}(2020)Galametz, Maury, Girart, {et~al.}}]{galametz2020observational}
Galametz, M., Maury, A., Girart, J.~M., {et~al.} 2020, \aap, 644, A47

\bibitem[{Galli {et~al.}(2006)Galli, Lizano, Shu, \& Allen}]{galli2006gravitational}
Galli, D., Lizano, S., Shu, F.~H., \& Allen, A. 2006, \apj, 647, 374

\bibitem[{Girart {et~al.}(2009)Girart, Beltr{\'a}n, Zhang, {et~al.}}]{girart2009magnetic}
Girart, J.~M., Beltr{\'a}n, M.~T., Zhang, Q., Rao, R., \& Estalella, R. 2009, Science, 324, 1408

\bibitem[{Girart {et~al.}(1999)Girart, Crutcher, \& Rao}]{girart1999detection}
Girart, J.~M., Crutcher, R.~M., \& Rao, R. 1999, \apjl, 525, L109

\bibitem[{Girart {et~al.}(2013)Girart, Frau, Zhang, {et~al.}}]{girart2013dr}
Girart, J.~M., Frau, P., Zhang, Q., {et~al.} 2013, \apj, 772, 69

\bibitem[{Girart {et~al.}(2006)Girart, Rao, \& Marrone}]{girart2006magnetic}
Girart, J.~M., Rao, R., \& Marrone, D.~P. 2006, Science, 313, 812

\bibitem[{Girart {et~al.}(2018)Girart, Fern{\'a}ndez-L{\'opez}, Li, {et~al.}}]{girart2018resolving}
Girart, J.~M., Fern{\'a}ndez-L{\'opez}, M., Li, Z.~Y., {et~al.} 2018, \apjl, 856, L27

\bibitem[{Goodman {et~al.}(1993)Goodman, Benson, Fuller, {et~al.}}]{goodman1993dense}
Goodman, A.~A., Benson, P.~J., Fuller, G.~A., \& Myers, P. C. 1999, \apj, 406, 528

\bibitem[{Hennebelle {et~al.}(2016)Hennebelle, Commer{\c{c}}on, Chabrier, {et~al.}}]{Hennebelle2016diskform}
Hennebelle, P., Commer{\c{c}}on, B., Chabrier, G., \& Marchand, P. 2016, \apjl, 830, L8

\bibitem[{Hennebelle {et~al.}(2020)Hennebelle, Commer{\c{c}}on, Lee, {et~al.}}]{Hennebelle2020diskform}
Hennebelle, P., Commer{\c{c}}on, B., Lee, Y.-N., \& Charnoz, S. 2020, \aap, 635, A67

\bibitem[{Hirano \& Machida(2019)Hirano, \& Machida}]{hirano2019Bmisalign}
Hirano, S., \& Machida, M.~N. 2019, \mnras, 485, 4667

\bibitem[{Hoang \& Lazarian(2009)Hoang, \& Lazarian}]{hoang2009grain}
Hoang, T., \& Lazarian, A. 2009, \apj, 697, 1316

\bibitem[{Hull {et~al.}(2020)Hull, Le Gouellec, Girart, {et~al.}}]{hull2020understanding}
Hull, C.~L.~H., Le Gouellec, V.~J.~M., Girart, J.~M., Tobin, J. J., \& Bourke, T. L. 2020, \apj, 892, 152

\bibitem[{Hull \& Plambeck(2015)Hull, \& Plambeck}]{hull2015stokes}
Hull, C.~L.~H., \& Plambeck, R.~L. 2015, JAI, 4, 1550005

\bibitem[{Hull \& Zhang(2019)Hull, \& Zhang}]{hull2019interferometric}
Hull, C.~L.~H., \& Zhang, Q. 2019, FrASS, 6, 3

\bibitem[{Hull {et~al.}(2013)Hull, Plambeck, Bolatto, {et~al.}}]{hull2013misalignment}
Hull, C.~L.~H., Plambeck, R.~L., Bolatto, A.~D., {et~al.} 2013, \apj, 768, 159

\bibitem[{Hull {et~al.}(2014)Hull, Plambeck, Kwon, {et~al.}}]{hull2014tadpol}
Hull, C.~L.~H., Plambeck, R.~L., Kwon, W., {et~al.} 2014, \apjs, 213, 13

\bibitem[{Hull {et~al.}(2018)Hull, Yang, Li, {et~al.}}]{hull2018alma}
Hull, C.~L.~H., Yang, H., Li, Z.-Y., {et~al.} 2018, \apj, 860, 82

\bibitem[{Joos {et~al.}(2012)Joos, Hennebelle \& Ciardi}]{joos2012protostellar}
Joos, M., Hennebelle, P., \& Ciardi, A. 2012, \aap, 543, A128

\bibitem[{Kataoka {et~al.}(2016)Kataoka, Muto, Momose, {et~al.}}]{kataoka2016grain}
Kataoka, A., Muto, T., Momose, M., Tsukagoshi, T., \& Dullemond, C. P. 2016, \apj, 820, 54

\bibitem[{Kataoka {et~al.}(2017)Kataoka, Tsukagoshi, Pohl, {et~al.}}]{kataoka2017evidence}
Kataoka, A., Tsukagoshi, T., Pohl, A., {et~al.} 2017, \apjl, 844, L5

\bibitem[{Kataoka {et~al.}(2015)Kataoka, Muto, Momose, {et~al.}}]{kataoka2015millimeter}
Kataoka, A., Muto, T., Momose, M., {et~al.} 2015, \apj, 809, 78

\bibitem[{Kounkel {et~al.}(2017)Kounkel, Hartmann, Loinard, {et~al.}}]{Kounkel2017dis}
Kounkel, M., Hartmann, L., Loinard, L., {et~al.} 2017, \apj, 834, 142

\bibitem[{Kwon {et~al.}(2019)Kwon, Stephens, Tobin, {et~al.}}]{kwon2019highly}
Kwon, W., Stephens, I.~W., Tobin, J.~J., {et~al.} 2019, \apj, 879, 25

\bibitem[{Lada \& Lada(2003)Lada, \& Lada}]{lada2003embedded}
Lada, C.~J., \& Lada, E.~A. 2003, \araa, 41, 57

\bibitem[{Le Gouellec {et~al.}(2019)Le Gouellec, Hull, Maury, {et~al.}}]{le2019characterizing}
Le Gouellec, V.~J.~M., Hull, C.~L.~H., Maury, A.~J., {et~al.} 2019, \apj, 885, 106

\bibitem[{Le Gouellec {et~al.}(2020)Le Gouellec, Maury, Guillet, {et~al.}}]{le2020IMS}
Le Gouellec, V.~J.~M., Maury, A.~J., Guillet, V. {et~al.} 2020, \aap, 644, A11

\bibitem[{Lee {et~al.}(2018)Lee, Li, Ching, {et~al.}}]{lee2018alma}
Lee, C.-F., Li, Z.-Y~J., Ching, T.~-C., \& Yang, H. 2018, \apj, 854, 56

\bibitem[{Li {et~al.}(2013)Li, Krasnopolsky, \& Shang}]{li2013misalignment}
Li, Z.-Y., Krasnopolsky, R., \& Shang, H. 2013, \apj, 774, 82

\bibitem[{Liu {et~al.}(2023)Liu, Takahashi, Machida, {et~al.}}]{liu2023omc3}
Liu, Y., Takahashi, S., Machida, M., {et~al.} 2023, arXiv:2312.13573

\bibitem[{Machida {et~al.}(2020)Machida, Hirano, \& Kitta}]{machida2020misalignment}
Machida, M.~N., Hirano, S., \& Kitta, H. 2020, \mnras, 491, 2180

\bibitem[{Machida {et~al.}(2007)Machida, Inutsuka, \& Matsumoto}]{machida2007magnetic}
Machida, M.~N., Inutsuka, S.-i., \& Matsumoto, T. 2007, \apj, 670, 1198

\bibitem[{Maury {et~al.}(2022)Maury, Hennebelle, \& Girart}]{maury2022}
Maury, A., Hennebelle, P., \& Girart, J.~M. 2022, FrASS, 9, 949223

\bibitem[{Maury {et~al.}(2018)Maury, Girart, Zhang, {et~al.}}]{maury2018magnetically}
Maury, A.~J., Girart, J.~M., Zhang, Q., {et~al.} 2018, \mnras, 477, 2760

\bibitem[{Myers {et~al.}(2018)Myers, Basu, \& Auddy}]{myers2018magnetic}
Myers, P.~C., Basu, S., \& Auddy, S. 2018, \apj, 868, 51

\bibitem[{Myers {et~al.}(2020)Myers, Stephens, Auddy, {et~al.}}]{myers2020magnetic}
Myers, P.~C., Stephens, I.~W., Auddy, S., {et~al.} 2020, \apj, 896, 163

\bibitem[{Ohashi {et~al.}(2018)Ohashi, Kataoka, Nagai, {et~al.}}]{ohashi2018two}
Ohashi, S., Kataoka, A., Nagai, H., {et~al.} 2018, \apj, 864, 81

\bibitem[{Pattle {et~al.}(2023)Pattle, Fissel, Tahani, {et~al.}}]{pattle2023ppvii}
Pattle, K., Fissel, L., Tahani, M., Liu, T., \& Ntormousi, E. 2023, in ASP Conf. Ser. 534, Protostars and Planets VII, ed. S. Inutsuka et al. (San Francisco, CA: ASP), 193

\bibitem[{Pineda {et~al.}(2020)Pineda, Segura-Cox, Caselli, {et~al.}}]{pineda2020protostellar}
Pineda, J.~E., Segura-Cox, D., Caselli, P., {et~al.} 2020, NatAs, 4, 1158

\bibitem[{Pudritz \& Ray(2019)Pudritz, \& Ray}]{pudritz2019role}
Pudritz, R.~E., \& Ray, T.~P. 2019, FrASS, 6, 54

\bibitem[{Qiu {et~al.}(2014)Qiu, Zhang, Menten, {et~al.}}]{qiu2014submillimeter}
Qiu, K., Zhang, Q., Menten, K.~M., {et~al.} 2014, \apjl, 794, L18

\bibitem[{Sadavoy {et~al.}(2018a)Sadavoy, Myers, Stephens, {et~al.}}]{sadavoy2018dusta}
Sadavoy, S.~I., Myers, P.~C., Stephens, I.~W., {et~al.} 2018a, \apj, 859, 165

\bibitem[{Sadavoy {et~al.}(2018b)Sadavoy, Myers, Stephens, {et~al.}}]{sadavoy2018dustb}
Sadavoy, S.~I., Myers, P.~C., Stephens, I.~W., {et~al.} 2018b, \apj, 869, 115

\bibitem[{Sanhueza {et~al.}(2021)Sanhueza, Girart, Padovani, {et~al.}}]{sanhueza2021gravity}
Sanhueza, P., Girart, J.~M., Padovani, M., {et~al.} 2021, \apjl, 915, L10

\bibitem[{Shimajiri {et~al.}(2013)Shimajiri, Sakai, Tsukagoshi, {et~al.}}]{shimajiri2013extensive}
Shimajiri, Y., Sakai, T., Tsukagoshi, T., {et~al.} 2013, \apjl, 774, L20

\bibitem[{Soler {et~al.}(2017)Soler, Ade, Angil{\`e}, {et~al.}}]{soler2017relation}
Soler, J.~D., Ade, P.~A.~R., Angil{\`e}, F.~E., {et~al.} 2017, \aap, 603, A64

\bibitem[{Stephens {et~al.}(2017a)Stephens, Yang, Li, {et~al.}}]{stephens2017alma}
Stephens, I.~W., Yang, H., Li, Z.-Y., {et~al.} 2017a, \apj, 851, 55

\bibitem[{Stephens {et~al.}(2017b)Stephens, Dunham, Myers, {et~al.}}]{stephens2017alignment}
Stephens, I.~W., Dunham, M.~M., Myers, P.~C., {et~al.} 2017b, \apj, 846, 16

\bibitem[{Stutz {et~al.}(2013)Stutz, Tobin, Stanke, {et~al.}}]{stutz2013discovery}
Stutz, A., Tobin, J., Stanke, T., {et~al.} 2013, in Protostars and Planets VI Posters, ed. H. Beuther, R. S. Klessen, C. P. Dullemond, \& T. Henning (Tucson, AZ: Univ. of Arizona Press)

\bibitem[{Tobin {et~al.}(2011)Tobin, Hartmann, Chiang, {et~al.}}]{tobin2011ism}
Tobin, J.~J., Hartmann, L., Chiang, H.-F., {et~al.} 2011, \apj, 740, 45

\bibitem[{Tobin {et~al.}(2020)Tobin, Sheehan, Megeath, {et~al.}}]{tobin2020vla}
Tobin, J.~J., Sheehan, P.~D., Megeath, S.~T., {et~al.} 2020, \apj, 890, 130

\bibitem[{Vaillancourt(2006)Vaillancourt}]{vaillancourt2006placing}
Vaillancourt, J.~E. 2006, \pasp, 118, 1340

\bibitem[{Wang {et~al.}(2022)Wang, {V{\"a}is{\"a}l{\"a}}, Shang, {et~al.}}]{wang2022magnetic}
Wang, W., {V{\"a}is{\"a}l{\"a}}, M.~S., Shang, H., {et~al.} 2022, \apj, 928, 85

\bibitem[{Yang {et~al.}(2016a)Yang, Li, Looney, {et~al.}}]{yang2016inclination}
Yang, H., Li, Z.-Y., Looney, L., \& Stephens, I. 2016a, \mnras, 456, 2794

\bibitem[{Yang {et~al.}(2016b)Yang, Li, Looney, {et~al.}}]{yang2016disc}
Yang, H., Li, Z.-Y., Looney, L., {et~al.} 2016b, \mnras, 460, 4109

\bibitem[{Yang {et~al.}(2017)Yang, Li, Looney, {et~al.}}]{yang2017scattering}
Yang, H., Li, Z.-Y., Looney, L., Girart, J. M., \& Stephens, I. W. 2017, \mnras, 472, 373

\bibitem[{Yang {et~al.}(2019)Yang, Li, Stephens, {et~al.}}]{yang2019does}
Yang, H., Li, Z.-Y., Stephens, I.~W., {et~al.} 2019, \mnras, 483, 2371

\bibitem[{Zhang {et~al.}(2014)Zhang, Qiu, Girart, {et~al.}}]{zhang2014magnetic}
Zhang, Q., Qiu, K., Girart, J.~M., {et~al.} 2014, \apj, 792, 116

\bibitem[{Zhao {et~al.}(2020)Zhao, Caselli, Li, {et~al.}}]{Zhao2020disk}
Zhao, B., Caselli, P., Li, Z.-Y., {et~al.} 2020, \mnras, 492, 3375

\end{thebibliography}

\appendix

\section{Weighted mean {\em B}-field position angle} \label{App:Bfield}

To trace the mean orientation of polarization for these protostars, we extract the size ranging from several hundreds to several thousands of au, encompassing all polarization measurements toward the targeted source, i.e. all Stokes {\em Q} and {\em U} pixels larger than 4$\sigma$ are enclosed.
We then calculate its position angle $\langle \theta \rangle$ by performing a total-intensity-weighted average, giving more weight to polarization directions in higher-density regions \citep{hull2013misalignment, hull2014tadpol}.
To guarantee adequate sampling of the derivatives in each pixel, the pixel size used here falls within a region of one-third to half of the beam FWHM \citep[e.g.,][]{soler2017relation}.
For example, in the case of HH212M, with major and minor axes of 0\farcs95 and 0\farcs70 respectively, and a beam FWHM of $\sim$ 0\farcs83, the pixel size should be within the range of 0\farcs28 $\sim$ 0\farcs42.
The calculation of intensity-weighted position angle of polarization $\langle \theta_{\rm Pol} \rangle_{I}$ is expressed as:
\begin{eqnarray}\label{eq1}
\langle \theta_{\rm Pol} \rangle_{I}=0.5\cdot {\rm arctan}  \biggl(\frac{D}{N}\biggr)
\end{eqnarray}
$N$ and $D$ are averaged values of Stokes {\em Q} and {\em U} weighted by the intensity:
\begin{eqnarray}\label{eq2}
N=\frac{\sum \limits_{ij} \biggl(Q(i,j) \cdot I(i,j)\biggr)}{\sum \limits_{ij}I(i,j)}, ~~~~~
D=\frac{\sum \limits_{ij} \biggl(U(i,j) \cdot I(i,j)\biggr)}{\sum \limits_{ij}I(i,j)}
\end{eqnarray}
where {\em i} and {\em j} are pixel numbers, $I(i,j)$, $Q(i,j)$, and $U(i,j)$ indicate the corresponding values in the $(i,j)$ pixel of the Stokes {\em I}, {\em Q}, {\em  U} images, respectively.
The error is estimated through error propagation:
\begin{eqnarray}\label{eq3}
\Delta \langle \theta_{\rm Pol} \rangle_{I} = \sqrt{\biggl(\frac{\partial \langle \theta_{\rm Pol} \rangle_{I}}{\partial N}\cdot \Delta {N}\biggr)^{2}+\biggl(\frac{\partial \langle \theta_{\rm Pol} \rangle_{I}}{\partial D}\cdot \Delta {D}\biggr)^{2}}
\end{eqnarray}
where
\begin{eqnarray}\label{eq4}
\Delta N = \sqrt{ \sum \limits_{ij} \biggl(\frac{\partial N}{\partial Q(i,j)}\cdot \Delta Q\biggr)^{2}+ \sum \limits_{ij} \biggl(\frac{\partial N}{\partial I(i,j)}\cdot \Delta I\biggr)^{2}} = \frac{\sqrt{(\Delta Q)^{2}\cdot\sum \limits_{ij} I^{2}(i,j)+ (\Delta I)^{2} \cdot\sum \limits_{ij} Q^{2}(i,j)}}{\sum \limits_{ij}I(i,j)}
\end{eqnarray}

\begin{eqnarray}\label{eq5}
\Delta D = \sqrt{ \sum \limits_{ij} \biggl(\frac{\partial D}{\partial U(i,j)}\cdot \Delta U\biggr)^{2}+ \sum \limits_{ij} \biggl(\frac{\partial D}{\partial I(i,j)}\cdot \Delta I\biggr)^{2}} = \frac{\sqrt{(\Delta U)^{2} \cdot\sum \limits_{ij} I^{2}(i,j)+ (\Delta I)^{2} \cdot\sum \limits_{ij} U^{2}(i,j)}}{\sum \limits_{ij}I(i,j)}
\end{eqnarray}
Here $\Delta I$, $\Delta Q$ and $\Delta U$ indicate the rms noise of the Stokes $I$, $Q$ and $U$ maps, respectively.
Then the Equation \ref{eq3} can be expressed as:
\begin{eqnarray}\label{eq6}
\Delta \langle \theta_{\rm Pol} \rangle_{I} = \frac{\sqrt{N^{2}\cdot(\Delta D)^{2}+D^{2}\cdot(\Delta N)^{2}}}{2(N^{2}+D^{2})}
\end{eqnarray}

The intensity-weighted method always makes sense when applied to point sources, but it may lead to misjudgment in certain extended cases, as the position angle in the extended regions also significantly contributes to the estimation, despite having much lower density.
For comparison, we also follow the approach discussed by \cite{galametz2020observational} and average the position angle of polarization by weighting it with its uncertainty $\sigma$.
The value of $\sigma$ is calculated by error propagation and described as:
\begin{eqnarray}\label{eq7}
\sigma(i,j) = \sqrt{\biggl(\frac{\partial \theta_{\rm Pol}(i,j)}{\partial Q(i,j)}\cdot \Delta {Q}\biggr)^{2}+\biggl(\frac{\partial \theta_{\rm Pol}(i,j)}{\partial U(i,j)}\cdot \Delta {U}\biggr)^{2}}
\end{eqnarray}
where $\theta_{\rm Pol}(i,j)=0.5\cdot\arctan(U(i,j)/Q(i,j))$, and the averaging position of polarization angle weighted by the uncertainty $\langle \theta_{\rm Pol} \rangle_{\sigma}$ is expressed by:
\begin{eqnarray}\label{eq8}
\langle \theta_{\rm Pol} \rangle_{\sigma}=0.5\cdot \arctan \biggl(\frac{K}{M}\biggr)
\end{eqnarray}
where $M$ and $K$ are average values of Stokes {\em Q} and {\em U} weighted by the uncertainty of the polarization position angle:
\begin{eqnarray}\label{eq9}
M = \frac{\sum\limits_{ij}\biggl(Q(i,j)/ \sigma^{2}(i,j)\biggr)}{\sum\limits_{ij}\biggl(1/\sigma^{2}(i,j)\biggr)}, ~~~~~
K = \frac{\sum\limits_{ij}\biggl(U_{i}/\sigma^{2}(i,j)\biggr)}{\sum\limits_{ij}\biggl(1/\sigma^{2}(i,j)\biggr)}
\end{eqnarray}

The error of uncertainty-weighted angle $\Delta \langle \theta_{\rm Pol} \rangle_{\sigma}$ is estimated by multiplying the internal uncertainty $iu$ derived from error propagation by the the square-root of the reduced chi-squared:
\begin{eqnarray}\label{eq10}
\Delta \langle \theta_{\rm Pol} \rangle_{\sigma}= iu\cdot \sqrt{\frac{\chi^{2}}{n-1}}
\end{eqnarray}
where $n$ is the total number of pixels, and
\begin{eqnarray}\label{eq11}
iu=\frac{1}{\sqrt{\sum\limits_{ij}(1/\sigma^{2}(i,j))}}, ~~~~~~
\chi^{2}=\sum\limits_{ij}\biggl(\frac{(\theta_{\rm Pol}(i,j)-\langle \theta_{\rm Pol} \rangle_{\sigma})^{2}}{\sigma^{2}(i,j)}\biggr)
\end{eqnarray}

Note that the weighted polarization angle $\langle \theta_{\rm Pol} \rangle$ should be rotated by 90$^{\circ}$ to infer the mean direction of the {\em B}-field $\langle \theta_{B} \rangle$.
The {\em B}-field position angles weighted by intensity $\langle \theta_{B} \rangle_{I}$ and uncertainty $\langle \theta_{B} \rangle_{\sigma}$ are available in Table \ref{Tab:PA}.

\section{Outflow parameters} \label{App:Outflow}

In many cases (47 protostars), blue- and red-shifted outflows can be clearly paired in a bipolar fashion.
However, there are 9 protostars that only have one clear outflow component, while the other is either very weak or overlaps with other emission from e.g. other outflows.
It should be mentioned that HOPS-373E, HOPS-398, and HOPS-402 do not exhibit clear outflows.

\begin{figure*}
\centering
\includegraphics[clip=true,trim=0cm 0cm 0cm 0cm,width=0.6\textwidth]{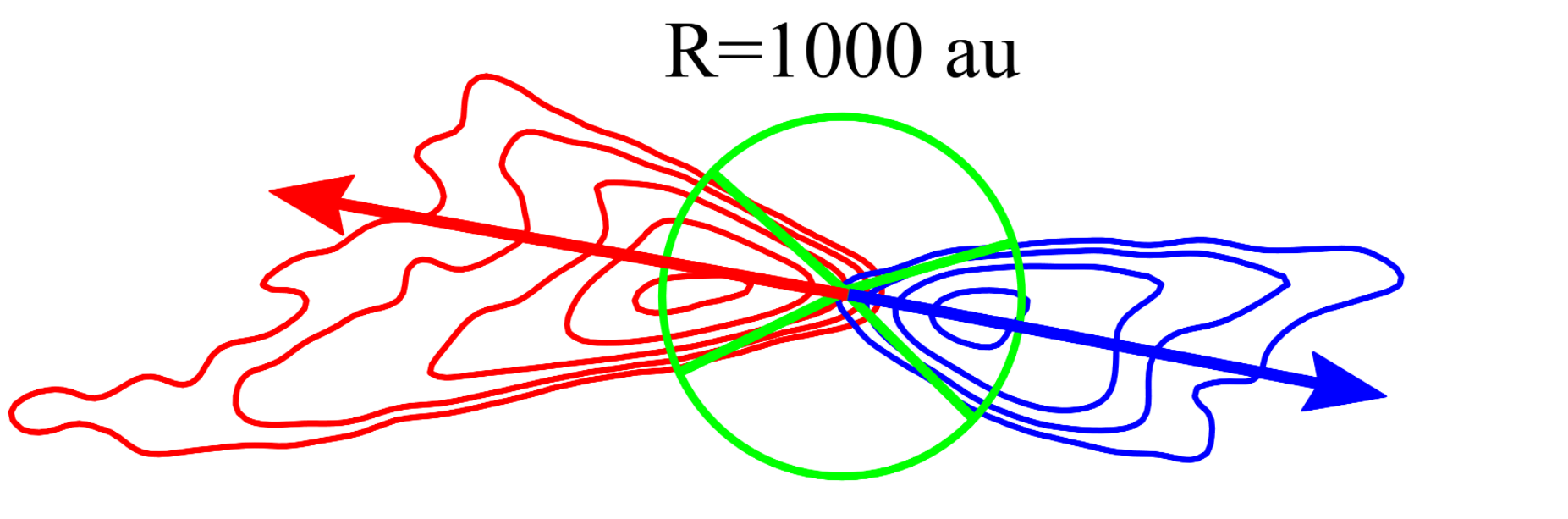}
\caption{Diagram to identify the position angle of outflows.} 
\label{Fig:Outflow}
\end{figure*}

To estimate the outflow direction, we measure its position angle using the following steps (see Figure \ref{Fig:Outflow}):
First, we connect the source center with the edges of the red-/blueshifted outflow lobe, and then we take the bisector of the open angle as the red-/blueshifted outflow position angle.
The outflow direction is determined by averaging the position angles of the redshifted and blueshifted components.
The edges of the lobes are defined using the 5$\sigma$ level of the emission.
For the monopolar cases, we take the position angle of the clear lobe as the outflow direction.
Uniformly, the radius from center to lobe edge is set to $R=1000$ au, as most sources have clear outflows at this scale.
The outflow position angle $\theta_{\rm Out}$ and its corresponding mean open angle $\Delta \theta_{\rm Out}$ for each source are listed in Table~\ref{Tab:PA} and Table~\ref{Tab:SelfScattering}.

We define the position angle for each of the outflow edges as $\theta_{\rm edge}=\arctan(y/x)$, where $x^{2}+y^{2}=R^{2}$.
Assuming $x$ and $y$ have an error of half beamsize $u_{3}$, i.e. $\Delta x=\Delta y = u_{3}$, then the error of $\theta_{\rm edge}$ is estimated by error propagation:
\begin{eqnarray}\label{eq12}
\Delta \langle \theta_{\rm edge} \rangle = \frac{u_{3}}{\sqrt{x^{2}+y^{2}}} =\frac{u_{3}}{R}
\end{eqnarray}
which is related to the location of the source within the OMCs region.
For monopolar and bipolar outflows, there are two (one lobe) and four (two lobes) independent measurements, thus the error of $\theta_{\rm Out}$ can be expressed:
\begin{eqnarray}\label{eq13}
\Delta \langle \theta_{\rm Out} \rangle =
\frac{\Delta \langle \theta_{\rm edge} \rangle}{\sqrt{2}} = 
\frac{u_{3}}{\sqrt{2}R}
\end{eqnarray}
for the cases with one clear lobe, and
\begin{eqnarray}\label{eq14}
\Delta \langle \theta_{\rm Out} \rangle =
\frac{\Delta \langle \theta_{\rm edge} \rangle}{2} = \frac{u_{3}}{2R}
\end{eqnarray}
for the cases with two clear lobes.

\section{C$^{17}$O opacity and velocity gradient} \label{App:VG}

The opacity of C$^{17}$O (3-2) is calculated by:
\begin{eqnarray}\label{eq15}
\tau_{\nu}=-\ln \biggl(1-\frac{T_{R}}{J_{\nu}(T_{EX})-J_{\nu}(T_{cmb})}\biggr)
\end{eqnarray}
where $T_{cmb}$=2.722 K is the cosmic microwave background Temperature, $J_{\nu}(T)$ is the Rayleigh-Jeans equivalent temperature of a black body at temperature $T$:
\begin{eqnarray}\label{eq16}
J_{\nu}(T) [\rm{K}]=\frac{h\nu}{k} \frac{[\rm{m}^{2}~\rm{kg}~\rm{s}^{-2}]}{[\rm{m}^{2}~\rm{kg}~\rm{s}^{-2}~\rm{K}^{-1}]}\cdot \frac{1}{e^{h\nu/kT}-1}
\end{eqnarray}

Assuming the Rayleigh–Jeans approximation is accurate, then the observable source radiation temperature $T_{R}$ is estimated as:
\begin{eqnarray}\label{eq17}
T_{R}=1222\cdot \frac{I}{\nu^{2} \theta_{\rm maj}\theta_{\rm min}} \cdot\frac{[\rm{mJy/beam}]}{[\rm{GHz^{2}\cdot arcsec^{2}}]}
\end{eqnarray}
where $\theta_{\rm maj}$ and $\theta_{\rm min}$ are the major and minor axes of the restoring beam, and $I$ is the peak of the C$^{17}$O spectral line.
The excitation temperature for an optically thick molecular line is estimated by:
\begin{eqnarray}\label{eq18}
T_{EX}=\frac{h\nu}{k}\cdot\ln\biggl(\frac{h\nu/k}{T_{R}+J_{\nu}(T_{cmb})}\biggr)
\end{eqnarray}

We have $^{12}{\rm CO}$ detection, which is usually optically thick in the OMCs \cite[e.g.,][]{shimajiri2013extensive, eisner2016proto}.
$T_{EX}(^{12}{\rm CO})$ ranges from 15.3 K to 50.4 K at envelope scales of 400 au.
Assuming $T_{EX}({\rm C}^{17}{\rm O})\approx 50$ K, we then obtained the opacity of C$^{17}$O for each source, with a range of 0.04 to 0.26, which is considered to be optically thin.

We estimate the velocity gradient by fitting the following function \citep{goodman1993dense, tobin2011ism}: 
\begin{eqnarray}\label{eq19}
v_{\rm lsr} = v_{\alpha} \Delta \alpha + v_{\delta} \Delta \delta + v_{0}
\end{eqnarray}
here $\Delta\alpha$ and $\Delta\delta$ are the offsets in right ascension and declination, and $v_{\alpha}$ and $v_{\delta}$ are the projections of the velocity gradient onto the $\alpha$ and $\delta$ axes. The parameter $v_{0}$ represents the systemic velocity with respect to the local standard of rest.
Then the direction of velocity gradient $\theta_{v}$ is calculated by
\begin{eqnarray}\label{eq20}
\theta_{v}=\arctan(v_{\delta}/v_{\alpha})
\end{eqnarray}
The uncertainty is from a least-squares fit of Eqation \ref{eq19} to the observed velocity field.

\pagebreak

\section{Tables} \label{App:Table}

\begin{longtable}{lcccccccccccc}
\caption{\label{Tab:parameters} Source properties for the entire sample. 
Column 2 to 13 present the Right Ascension (R.A.), Declination (Dec.), Beam Size, Position Angle (P.A.), Distance ({\em D}), Evolutionary Stage, Bolometric Luminosity ($L_{\rm{bol}}$), Bolometric Temperature ($T_{\rm{bol}}$), error of total intensity map ($\sigma_{I}$), error of $^{12}$CO image cube ($\sigma_{^{12}{\rm CO}}$), error of C$^{17}$O image cube ($\sigma_{\rm{C}^{17}\rm{O}}$), and Type of Stoke {\em I}, respectively.
The values {\em D}, evolutionary stage, $L_{\rm{bol}}$, and $T_{\rm{bol}}$ are obtained from \cite{tobin2020vla} and \cite{federman2023300}.
The Stokes {\em I} type are classified by comparing the ratio ({\em R}) of the area enclosed by the intensity contour at 0.2 times the maximum to the restoring beam size.
Sources with $R<1$, $1\le R\le 3$, and $R>3$ are categorized as ``Unresolved'', ``Compact'', and ``Extended'', sources are classified as either ``Binary'' or ``Multiple'' when there are two or more well-resolved components within 10\arcsec ($\sim$ 4000 au) from the peak intensity.
}\\
\hline\hline
\multirow{2}*{Name} & R.A. & Dec. & Beam & P.A. & {\em D} & \multirow{2}*{Class} & $L_{\rm{bol}}$ & $T_{\rm{bol}}$ & $\sigma_{I}$ & $\sigma_{^{12}{\rm CO}}$ & $\sigma_{\rm{C}^{17}\rm{O}}$ & Type \\
~ & (h:m:s) & (d:m:s) & ($^{\arcsec} \times ^{\arcsec}$) & ($^{\circ}$) & (pc) & ~ & (L$_{\odot}$) & (K) & (mJy) & (mJy) & (mJy) & Stokes {\em I}  \\
\hline
\endfirsthead
\caption{Source properties for the entire sample.}\\
\hline\hline
\multirow{2}*{Name} & R.A. & Dec. & Beam & P.A. & {\em D} & \multirow{2}*{Class} & $L_{\rm{bol}}$ & $T_{\rm{bol}}$ & $\sigma_{I}$ & $\sigma_{^{12}{\rm CO}}$ & $\sigma_{\rm{C}^{17}\rm{O}}$ & Type \\
~ & (h:m:s) & (d:m:s) & ($^{\arcsec} \times ^{\arcsec}$) & ($^{\circ}$) & (pc) & ~ & (L$_{\odot}$) & (K) & (mJy) & (mJy) & (mJy) & Stokes {\em I}  \\
\hline
\endhead
\hline
\endfoot
HH212M   & 05:43:51.41 & -01:02:53.25 & 0.95$\times$0.70 & -70.0 & 429.2 & 0 & 14.00 & 53.0 & 0.10 & 30.0 & 2.2 & Extended  \\
HH270IRS & 05:51:22.72 &  02:56:04.95 & 0.93$\times$0.71 & -79.0 & 460.0 & I & /& / & 0.13 & 95.0 & 1.8 & Compact   \\
HOPS-10  & 05:35:09.05 & -05:58:26.94 & 0.92$\times$0.66 & -68.8 & 388.2 & 0 & 3.33 & 46.2 & 0.09 & 25.0 & 2.7 & Compact   \\
HOPS-11  & 05:35:13.43 & -05:57:57.96 & 0.90$\times$0.66 & -70.2 & 388.3 & 0 & 9.00 & 48.8 & 0.11 & 53.0 & 2.7 & Compact   \\
HOPS-12E & 05:35:08.95 & -05:55:55.04 & 0.82$\times$0.66 & -78.8 & 388.6 & 0 & / & / & 0.11 & 35.0 & 2.4 & Binary   \\
HOPS-12W & 05:35:08.63 & -05:55:54.70 & 0.92$\times$0.66 & -68.6 & 388.6 & 0 & 7.31 & 42.0 & 0.11 & 46.0 & 2.6 & Binary   \\
HOPS-50  & 05:34:40.92 & -05:31:44.79 & 0.89$\times$0.66 & -70.3 & 391.5 & 0 & 4.20 & 51.4 & 0.08 & 14.0 & 2.4 & Compact  \\
HOPS-53  & 05:33:57.40 & -05:23:30.05 & 0.82$\times$0.66 & -78.9 & 390.5 & 0 & 26.42 & 45.9 & 0.09 & 44.0 & 3.1 & Extended  \\
HOPS-60  & 05:35:23.29 & -05:12:03.50 & 0.82$\times$0.66 & -79.1 & 392.8 & 0  & 21.93 & 54.1 & 0.12 & 79.0 & 2.3 & Compact  \\
HOPS-78  & 05:35:25.97 & -05:05:43.43 & 0.82$\times$0.66 & -79.2 & 392.8 & 0  & 8.93 & 38.1 & 0.13 & 40.0 & 2.6 & Extended   \\
HOPS-81  & 05:35:28.02 & -05:04:57.41 & 0.90$\times$0.67 & -69.6 & 392.8 & 0  & 1.24 & 40.1 & 0.08 & 41.0 & 1.7 & Compact  \\
HOPS-84  & 05:35:26.57 & -05:03:55.20 & 0.91$\times$0.66 & -68.7 & 392.8 & I & 49.11 & 90.8 & 0.17 & 15.0 & 2.6 & Compact  \\
HOPS-87N  & 05:35:23.42 & -05:01:30.62 & 0.82$\times$0.66 & -79.6 & 392.7 & 0  & 36.49 & 38.1 & 0.36 & 54.0 & 2.4 & Binary   \\
HOPS-87S  & 05:35:23.67 & -05:01:40.32 & 0.82$\times$0.66 & -79.6 & 392.7 & 0  & / & / & 0.36 & 54.0 & 2.4 & Binary   \\
HOPS-88  & 05:35:22.47 & -05:01:14.38 & 0.89$\times$0.67 & -70.9 & 392.7 & 0  & 15.81 & 42.4 & 0.18 & 119.0 & 2.2 & Extended  \\
HOPS-96  & 05:35:29.72 & -04:58:48.68 & 0.82$\times$0.66 & -79.0 & 392.7 & 0  & 6.19 & 35.6 & 0.11 & 15.0 & 1.8 & Compact  \\
HOPS-124 & 05:39:19.91 & -07:26:11.27 & 0.91$\times$0.65 & -69.5 & 398.0 & 0  & 58.29 & 44.8 & 0.37 & 59.0 & 3.0 & Compact  \\
HOPS-153 & 05:37:57.03 & -07:06:56.32 & 0.89$\times$0.66 & -71.0 & 387.9 & 0  & 4.43 & 39.4 & 0.07 & 22.0 & 2.0 & Extended   \\
HOPS-164 & 05:37:00.43 & -06:37:10.96 & 0.89$\times$0.66 & -70.6 & 385.0 & 0  & 0.58 & 50.0 & 0.07 & 36.0 & 2.1 & Compact  \\
HOPS-168 & 05:36:18.95 & -06:45:23.63 & 0.81$\times$0.66 & -80.2 & 383.3 & 0  & 48.07 & 54.0 & 0.10 & 64.0 & 2.5 & Extended \\
HOPS-169 & 05:36:36.17 & -06:38:54.46 & 0.81$\times$0.66 & -80.1 & 384.0 & 0  & 3.91 & 32.5 & 0.10 & 49.0 & 2.1 & Extended  \\
HOPS-182 & 05:36:18.80 & -06:22:10.29 & 0.82$\times$0.66 & -80.2 & 385.1 & 0  & 71.12 & 51.9 & 0.07 & 74.0 & 2.1 & Extended \\
HOPS-203N& 05:36:22.87 & -06:46:06.68 & 0.82$\times$0.66 & -79.9 & 383.5 & 0  & 20.44 & 43.7 & 0.09 & 38.0 & 1.9 & Multiple   \\
HOPS-203S& 05:36:22.90 & -06:46:09.59 & 0.91$\times$0.65 & -69.3 & 383.5 & 0  & / & / & 0.12 & 36.0 & 2.6 & Multiple  \\
HOPS-224 & 05:41:32.07 & -08:40:09.87 & 0.88$\times$0.66 & -70.9 & 440.3 & 0  & 2.99 & 48.6 & 0.09 & 36.0 & 2.0 & Compact  \\
HOPS-247 & 05:41:26.19 & -07:56:51.95 & 0.88$\times$0.66 & -71.3 & 430.9 & 0  & 3.09 & 42.8 & 0.09 & 18.0 & 2.4 & Compact   \\
HOPS-250 & 05:40:48.85 & -08:06:57.16 & 0.91$\times$0.65 & -69.7 & 428.5 & 0  & 6.79 & 69.4 & 0.10 & 74.0 & 2.5 & Compact  \\
HOPS-288 & 05:39:56.01 & -07:30:27.67 & 0.81$\times$0.66 & -81.2 & 405.5 & 0  & 135.47 & 48.6 & 0.22 & 63.0 & 2.2 & Extended  \\
HOPS-303 & 05:42:02.65 & -02:07:45.99 & 0.94$\times$0.70 & -71.4 & 410.0 & 0  & 1.49 & 43.2 & 0.10 & 19.0 & 2.6 & Extended  \\
HOPS-310 & 05:42:27.68 & -01:20:01.40 & 0.93$\times$0.69 & -81.1 & 414.3 & 0  & 13.83 & 51.8 & 0.13 & 44.0 & 2.5 & Compact  \\
HOPS-317N& 05:46:08.60 & -00:10:38.54 & 0.94$\times$0.69 & -81.9 & 427.1 &  0 & 4.76 & 47.5 & 0.23 & 85.0 & 2.1 & Binary  \\
HOPS-317S& 05:46:08.38 & -00:10:43.64 & 0.94$\times$0.69 & -81.9 & 427.1 & 0  & / & / & 0.23 & 75.0 & 2.1 & Binary  \\
HOPS-325 & 05:46:39.20 & 00:01:12.25 & 0.96$\times$0.70 & -69.2 & 428.5 & 0  & 6.2 & 49.2 & 0.11 & 94.0 & 4.0 & Extended  \\
HOPS-340 & 05:47:01.32 & 00:26:23.00 & 0.96$\times$0.70 & -68.6 & 430.9 & 0  & 1.85 & 40.6 & 0.08 & 32.0 & 2.4 & Binary  \\
HOPS-341 & 05:47:00.92 & 00:26:21.45 & 0.96$\times$0.70 & -68.3 & 430.9 & 0  & 2.07 & 39.4 & 0.08 & 43.0 & 2.5 & Binary  \\
HOPS-354 & 05:54:24.27 & 01:44:19.82 & 0.93$\times$0.70 & -81.2 & 355.4 & 0  & 6.57 & 34.8 & 0.07 & 19.0 & 1.4 & Extended   \\
HOPS-358 & 05:46:07.26 & -00:13:30.30 & 0.95$\times$0.70 & -70.2 & 426.8 & 0  & 24.96 & 41.7 & 0.13 & 44.0 & 2.6 & Extended   \\
HOPS-359 & 05:47:24.85 & 00:20:59.34 & 0.94$\times$0.71 & -70.8 & 429.4 & 0  & 10.00 & 36.7 & 0.13 & 8.0 & 1.8 & Extended  \\
HOPS-361N& 05:47:04.64 & 00:21:47.77 & 0.94$\times$0.69 & -81.5 & 430.4 & 0  & / & / & 0.56 & 70.0 & 3.2 & Binary  \\
HOPS-361S& 05:47:04.79 & 00:21:42.74 & 0.94$\times$0.69 & -81.8 & 430.4 & 0  & 478.99 & 69.0 & 1.15 & 130.0 & 3.0 & Binary  \\
HOPS-370 & 05:35:27.64 & -05:09:34.45 & 0.91$\times$0.66 & -68.7 & 392.8 & I & 360.86 & 71.5 & 0.21 & 102.0 & 3.6 & Extended  \\
HOPS-373W & 05:46:30.91 & -00:02:35.20 & 0.95$\times$0.70 & -69.6 & 428.1 & 0 & 5.32 & 36.9 & 0.12 & 67.0 & 2.4 & Binary  \\
HOPS-373E & 05:46:31.11 & -00:02:33.10 & 0.95$\times$0.70 & -69.6 & 428.1 & 0 & / & / & 0.12 & 67.0 & / & Binary  \\
HOPS-383 & 05:35:29.79 & -04:59:50.43 & 0.90$\times$0.67 & -69.9 & 392.8 & 0  & 7.83 & 45.8 & 0.07 & 28.0 & 2.0 & Compact   \\
HOPS-384 & 05:41:44.14 & -01:54:46.05 & 0.93$\times$0.69 & -80.6 & 409.5 & 0  & 1477.95 & 51.9 & 0.30 & 24.0 & 4.2 & Extended  \\
HOPS-395 & 05:39:17.09 & -07:24:24.64 & 0.81$\times$0.66 & -79.9 & 397.2 & 0  & 0.5 & 31.7 & 0.09 & 15.0 & 2.8 & Extended  \\
HOPS-398 & 05:41:29.42 & -02:21:16.44 & 0.96$\times$0.69 & -70.0 & 408.0 & 0  & 1.01 & 23.0 & 0.11 & 5.0 & / & Extended  \\
HOPS-399 & 05:41:24.94 & -02:18:06.71 & 0.93$\times$0.68 & -80.6 & 407.9 & 0  & 6.34 & 31.1 & 0.21 & 67.0 & 2.0 & Extended  \\
HOPS-400 & 05:42:45.26 & -01:16:13.94 & 0.93$\times$0.69 & -81.2 & 415.4 & 0  & 2.94 & 35.0 & 0.15 & 50.0 & 1.5 & Extended  \\
HOPS-401 & 05:46:07.73 & -00:12:21.36 & 0.95$\times$0.70 & -69.6 & 426.9 & 0  & 0.61 & 26.0 & 0.08 & 16.0 & 2.3 & Extended  \\
HOPS-402 & 05:46:10.04 & -00:12:16.97 & 0.94$\times$0.69 & -81.5 & 426.9 & 0  & 0.55 & 24.2 & 0.10 & 4.0 & / & Compact  \\
HOPS-403 & 05:46:27.91 & -00:00:52.20 & 0.94$\times$0.69 & -81.7 & 428.2 & 0  & 4.14 & 43.9 & 0.14 & 60.0 & 1.4 & Extended  \\
HOPS-404 & 05:48:07.72 & 00:33:51.76 & 0.94$\times$0.69 & -81.2 & 430.1 & 0  & 0.95 & 26.1 & 0.10 & 6.0 & 2.3 & Compact  \\
HOPS-407 & 05:46:28.26 & 00:19:27.90 & 0.97$\times$0.70 & -68.3 & 419.1 & 0  & 0.71 & 26.8 & 0.08 & 16.0 & 2.2 & Extended  \\
HOPS-408 & 05:39:30.90 & -07:23:59.80 & 0.89$\times$0.66 & -71.0 & 398.9 & 0  & 0.52 & 37.9 & 0.06 & 18.0 & 2.1 & Extended  \\
HOPS-409 & 05:35:21.37 & -05:13:17.93 & 0.89$\times$0.67 & -71.0 & 392.8 & 0  & 8.18 & 28.4 & 0.13 & 40.0 & 2.0 & Compact  \\
OMC1N-4-5-ES & 05:35:15.97 & -05:20:14.28 & 0.91$\times$0.66 & -68.8 & 392.8 & /  & / & / & 0.19 & 46.0 & 3.5 & Binary  \\
OMC1N-4-5-EN & 05:35:16.05 & -05:20:05.78 & 0.91$\times$0.66 & -68.8 & 392.8 & /  & / & / & 0.19 & 46.0 & 2.6 & Binary  \\
OMC1N-4-5-W & 05:34:14.26 & -05:20:11.65 & 0.91$\times$0.69 & -68.9 & 392.8 & Starless  & / & / & 0.12 & / & / & Unresolved   \\
OMC1N-6-7 & 05:35:15.70 & -05:20:39.35 & 0.91$\times$0.66 & -69.4 & 392.8 & / & / & / & 0.25 & 98.0 & 3.6 & Extended  \\
OMC1N-8-N & 05:35:18.20 & -05:20:48.62 & 0.91$\times$0.66 & -69.5 & 392.8 & / & / & / & 0.19 & 31.0 & 3.0 & Multiple  \\
OMC1N-8-S & 05:35:18.01 & -05:20:55.82 & 0.91$\times$0.66 & -69.5 & 392.8 & / & / & / & 0.19 & 31.0 & 3.0 & Multiple  \\
\hline
\end{longtable}

\begin{table*}[h]
\caption{Spectral setup. The fourth spectral window includes C$^{17}$O, which is marked by an asterisk symbol.}
\begin{center}
\label{Tab:spectralsetup}
\begin{tabular}{l c c c c c}
\hline
\hline
Spectral window & Channels & Rest Frequency & Bandwidth & Spectral resolution & Velocity resolution \\
~ & ~ & (GHz) & (MHz) & (MHz) & (\kms) \\
\hline
1: CO (3--2) & 1920 & 345.79599 & 468.75 & 0.24 & 0.21 \\
2: continuum & 1920 & 348.50000 & 1875.00 & 0.98 & 0.84 \\
3: continuum & 1920 & 334.50000 & 1875.00 & 0.98 & 0.88 \\
4: continuum* & 1920 & 336.50000 & 1875.00 & 0.98 & 0.87 \\
\hline
\end{tabular}
\end{center}
\end{table*}

\begin{longtable}{lcccccccccccc}
\caption{\label{Tab:PA} Parameters of the entire sample.
Column 2 to 11 present the outflow position angle ($\theta_{\rm Out}$), outflow mean open angle ($\Delta \theta_{\rm Out}$), intensity-weighted {\em B}-field position angle within the field of view ($\langle \theta_{B_{1}} \rangle_{I}$) and within 400--1000 au ($\langle \theta_{B_{2}} \rangle_{I}$), uncertainty-weighted {\em B}-field position angle within 400--1000 au ($\langle \theta_{B} \rangle_{\sigma}$), position angle of velocity gradient ($\theta_{v}$), absolute velocity gradient ($\vert \nabla v_{\rm{abs}} \vert$), disk orientation along the major axis ($\theta_{\rm Disk}$), the inclination angle of the disk ($\theta_{\rm Inc}$), and the type of \textit{B}-field morphology, respectively.
$\theta_{\rm Disk}$ and $\theta_{\rm Inc}$ are obtained from \cite{tobin2020vla}.
The type of \textit{B}-field is classified as Standard Hourglass (Std-hourglass), Rotated Hourglass (Rot-hourglass), Spiral, Complex and Not enough data.
}\\
\hline\hline
\multirow{2}*{Name} &  $\theta_{\rm Out}$ & $\Delta \theta_{\rm Out}$ & $\langle \theta_{B_{1}} \rangle_{I}$ & $\langle \theta_{B_{2}} \rangle_{I}$ & $\langle \theta_{B} \rangle_{\sigma}$ & $\theta_{v}$ & $\vert \nabla v_{\rm{abs}} \vert$ & $\theta_{\rm Disk}$ & $\theta_{\rm Inc}$ & Type \\
~ & ($^{\circ}$) & ($^{\circ}$) & ($^{\circ}$) & ($^{\circ}$) & ($^{\circ}$) & ($^{\circ}$) & (km s$^{-1}$ arcsec $^{-1}$) & ($^{\circ}$) & ($^{\circ}$) &\textit{B}-field \\
\hline
\endfirsthead
\caption{Parameters of the entire sample.}\\
\hline\hline
\multirow{2}*{Name} &  $\theta_{\rm Out}$ & $\Delta \theta_{\rm Out}$ & $\langle \theta_{B_{1}} \rangle_{I}$ & $\langle \theta_{B_{2}} \rangle_{I}$ & $\langle \theta_{B} \rangle_{\sigma}$ & $\theta_{v}$ & $\nabla v_{\rm{abs}}$ & $\theta_{\rm Disk}$ & $\theta_{\rm Inc}$ & Type\\
~ & ($^{\circ}$) & ($^{\circ}$) & ($^{\circ}$) & ($^{\circ}$) & ($^{\circ}$) & ($^{\circ}$) & (km s$^{-1}$ arcsec $^{-1}$) & ($^{\circ}$) & ($^{\circ}$) & \textit{B}-field \\
\hline
\endhead
\hline
\endfoot
\textbf{Perp-Type} & ~ & ~ & ~ & ~ & ~ & ~ & ~ & ~ & ~ \\
\hline
HH270IRS & 176.1$\pm$6.6 & 92.7 & 76.9 & 68.9$\pm$0.8  & 64.4$\pm$7.0  & 64.6$\pm$1.8  & 1.2 & 87.4 & 58.8 & Rot-hourglass\\
HOPS-78  & 81.0$\pm$5.6  & 69.6 & 176.3 & 172.3$\pm$0.5 & 175.0$\pm$3.5 & 153.4$\pm$6.6 & 1.3 & 171.1 & 71.4 & Rot-hourglass \\
HOPS-88  & 81.1$\pm$5.6 & 90.4 & 31.5 & 32.8$\pm$1.4  & 27.9$\pm$5.1  & 149.3$\pm$4.0 & 1.3 & 166.9 & 25.8 & Complex \\
HOPS-96  & 49.1$\pm$5.6 & 75.1 & 113.3 & 100.0$\pm$1.6 & 101.5$\pm$2.6 & 123.2$\pm$6.8 & 1.1 &  134.6 & 23.6 & Spiral \\
HOPS-124 & 75.4$\pm$5.7 & 66.6 & 127.1 & 153.7$\pm$3.0 & 155.9$\pm$9.5 & 164.0$\pm$3.6 & 2.8  & 3.3 & 44.4 & Spiral \\
HOPS-288 & 39.4$\pm$5.8 & 47.6 & 124.2 & 122.8$\pm$0.4 & 119.4$\pm$3.5 & 118.3$\pm$3.8 & 3.7 & 145.6 & 64.6 & Rot-hourglass \\
HOPS-409 & 19.0$\pm$5.6 & 51.8 & 113.9 & 102.3$\pm$1.0 & 92.3$\pm$2.3 & 115.7$\pm$2.9 & 2.0 & 115.8 & 69.3 & Rot-hourglass \\
OMC1N-4-5-ES& 53.9$\pm$5.6 & 72.2 & 155.9 & 164.3$\pm$0.7 & 160.6$\pm$2.3 & 153.4$\pm$13.8 & 1.1 & / & /  & Rot-hourglass  \\
\hline
\textbf{Nonperp-Type} & ~ & ~ & ~ & ~ & ~ & ~ & ~ & ~ & ~ \\
\hline
HOPS-182 & 57.4$\pm$5.5 & 82.3 & 89.5 & 99.1$\pm$0.4  & 90.1$\pm$3.0  & 74.9$\pm$2.2  & 1.2  & 150.8 & 60.7 & Spiral \\
HOPS-310 & 136.5$\pm$5.9 & 75.0 & 48.9 & 49.3$\pm$0.7  & 46.6$\pm$2.6  & 116.7$\pm$3.9 & 2.2 & 45.7 & 58.9 & Rot-hourglass \\
HOPS-341 & 63.6$\pm$6.2 & 65.4 & 8.2 & 14.3$\pm$8.3   & 17.1$\pm$22.2  & 177.1$\pm$2.0 & 1.2 & 145.8 & 53.1 & Not enough data \\
HOPS-361N& 26.1$\pm$6.2 & 59.7 & 42.0 & 53.6$\pm$0.7  & 71.2$\pm$6.4  & 166.4$\pm$2.0 & 3.3 & 130.1 & 66.2 & Spiral \\
HOPS-361S& 26.6$\pm$8.7 & 79.8 & 170.4 & 178.3$\pm$2.3 & 166.4$\pm$6.7 & 147.7$\pm$5.0 & 1.2  & 14.4  & 54.8 & Spiral \\
HOPS-404 & 134.9$\pm$17.4 & 89.8 & 70.9 & 90.9$\pm$0.9  & 102.9$\pm$6.2  & 99.0$\pm$4.2 & 1.1 & 144.2 & 0  & Spiral \\
\hline
\textbf{Unres-Type} & ~ & ~ & ~ & ~ & ~ & ~ & ~ & ~ & ~ \\
\hline
HOPS-11  & 137.9$\pm$5.6 & 42.7 & 145.6 & 150.6$\pm$2.0 & 151.9$\pm$7.9 &  ~ & $\lesssim 1.0$ & 94.4 & 22.6 & Std-hourglass\\
HOPS-12W & 127.9$\pm$7.9 & 124.5 & 46.6 & 21.3$\pm$1.4  & 23.2$\pm$6.4  & ~ & $\lesssim 1.0$  & 103.3 & 33.6 & Complex \\
HOPS-87N  & 173.4$\pm$7.0 & 62.7 & 37.5 & 36.1$\pm$0.1  & 42.5$\pm$2.0  & ~ & $\lesssim 1.0$  & 11.1 & 13.5 & Std-hourglass \\
HOPS-164 & 64.2$\pm$5.5 & 56.4 & 170.2 & 168.3$\pm$4.3 & 166.0$\pm$5.0 & ~ & $\lesssim 1.0$  & 150.8 & 50.5 & Not enough data \\
HOPS-168 & 168.4$\pm$5.5 & 89.4 & 65.0 & 64.0$\pm$0.9  & 51.8$\pm$3.6  & ~& $\lesssim 1.0$  &  127.1 & 46.2 & Rot-hourglass \\
HOPS-169 & 174.2$\pm$5.5 & 43.1 & 66.9 & 48.4$\pm$1.2  & 45.5$\pm$4.6  & ~ & $\lesssim 1.0$  & 61.5 & 40.1 & Rot-hourglass \\
HOPS-224 & 79.7$\pm$6.3 & 62.0 & 65.9 & 71.4$\pm$1.7  & 73.3$\pm$5.4 & ~ & $\lesssim 1.0$  & 170.4 & 51.6 & Std-hourglass \\
HOPS-303 & 66.5$\pm$5.9 & 68.5 & 95.6 & 118.6$\pm$2.4  & 120.4$\pm$10.7  & ~ & $\lesssim 1.0$  & 164.0 & 37.7 & Spiral \\
HOPS-317N& 39.4$\pm$6.3 & 94.3 & 75.4 & 79.7$\pm$3.5   & 80.2$\pm$1.3 & ~ & $\lesssim 1.0$  & 48.3 & 25.8 & Not enough data \\
HOPS-317S& 87.2$\pm$17.3 & 78.7 & 7.3  & 10.1$\pm$0.3   & 3.8$\pm$5.2 & ~ & $\lesssim 1.0$ & 92.2 & 45.4 & Rot-hourglass \\
HOPS-325 & 17.1$\pm$6.1 & 56.4 & 157.3 & 154.9$\pm$2.1 & 146.1$\pm$5.9 &  ~  & $\lesssim 1.0$   &  119.0 & 25.8 & Rot-hourglass \\
HOPS-359 & 71.6$\pm$2.5 & 23.6 & 50.6 & 49.7$\pm$0.3  & 56.0$\pm$6.7  & ~ & $\lesssim 1.0$  &  3.6 & 28.1 & Std-hourglass \\
HOPS-370 & 16.7$\pm$5.6 & 114.5 & 110.5 & 108.4$\pm$0.3 & 109.9$\pm$2.0 & ~ & $\lesssim 1.0$  & 109.7 & 71.1 & Rot-hourglass \\
HOPS-373W & 89.2$\pm$6.1 & 72.4 & 35.0  & 30.9$\pm$1.0  & 25.1$\pm$4.6  & ~ & $\lesssim 1.0$ &  144.3 & 25.8 & Std-hourglass \\
HOPS-384 & 104.5$\pm$9.2 & 50.0 & 6.4 & 172.1$\pm$0.2 & 172.4$\pm$2.6 & ~ & $\lesssim 1.0$  & 60.2 & 45.6 & Spiral \\
HOPS-395 & 1.4$\pm$8.0  & 54.4 & 44.3 & 2.6$\pm$1.0  & 1.5$\pm$4.1  & ~ &  $\lesssim 1.0$ &  79.9 & 0 & Complex \\
HOPS-399 & 163.9$\pm$5.8 & 78.0 & 101.5 & 96.4$\pm$0.1  & 96.3$\pm$3.6  & ~ & $\lesssim 1.0$   & 143.6 & 37.9 & Complex \\
HOPS-400 & 81.1$\pm$6.0 & 83.8 & 90.1 & 93.3$\pm$0.2  & 95.4$\pm$2.7  & ~ & $\lesssim 1.0$  & 18.5 & 21.3 & Std-hourglass \\
HOPS-401 & 69.7$\pm$4.3 & 30.8 & 144.4 & 148.4$\pm$1.7 & 148.5$\pm$6.3 & ~ & $\lesssim 1.0$  & 176.9 & 40.4 & Rot-hourglass \\
HOPS-403 & 62.2$\pm$6.1 & 70.5 & 134.3 & 107.9$\pm$1.6 & 104.5$\pm$8.8 & ~ & $\lesssim 1.0$  & 64.0 & 13.0 & Spiral \\
HOPS-407 & 98.9$\pm$7.5 & 39.0 & 88.8 & 88.2$\pm$0.3  & 82.4$\pm$2.3  & ~ & $\lesssim 1.0$  & 173.6 & 38.9 & Std-hourglass \\
HOPS-408 & 85.7$\pm$8.2 & 83.5 & 17.3 & 10.5$\pm$1.0  & 10.2$\pm$1.5  & ~ &  $\lesssim 1.0$  & 157.5 & 31.0 & Not enough data \\
OMC1N-4-5-EN & 113.7$\pm$5.6 & 35.9 & 76.4 & 74.5$\pm$0.6  & 71.2$\pm$2.8  &  ~ & $\lesssim 1.0$  &  / & / & Complex \\
OMC1N-6-7  & 72.7$\pm$5.6 & 103.4 & 18.7 & 12.8$\pm$1.5  & 17.2$\pm$5.8  & ~ & $\lesssim 1.0$  & / & / & Rot-hourglass  \\
OMC1N-8-N  & 130.9$\pm$5.6 & 61.2 & 117.8 & 122.3$\pm$0.7 & 125.8$\pm$2.5 & ~ & $\lesssim 1.0$ & / & / & Std-hourglass \\
\hline
\textbf{Other} & ~ & ~ & ~ & ~ & ~ & ~ & ~ & ~ & ~ \\
\hline
HOPS-250 & 134.5$\pm$6.1 & 98.6 & 142.2 & / & / &   & $\lesssim 1.0$ & 43.4 & 53.6 & Not enough data \\
HOPS-373E & / & / & 145.0  & 149.1$\pm$1.0  & 154.9$\pm$4.6  & / & / &  144.3 & 25.8 & Complex \\
HOPS-398 & / & / & 92.5 & 52.4$\pm$1.6  & 20.0$\pm$9.0  & / & / & 10.6 & 30.3 & Complex  \\
HOPS-402 & / & / & 90.3 & 90.9$\pm$0.7 & 90.6$\pm$1.0 & / & / & 58.7 & 47.6 & Not enough data \\ 
HOPS-87S & 58.7$\pm$8.0 & 86.5 & / & / & / & / & / & / & / & / \\
OMC1N-8-S & 120.2$\pm$7.0 & 99.3 & / & / & / & / & / & / & / & / \\ 
OMC1N-4-5-W & / & /& / & / & / & /& / & / & /  & / \\
\end{longtable}

\begin{table}[ht]
\begin{center}
\caption{Self-scattering sources.
Column 2 to 6 present the outflow position angle ($\theta_{\rm Out}$), outflow mean open angle ($\Delta \theta_{\rm Out}$), intensity-weighted position angle of polarization within inner 400 au scale ($\langle \theta_{\rm Pol} \rangle_{I})$, disk orientation along the major axis ($\theta_{\rm Disk}$), and the inclination angle of the disk ($\theta_{\rm Inc}$), respectively.
$\theta_{\rm Disk}$ and $\theta_{\rm Inc}$ are obtained from \cite{tobin2020vla}.
}
\label{Tab:SelfScattering}
\begin{tabular}{l c c c c c c c c c c c}
\hline
\hline
\multirow{2}*{Name} &  $\theta_{\rm Out}$ & $\Delta \theta_{\rm Out}$ & $\langle \theta_{\rm Pol} \rangle_{I}$ & $\theta_{\rm Disk}$ & $\theta_{\rm Inc}$\\
~ & (\arcdeg) & (\arcdeg) & (\arcdeg) & (\arcdeg) & (\arcdeg)  \\
\hline
HH212M   & 23.5$\pm$6.1  & 51.2  & 27.9$\pm$0.4  & 118.6 & 63.0  \\
HOPS-10  & 37.6$\pm$5.6  & 75.2  & 30.6$\pm$2.3  & 116.6 & 60.0  \\
HOPS-12E & 152.5$\pm$5.6 & 41.7  & 147.7$\pm$2.9 & 76.6  & 38.9  \\
HOPS-50  & 164.1$\pm$7.9 & 115.5 & 147.3$\pm$0.8 & 68.3  & 58.7  \\
HOPS-53  & 14.4$\pm$5.6  & 69.9  & 37.5$\pm$2.1  & 128.7 & 44.4  \\
HOPS-60  & 61.1$\pm$5.6  & 86.3  & 74.7$\pm$0.6  & 159.0 & 54.6  \\
HOPS-81  & 30.0$\pm$5.6  & 72.8  & 52.7$\pm$3.1  & 122.0 & 53.1  \\
HOPS-84  & 83.7$\pm$5.6  & 62.4  & 78.1$\pm$0.4  & 167.5 & 62.5  \\
HOPS-153 & 127.0$\pm$5.6 & 44.6  & 120.8$\pm$0.6 & 32.4  & 75.1  \\
HOPS-203N& 139.2$\pm$5.5 & 50.6  & 141.1$\pm$0.6 & 50.2  & 67.1  \\
HOPS-203S& 97.3$\pm$7.8  & 50.2  & 92.9$\pm$1.5  & 175.4 & 53.1  \\
HOPS-247 & 110.0$\pm$7.7 & 91.9  & 105.3$\pm$0.5 & 17.9  & 45.9  \\
HOPS-340 & 7.2$\pm$6.2   & 55.6  & 24.9$\pm$3.6  & 99.7  & 50.5  \\
HOPS-354 & 52.4$\pm$5.1  & 100.3 & 57.2$\pm$3.2  & 158.6 & 29.0  \\
HOPS-358 & 156.2$\pm$6.1 & 57.5  & 171.5$\pm$0.7 & 81.2  & 74.7  \\
HOPS-383 & 139.2$\pm$5.6 & 52.0  & 130.3$\pm$1.2 & 49.7  & 48.2  \\
\hline
\end{tabular}
\end{center}
\end{table}

\pagebreak
\section{Figures} \label{App:Figure} 

\begin{figure*}[ht]
\centering
\caption{Self-scattering dominated protostars. 
First column: 870~$\mu$m dust polarization intensity in color scale overlaid with redshifted and blueshifted outflow lobes (obtained from the $^{12}$CO (3--2) line), the polarization segments, and the dust continuum emission (Stokes {\em I}) contours.
Blue contours indicate the blueshifted outflow, while red contours are redshifted outflow, with counter levels set at 5 times the outflow {\em rms} × (1, 2, 4, 8, 16, 32).
The magenta segments represent the polarization.
The regions of polarization intensity less than 3$\sigma$ have been masked.
Second column: the velocity field in color scale (obtained from the C$^{17}$O (3--2) line) overlaid with the polarization segments and Stokes {\em I} contours.
Third column: an enlarged perspective of 1000 au of the second column.
The magenta arrow indicates the mean polarization direction weighted by the intensity within inner region of 400 au.
In the second and third panels, the red and blue arrows indicate the mean direction of the red-shifted and blue-shifted outflows.
For the velocity field, regions with an S/N less than 4 have been flagged.
In all panels, the black contour levels for the Stokes {\textit I} image are 10 times the {\em rms} × (1, 2, 4, 8, 16, 32, 64, 128, 256, 512).
The black dotted square in the first column corresponds to 2000 au scale, while the black dashed circles correspond to scales of 1000 au, respectively.
Black solid circle indicates the disk size of the protostars, obtained from \cite{tobin2020vla}.}
\includegraphics[clip=true,trim=0cm 1cm 0cm 2cm,width=0.49 \textwidth]{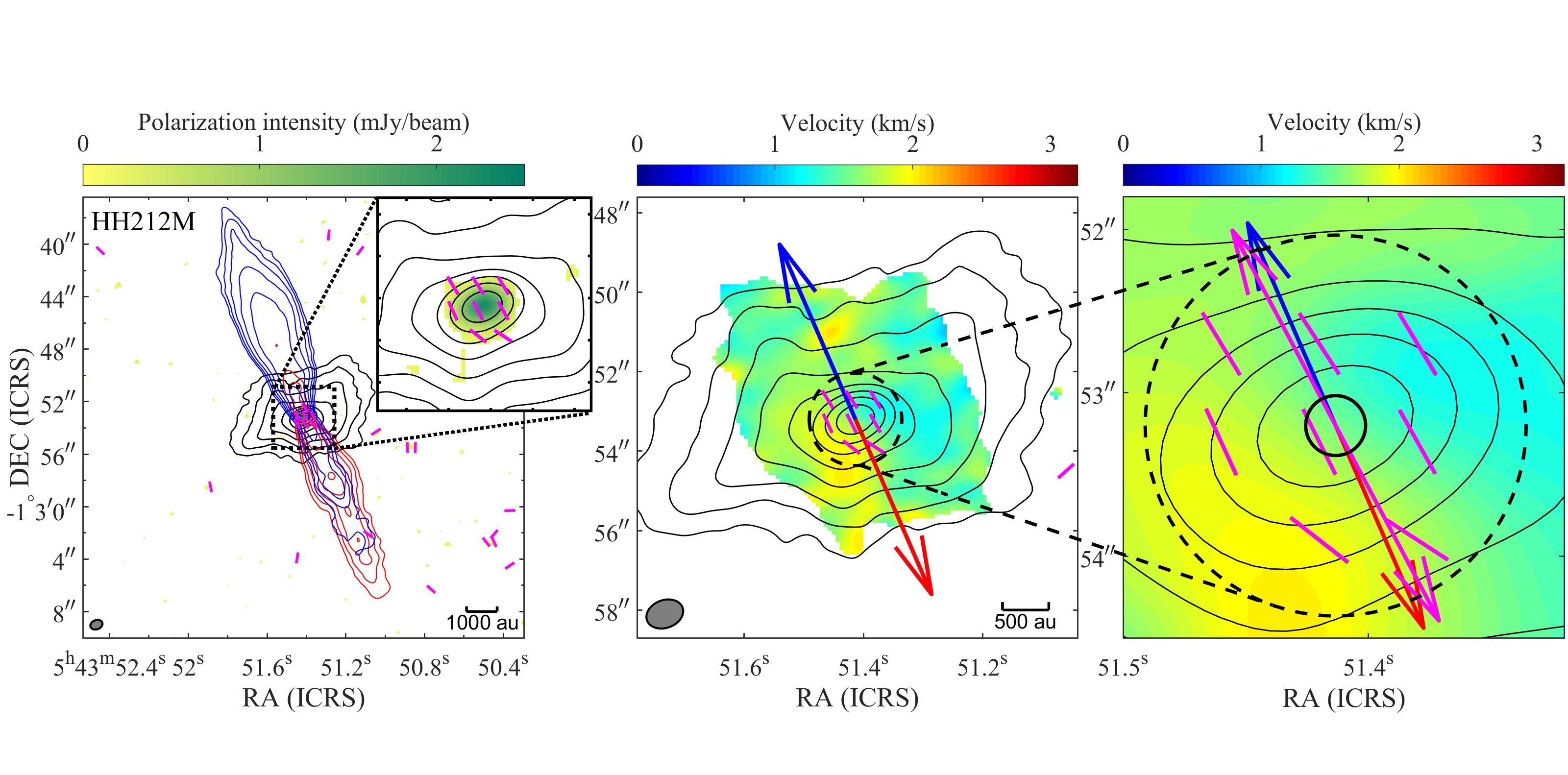}
\includegraphics[clip=true,trim=0cm 1cm 0cm 2cm,width=0.49 \textwidth]{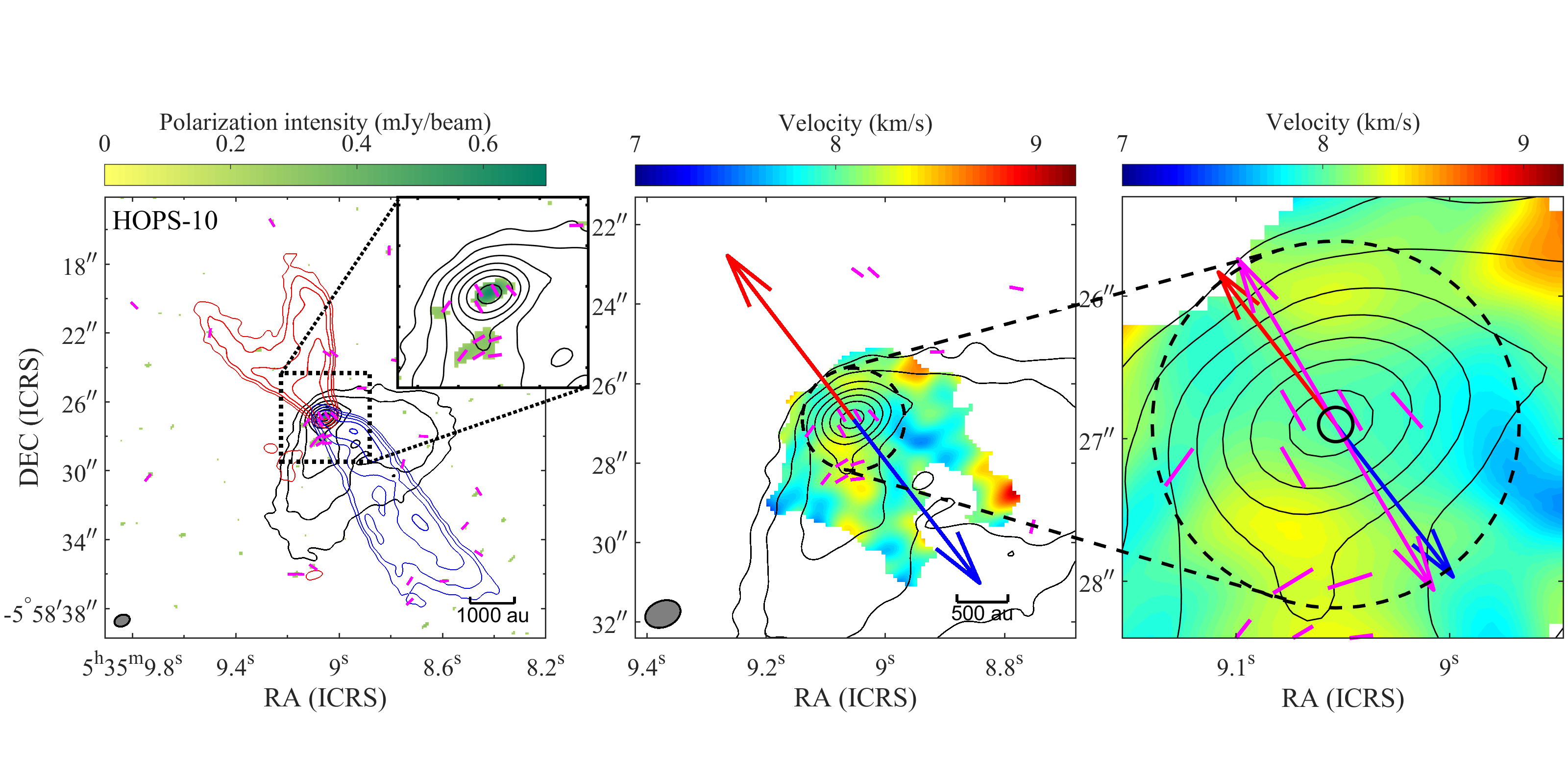}
\includegraphics[clip=true,trim=0cm 1cm 0cm 2cm,width=0.49 \textwidth]{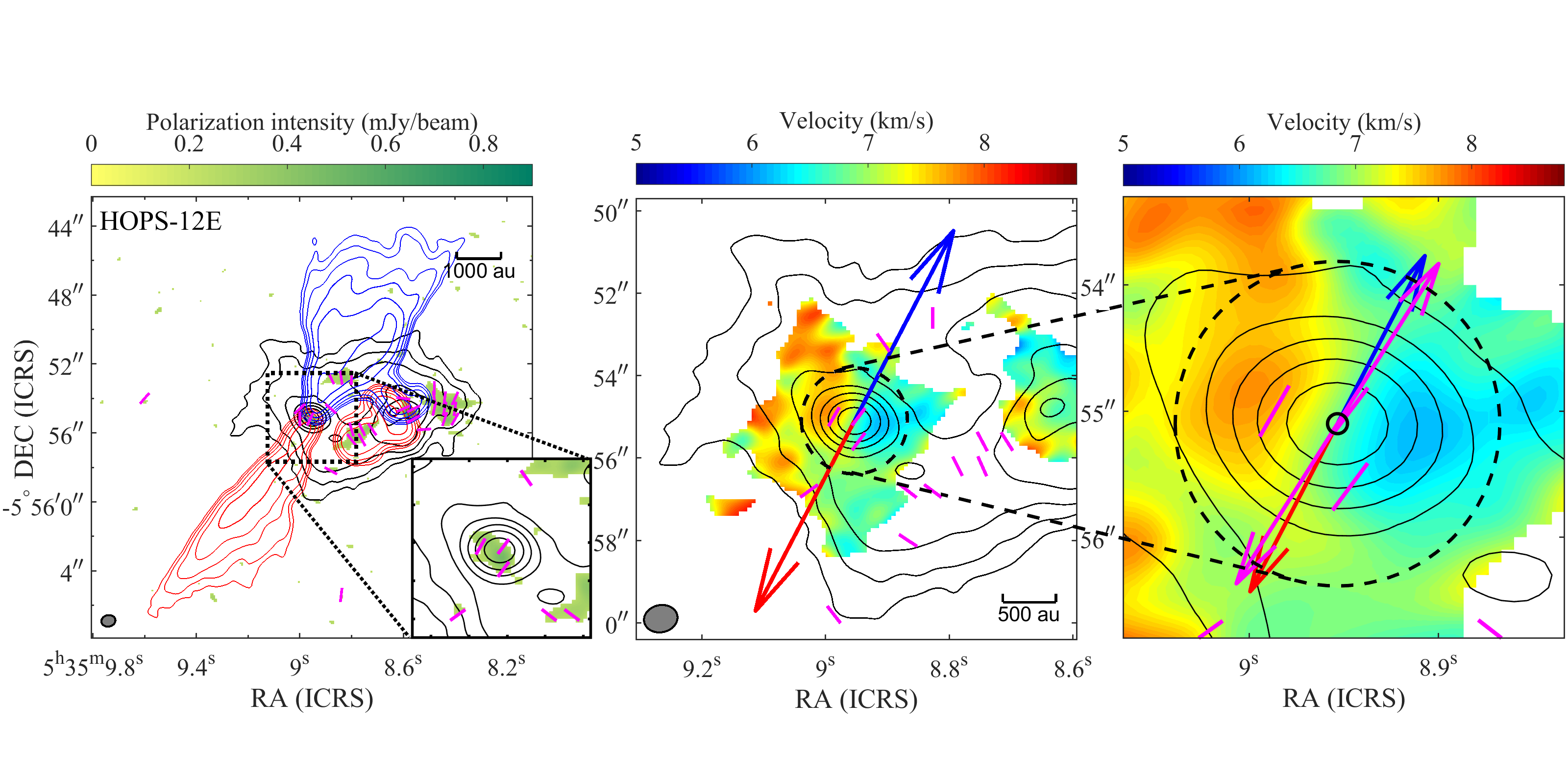}
\includegraphics[clip=true,trim=0cm 1cm 0cm 2cm,width=0.49 \textwidth]{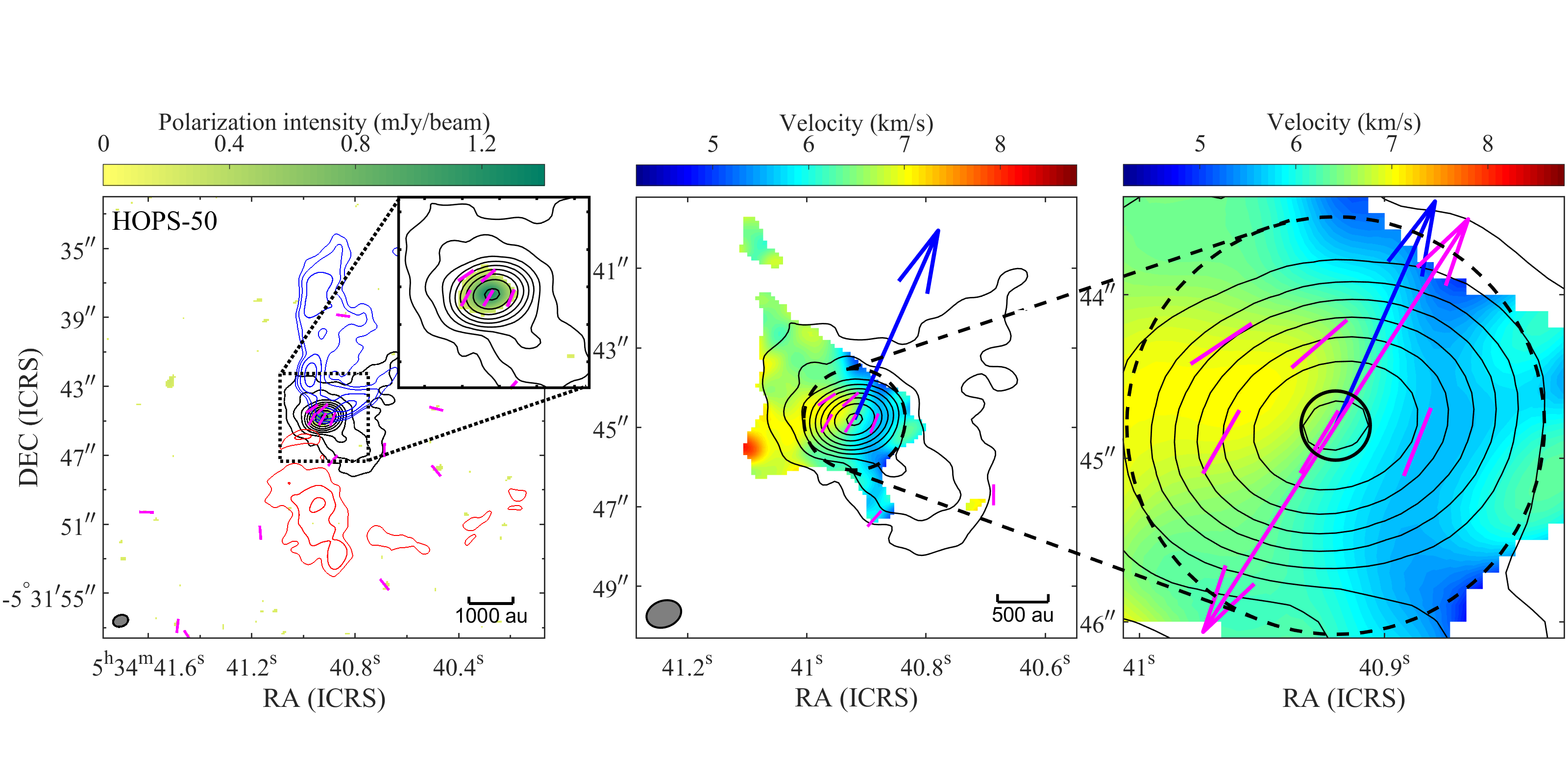}
\includegraphics[clip=true,trim=0cm 1cm 0cm 2cm,width=0.49 \textwidth]{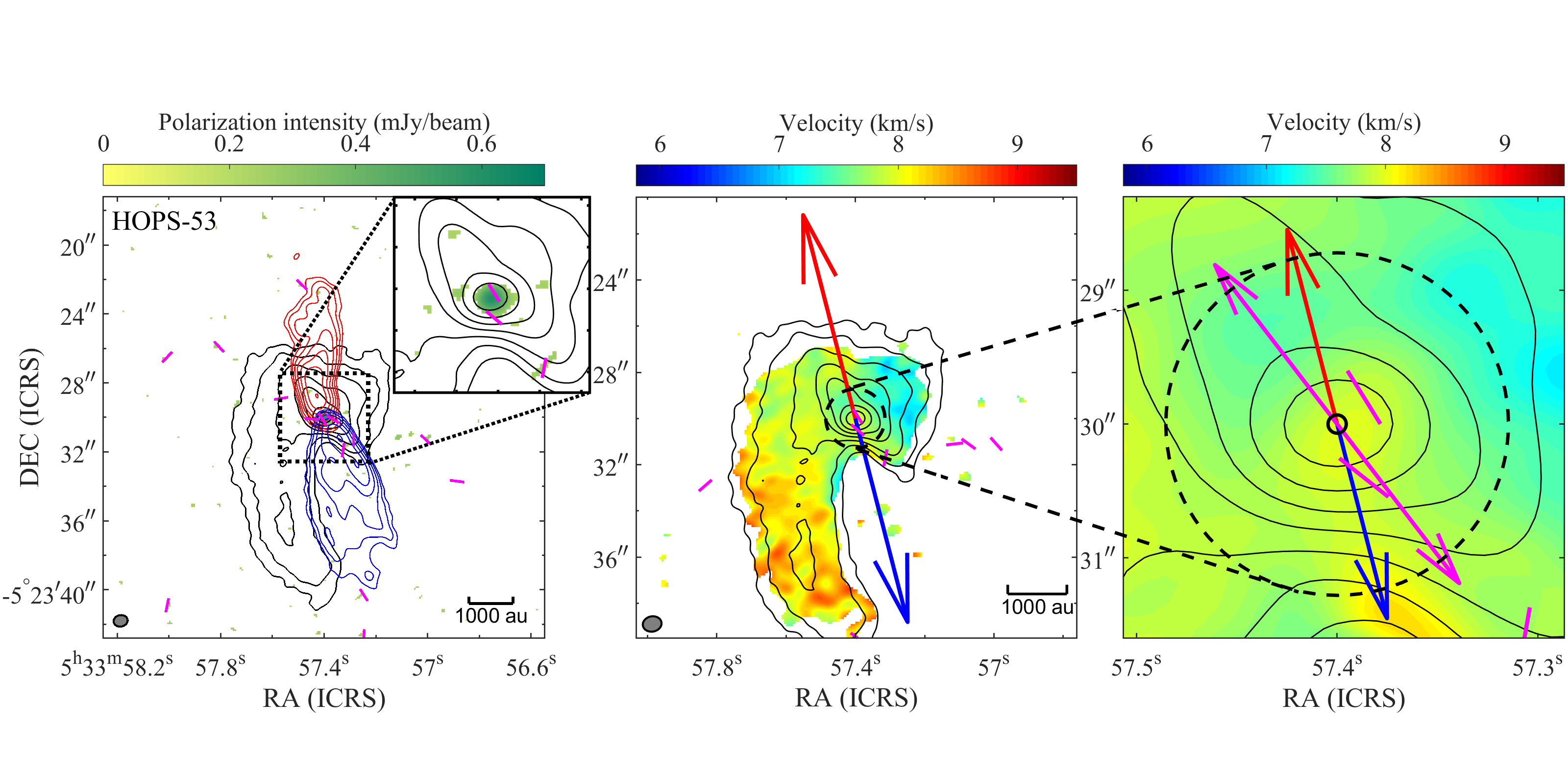}
\includegraphics[clip=true,trim=0cm 1cm 0cm 2cm,width=0.49 \textwidth]{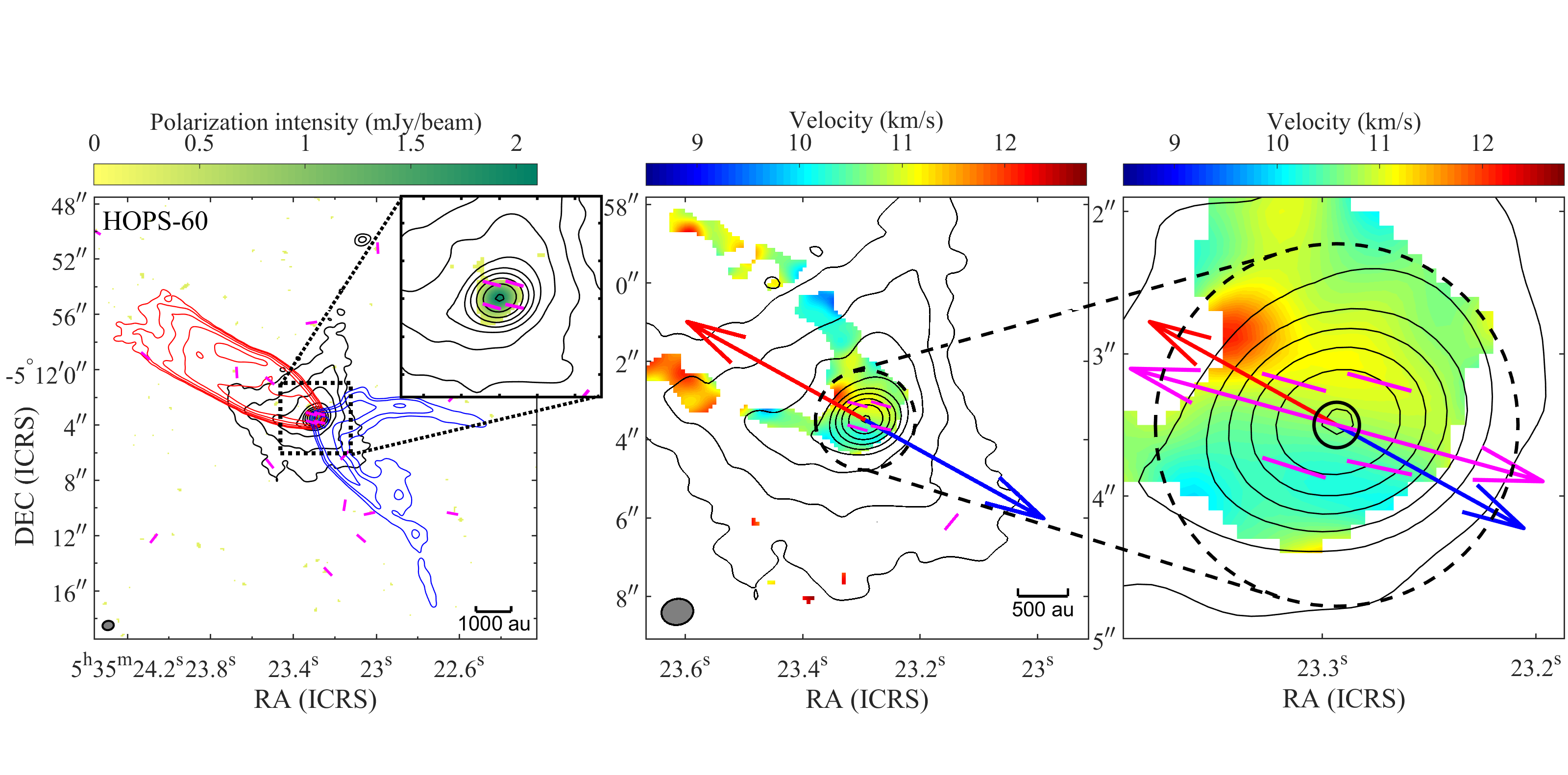}
\includegraphics[clip=true,trim=0cm 1cm 0cm 2cm,width=0.49 \textwidth]{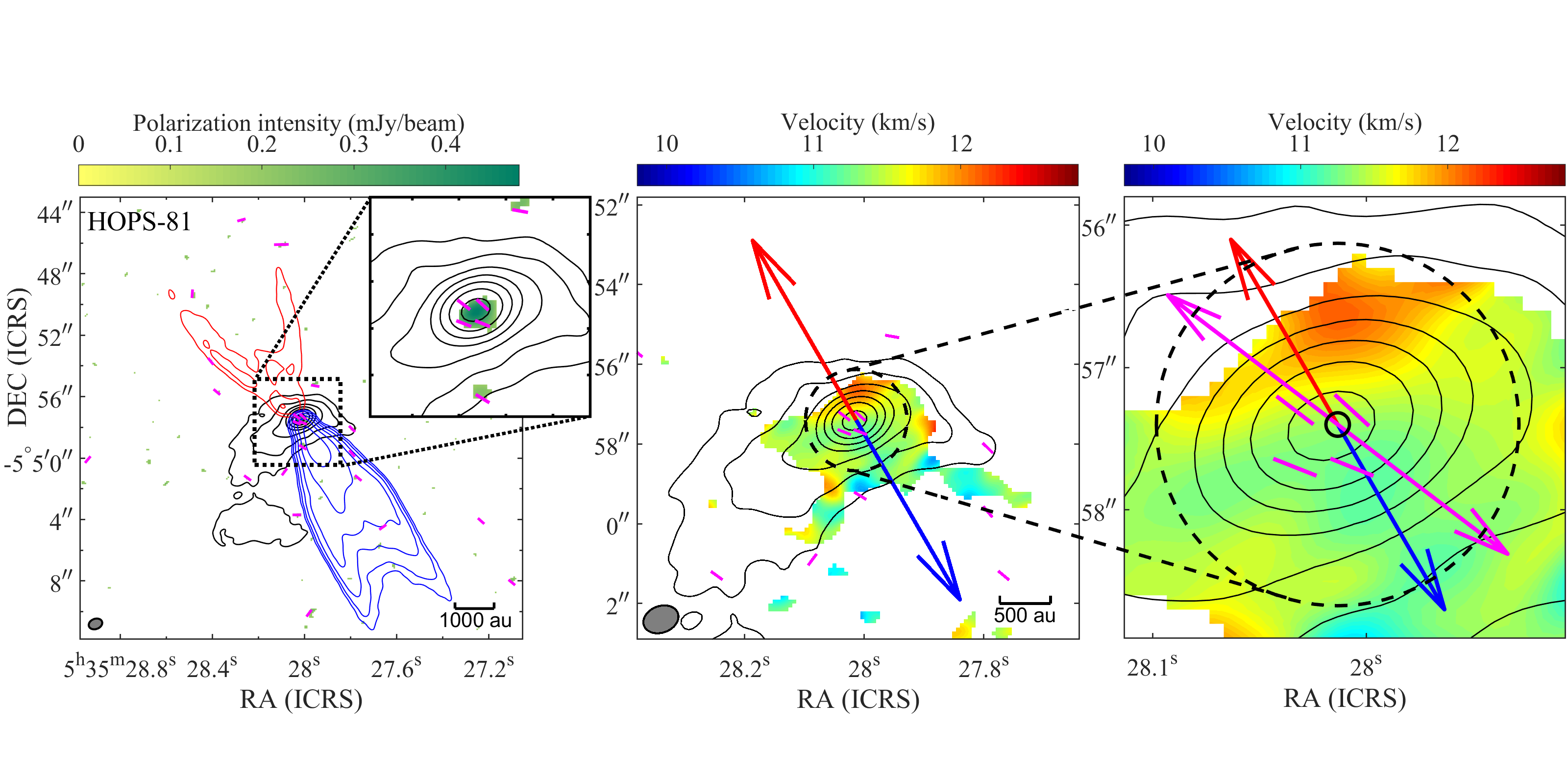}
\includegraphics[clip=true,trim=0cm 1cm 0cm 2cm,width=0.49 \textwidth]{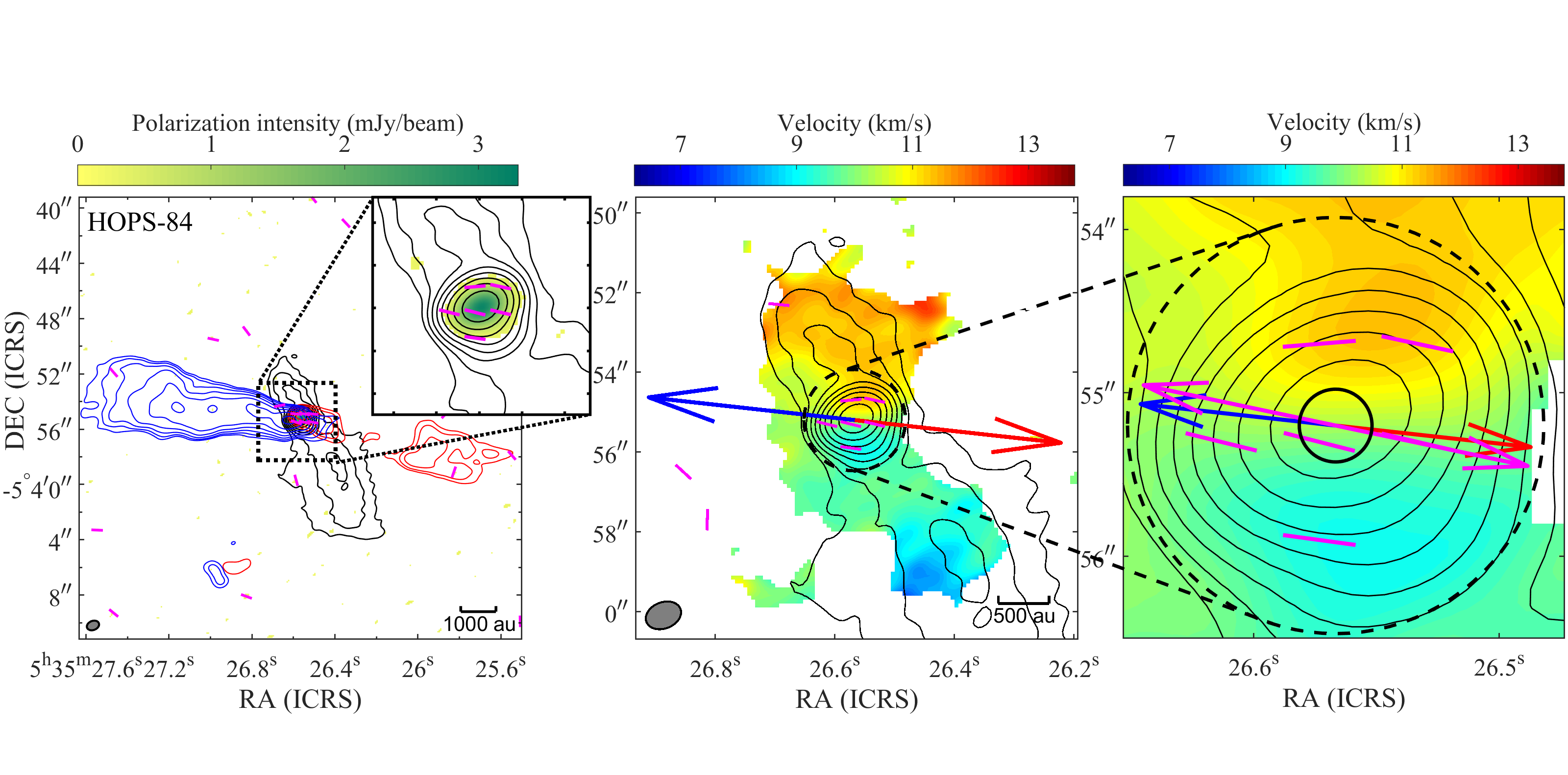}
\end{figure*}

\begin{figure*}
\centering
\text{\textbf{Figure 7.} Self-scattering dominated protostars.}\\
\includegraphics[clip=true,trim=0cm 1cm 0cm 2cm,width=0.49 \textwidth]{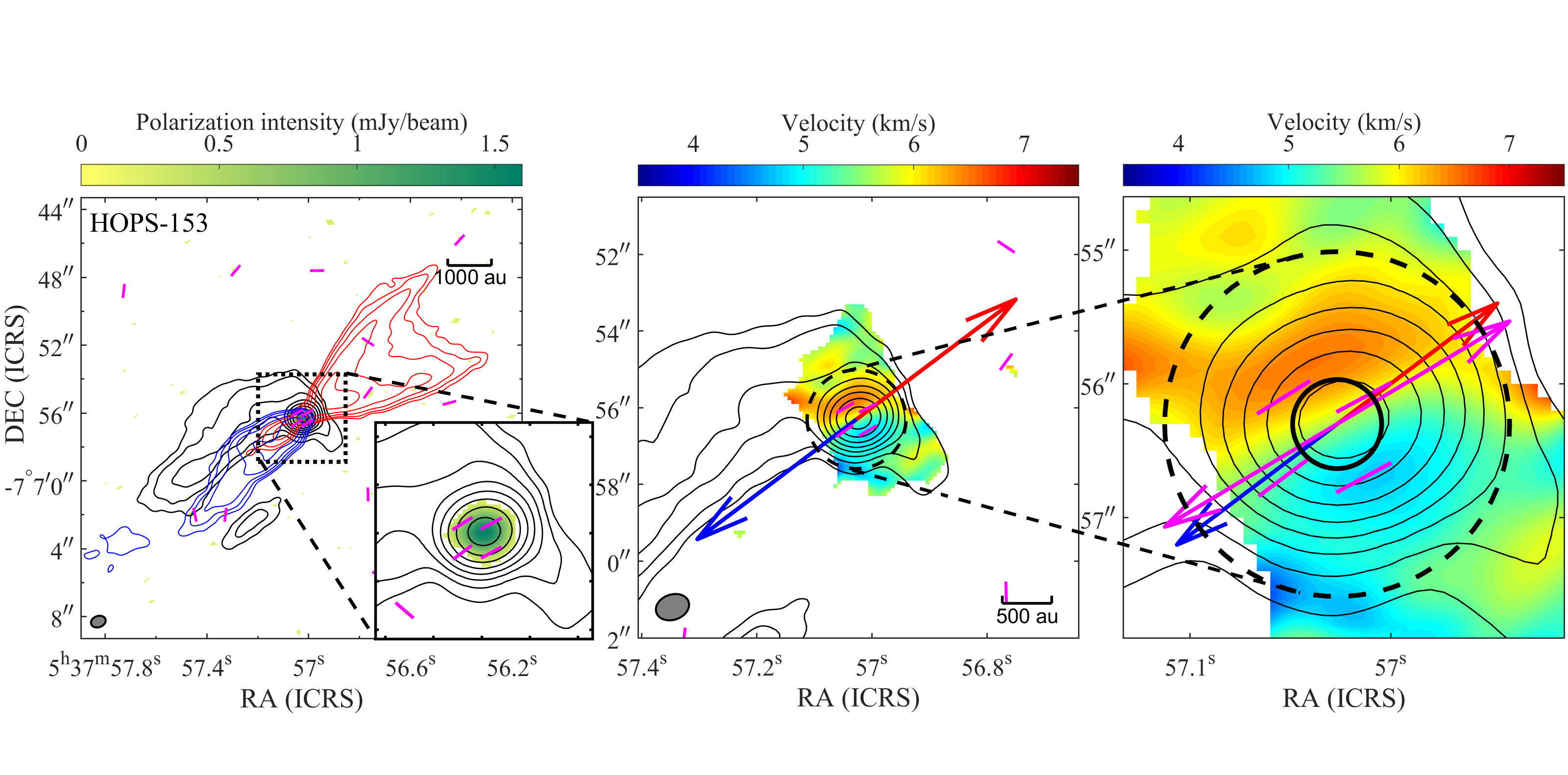}
\includegraphics[clip=true,trim=0cm 1cm 0cm 2cm,width=0.49 \textwidth]{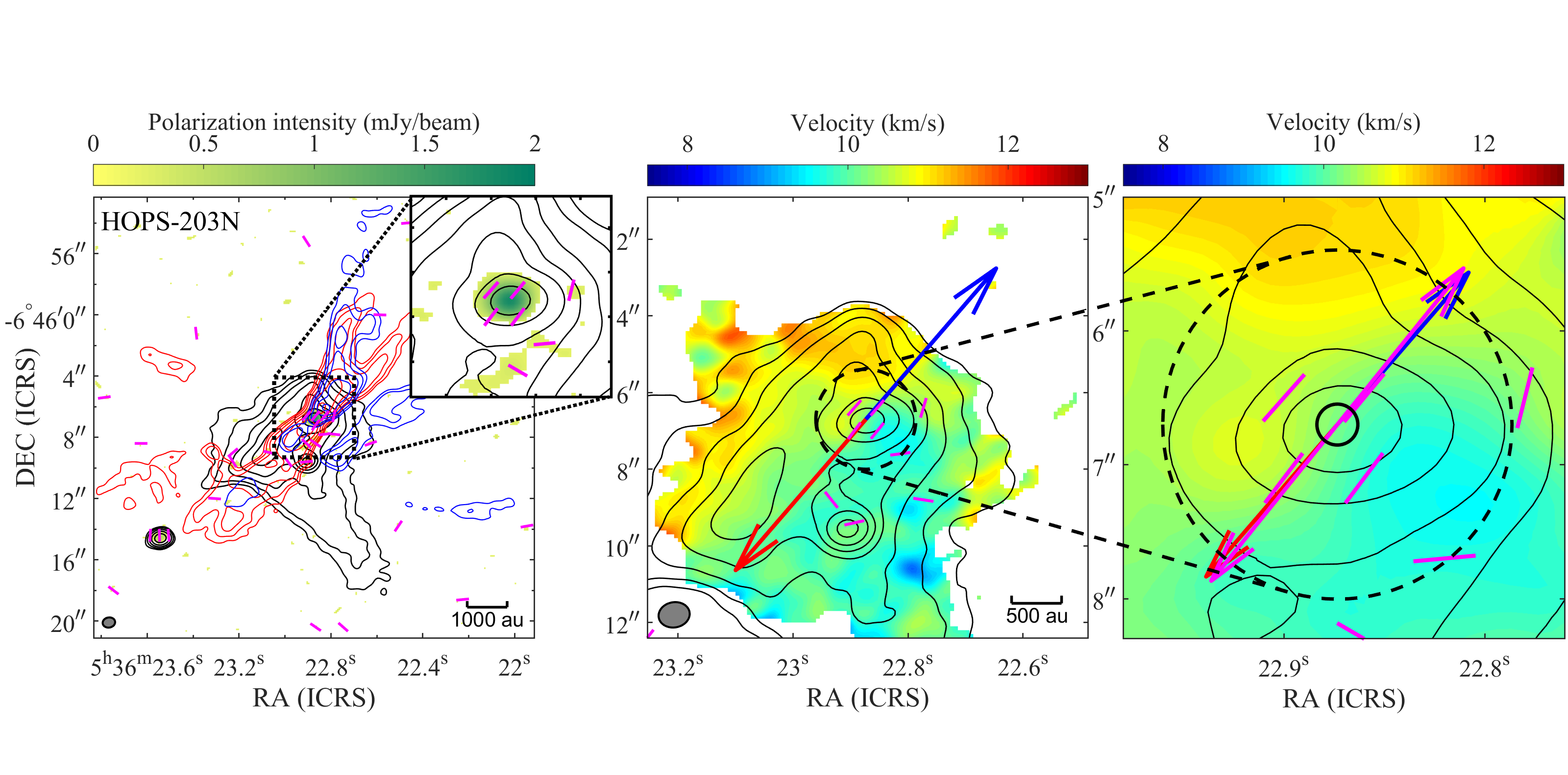}
\includegraphics[clip=true,trim=0cm 1cm 0cm 2cm,width=0.49 \textwidth]{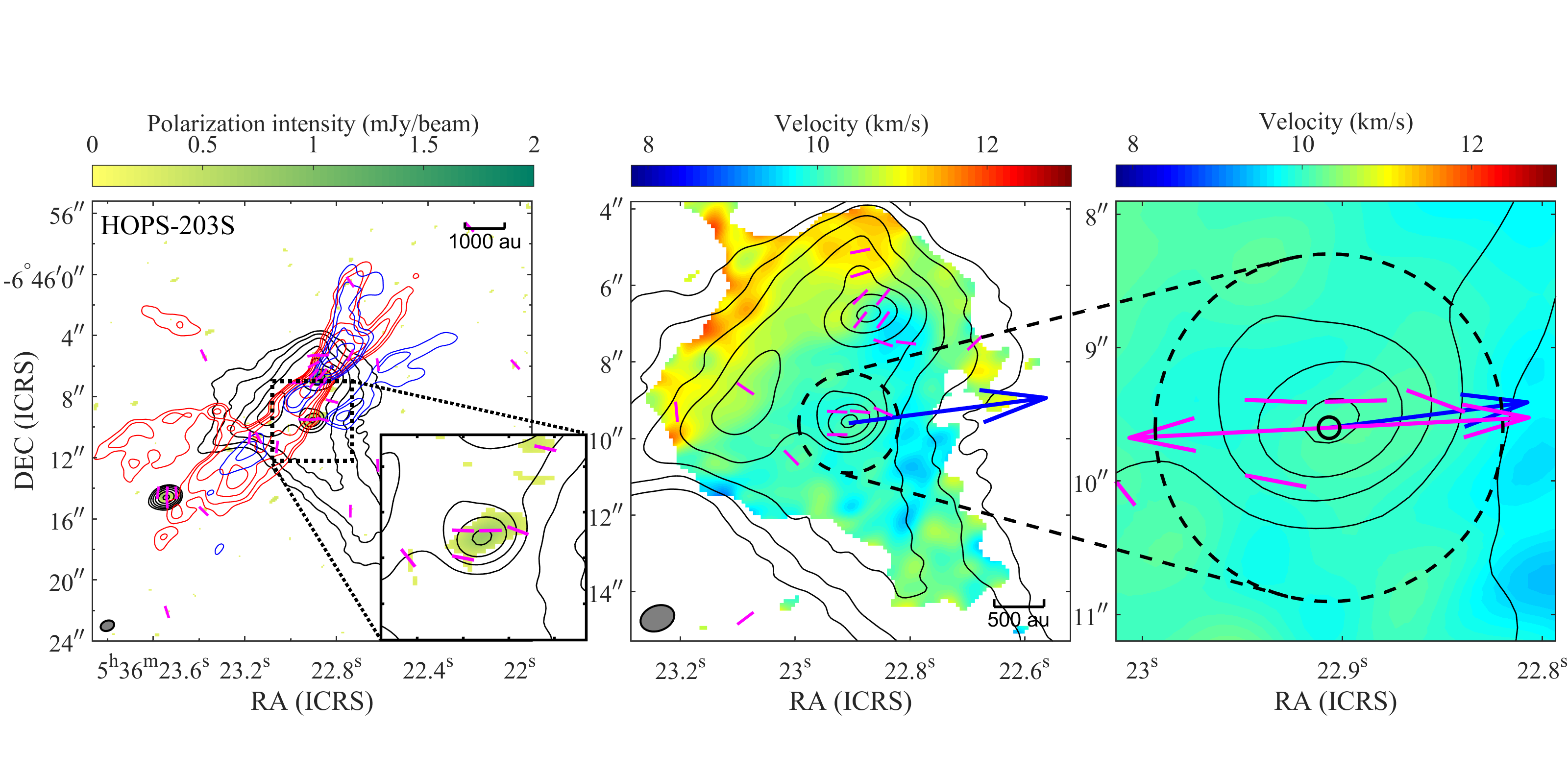}
\includegraphics[clip=true,trim=0cm 1cm 0cm 2cm,width=0.49 \textwidth]{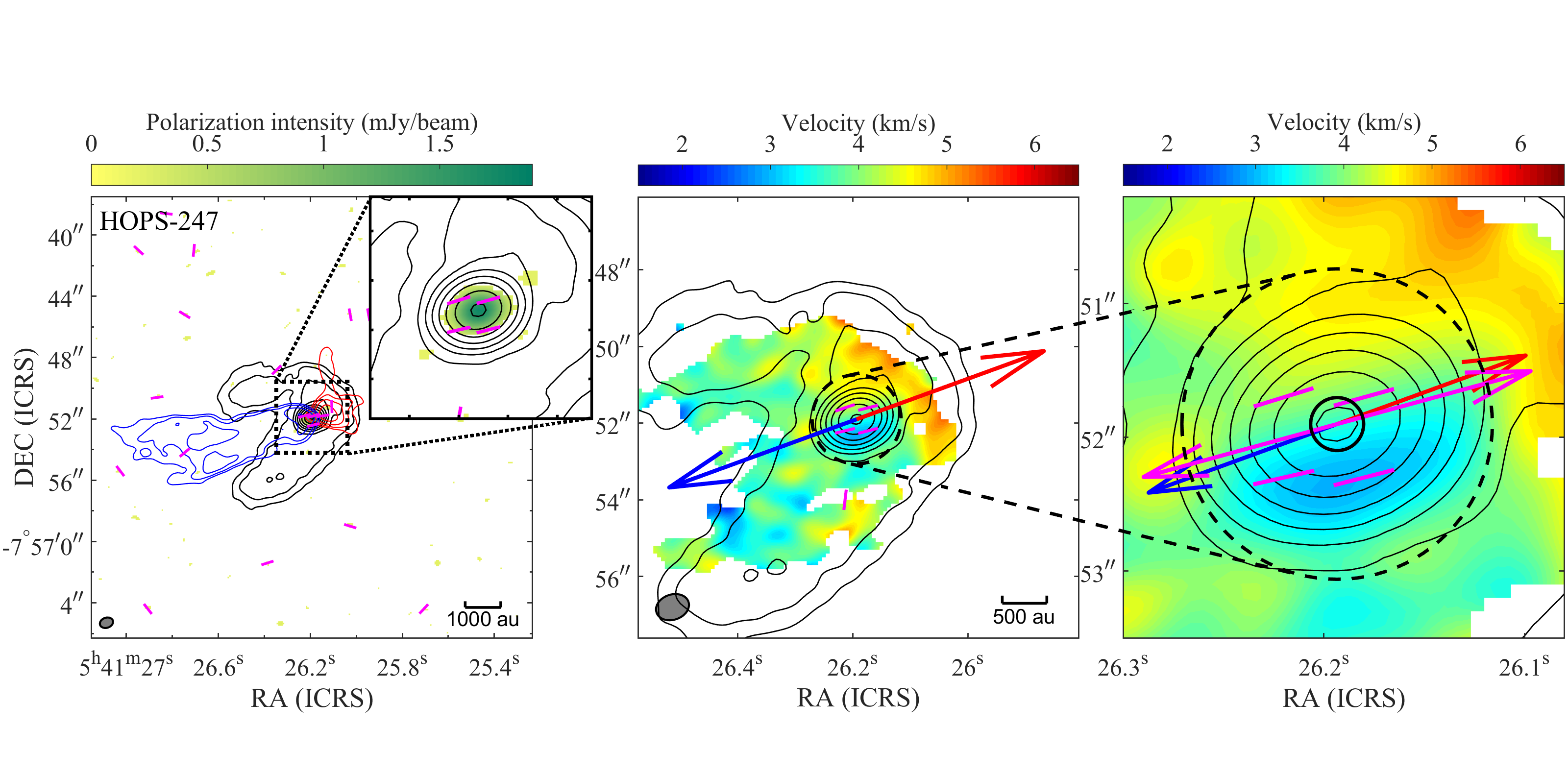}
\includegraphics[clip=true,trim=0cm 1cm 0cm 2cm,width=0.49 \textwidth]{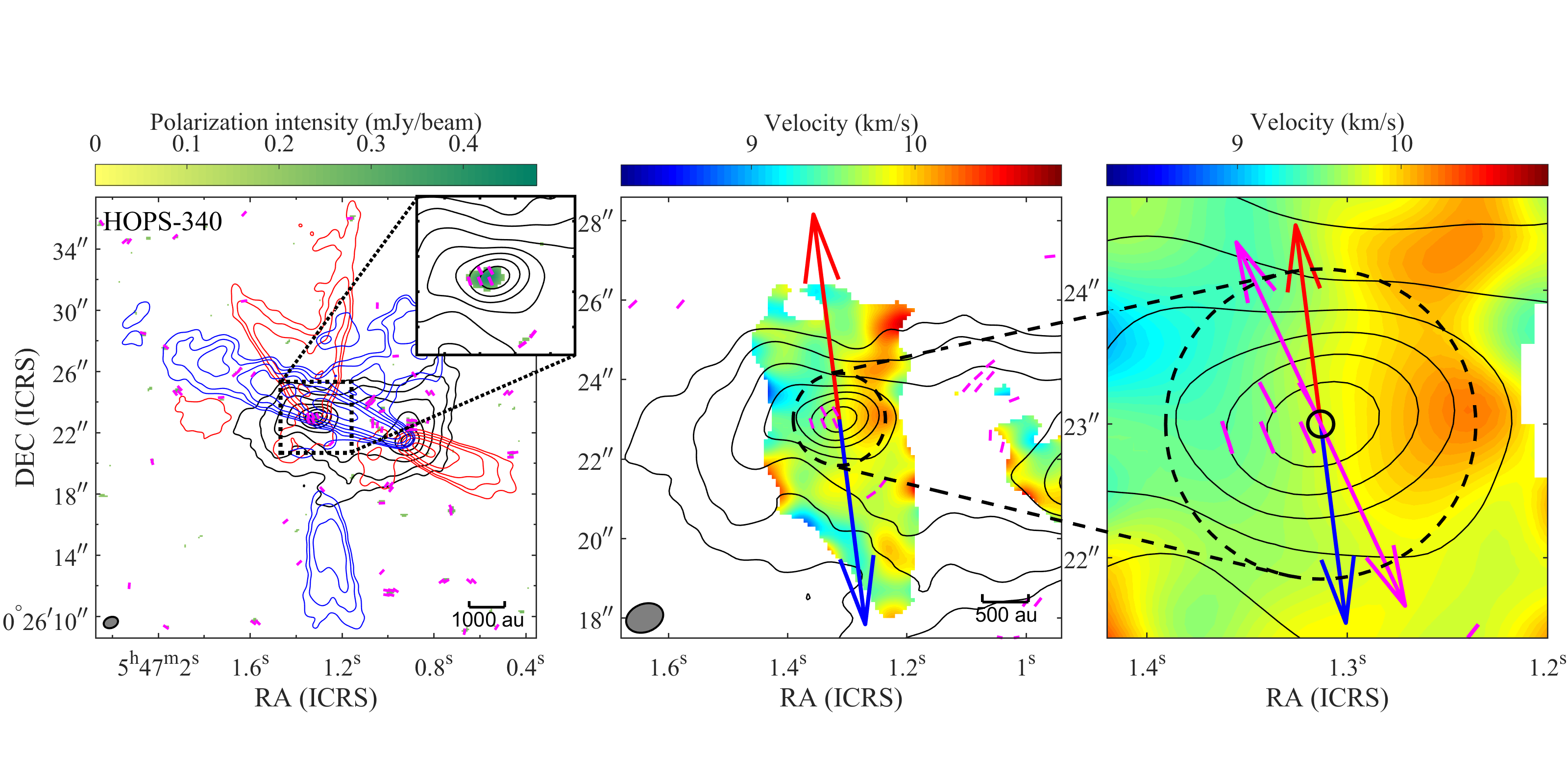}
\includegraphics[clip=true,trim=0cm 1cm 0cm 2cm,width=0.49 \textwidth]{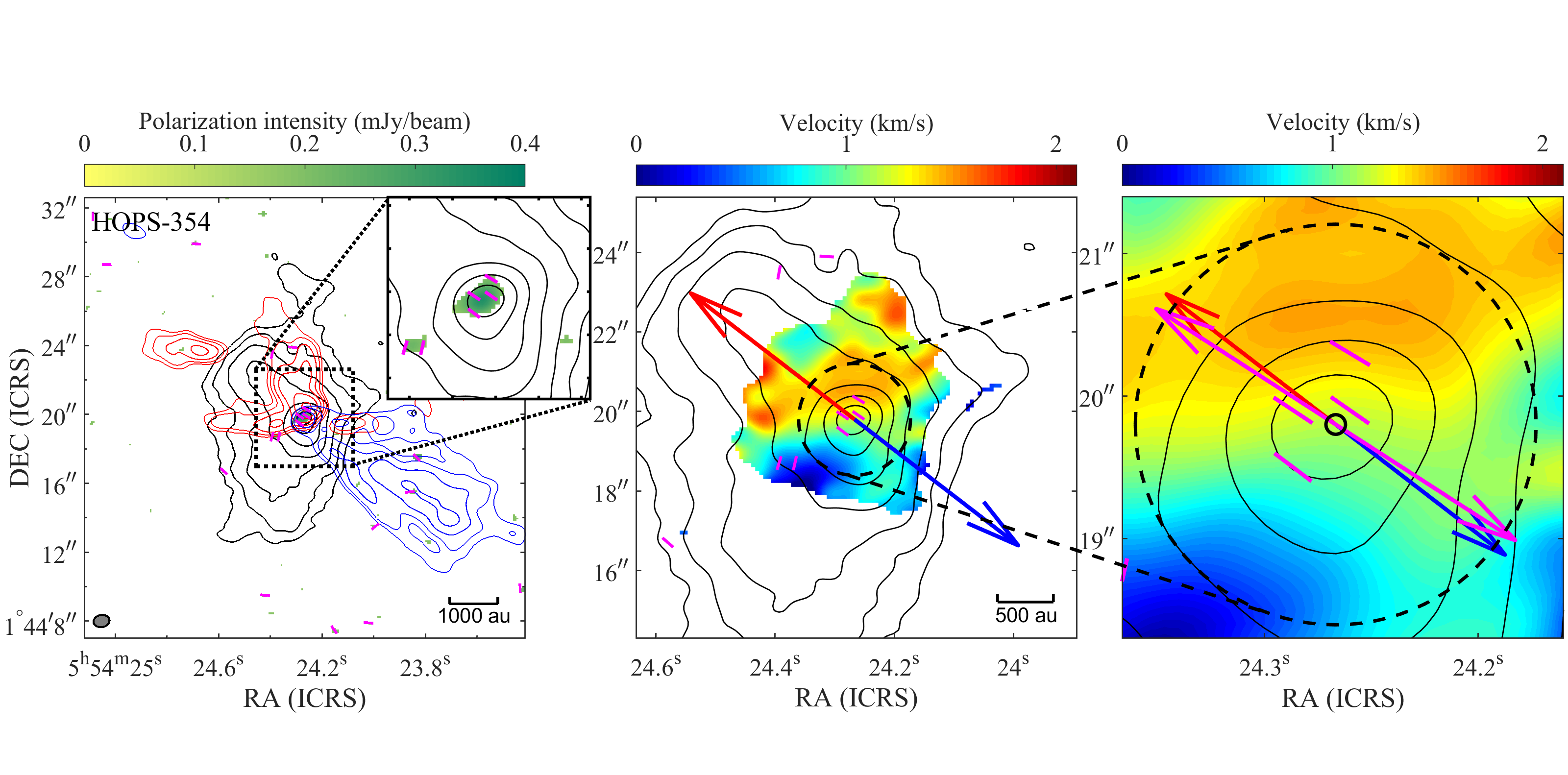}
\includegraphics[clip=true,trim=0cm 1cm 0cm 2cm,width=0.49 \textwidth]{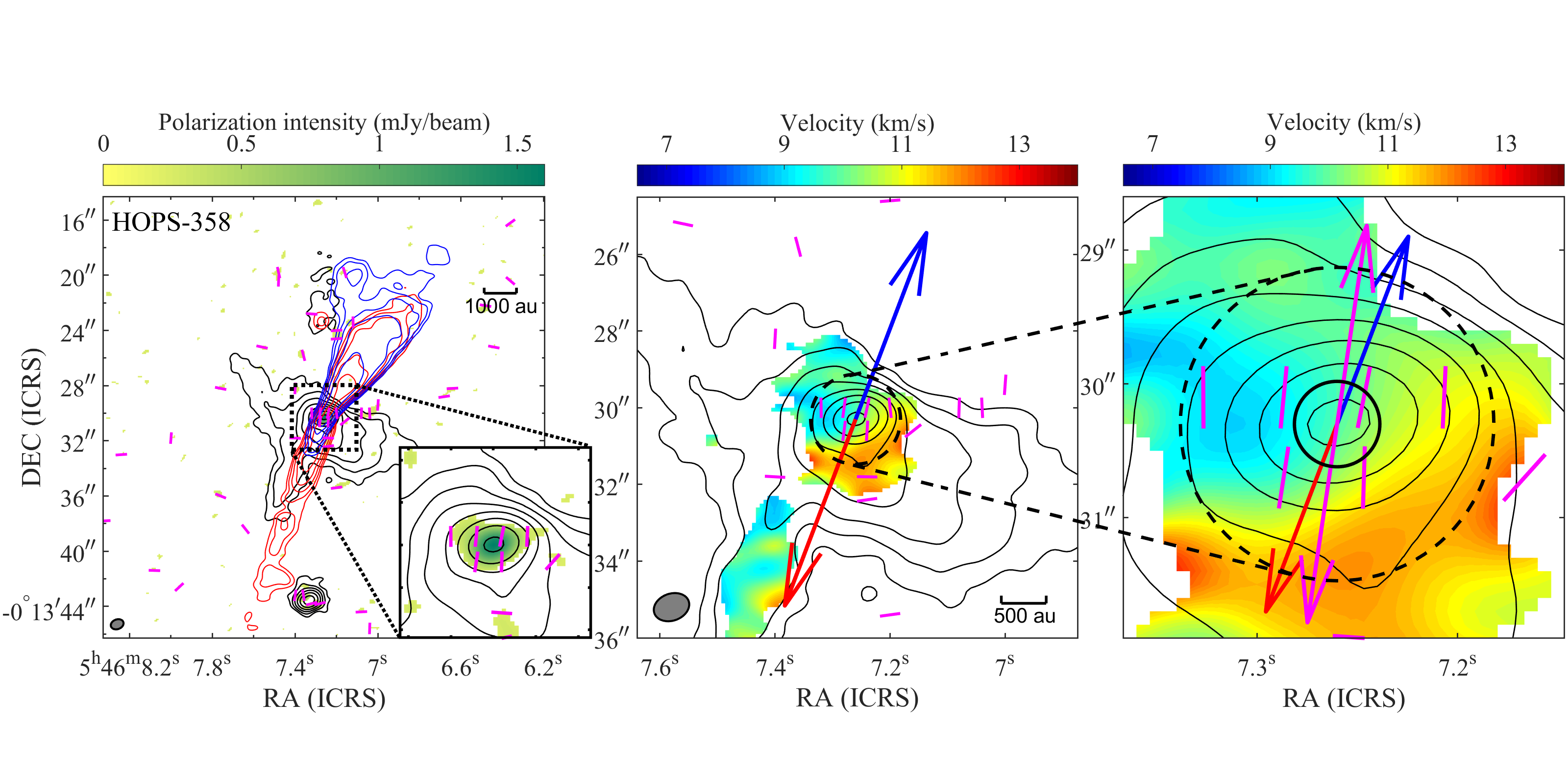}
\includegraphics[clip=true,trim=0cm 1cm 0cm 2cm,width=0.49 \textwidth]{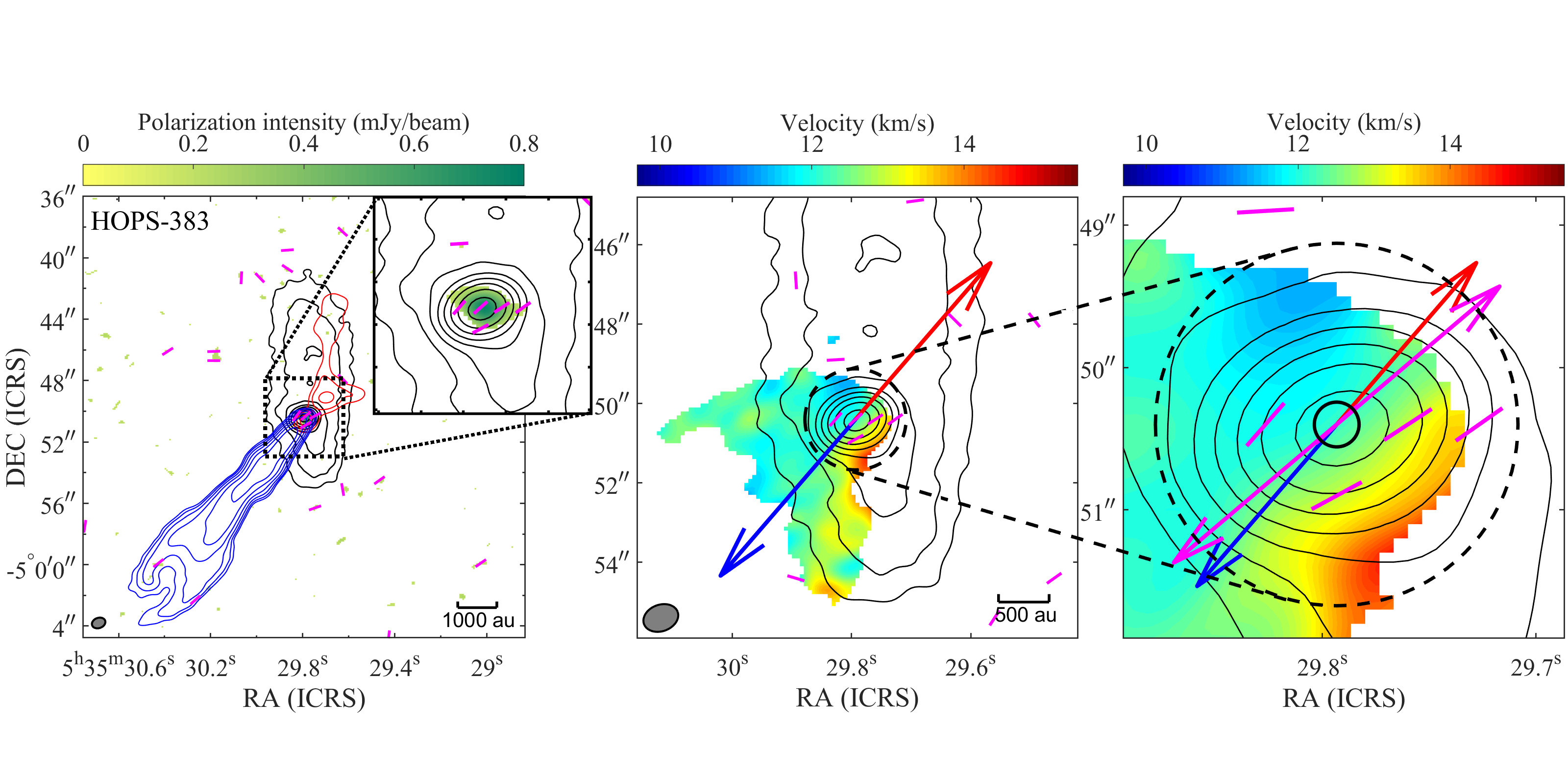}\\
\end{figure*}

\begin{figure*}
\centering
\caption{Perp-Type: velocity gradient direction $\perp$ outflow direction ($67.5^{\circ}\le \vert \theta_{\rm Out}-\theta_{v} \vert \le 90^{\circ}$). 
First column: 870~$\mu$m dust polarization intensity in color scale overlaid with the redshifted and blueshifted outflow lobes (obtained from the $^{12}$CO (3--2) line), polarization segments, and dust continuum emission (Stokes {\em I}) contours.
Blue contours indicate the blueshifted outflow, while red contours are redshifted outflow, with counter levels set at 5 times the outflow {\em rms} × (1, 2, 4, 8, 16, 32).
The magenta segments represent the polarization.
The regions of polarization intensity less than 3$\sigma$ have been masked.
Second column: the velocity field in color scale (obtained from the C$^{17}$O (3--2) line) overlaid with the {\textit B}-field segments (i.e., polarization rotated by 90\arcdeg) and Stokes {\em I} contours.
Third column: an enlarged perspective of 1000 au of the second column.
In the second and third panels,
the black segments represent the {\textit B}-fields.
The red and blue arrows indicate the mean direction of the red-shifted and blue-shifted outflows.
For the velocity field, regions with an S/N less than 4 have been flagged.
The grey arrows in the second and third columns indicate the mean \textit{B}-field directions weighted by the intensity including all the polarization segments, and weighted by the uncertainty within annular region of 400--1000 au, respectively.
In all panels, the black contour levels for the Stokes {\textit I} image are 10 times the {\em rms} × (1, 2, 4, 8, 16, 32, 64, 128, 256, 512).
The black dotted square in the first column corresponds to 2000 au scale, while the black dashed, and solid circles correspond to scales of 1000 au, and 400 au, respectively.
}
\includegraphics[clip=true,trim=0cm 1cm 0cm 2cm,width=0.49 \textwidth]{HH270IRS.png}
\includegraphics[clip=true,trim=0cm 1cm 0cm 2cm,width=0.49 \textwidth]{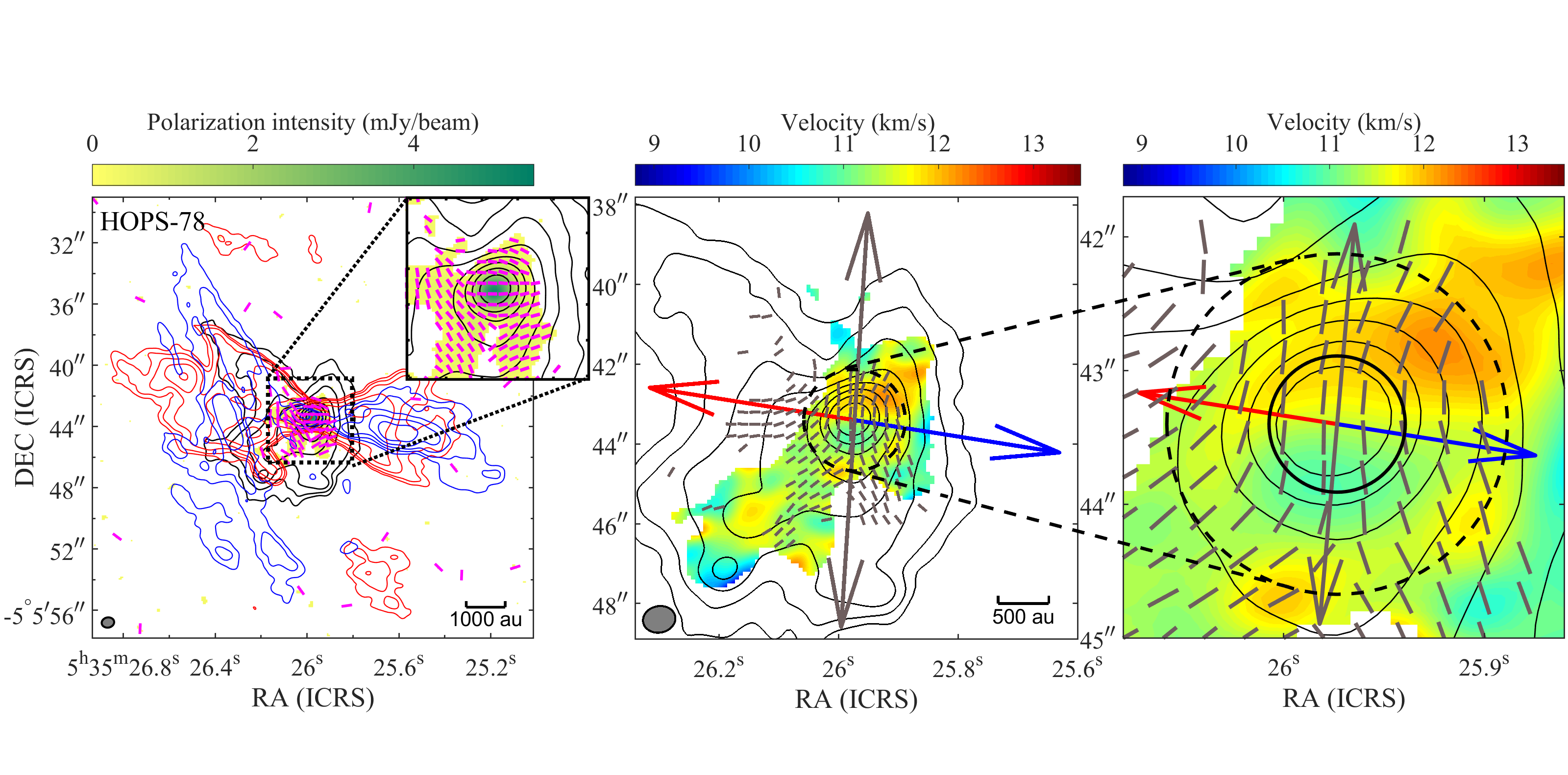}
\includegraphics[clip=true,trim=0cm 1cm 0cm 2cm,width=0.49 \textwidth]{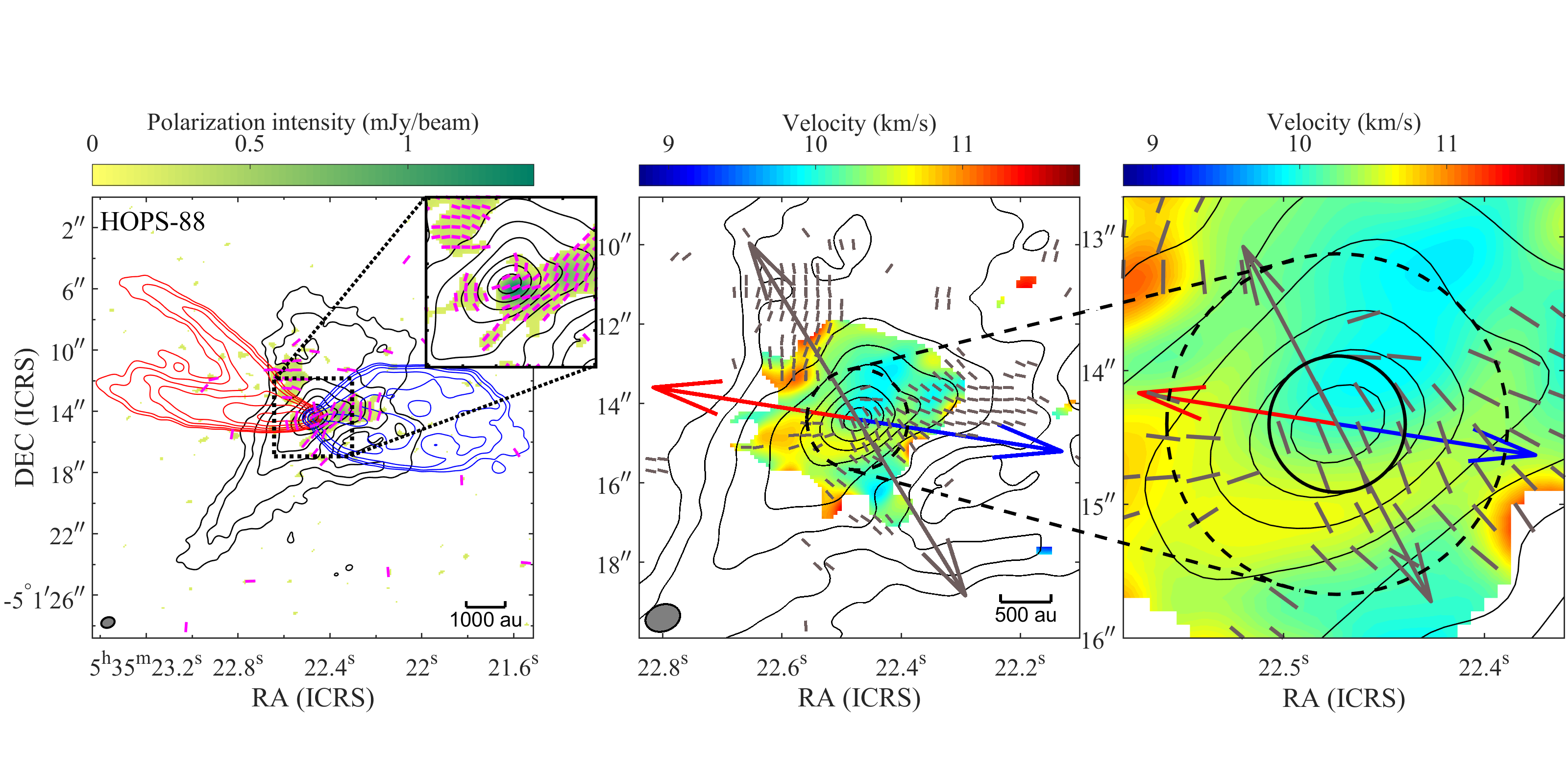}
\includegraphics[clip=true,trim=0cm 1cm 0cm 2cm,width=0.49 \textwidth]{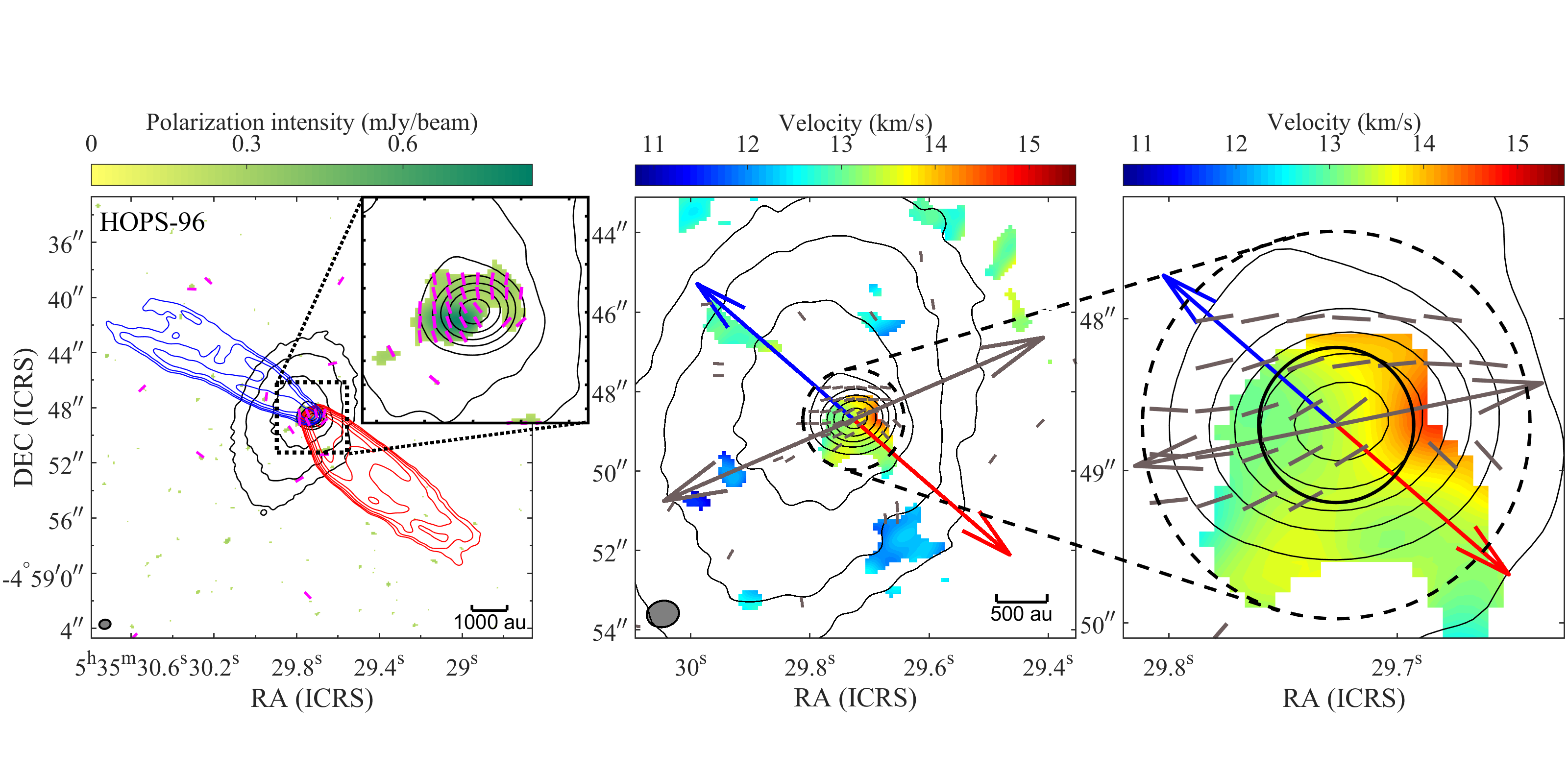}
\includegraphics[clip=true,trim=0cm 1cm 0cm 2cm,width=0.49 \textwidth]{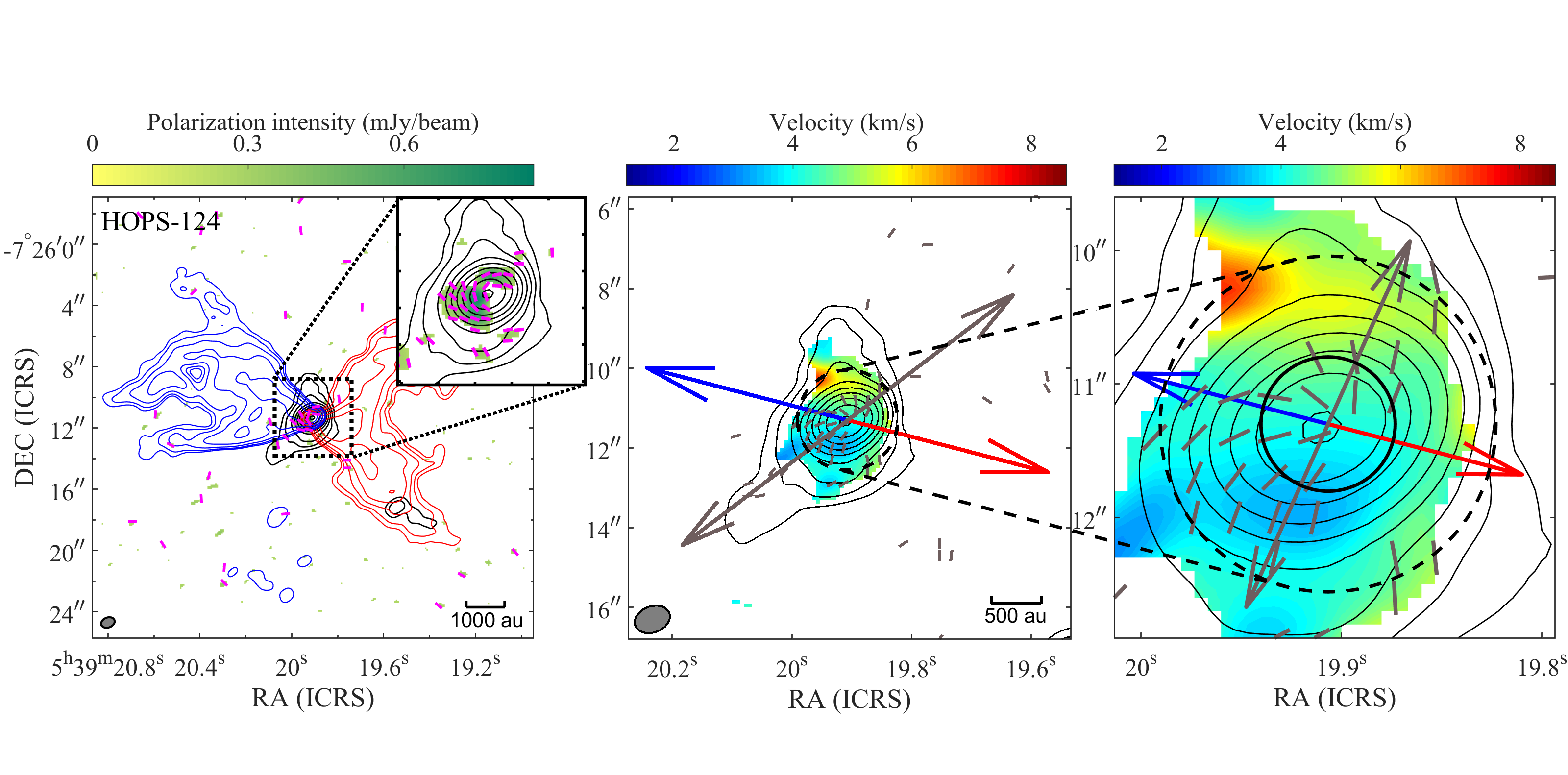}
\includegraphics[clip=true,trim=0cm 1cm 0cm 2cm,width=0.49 \textwidth]{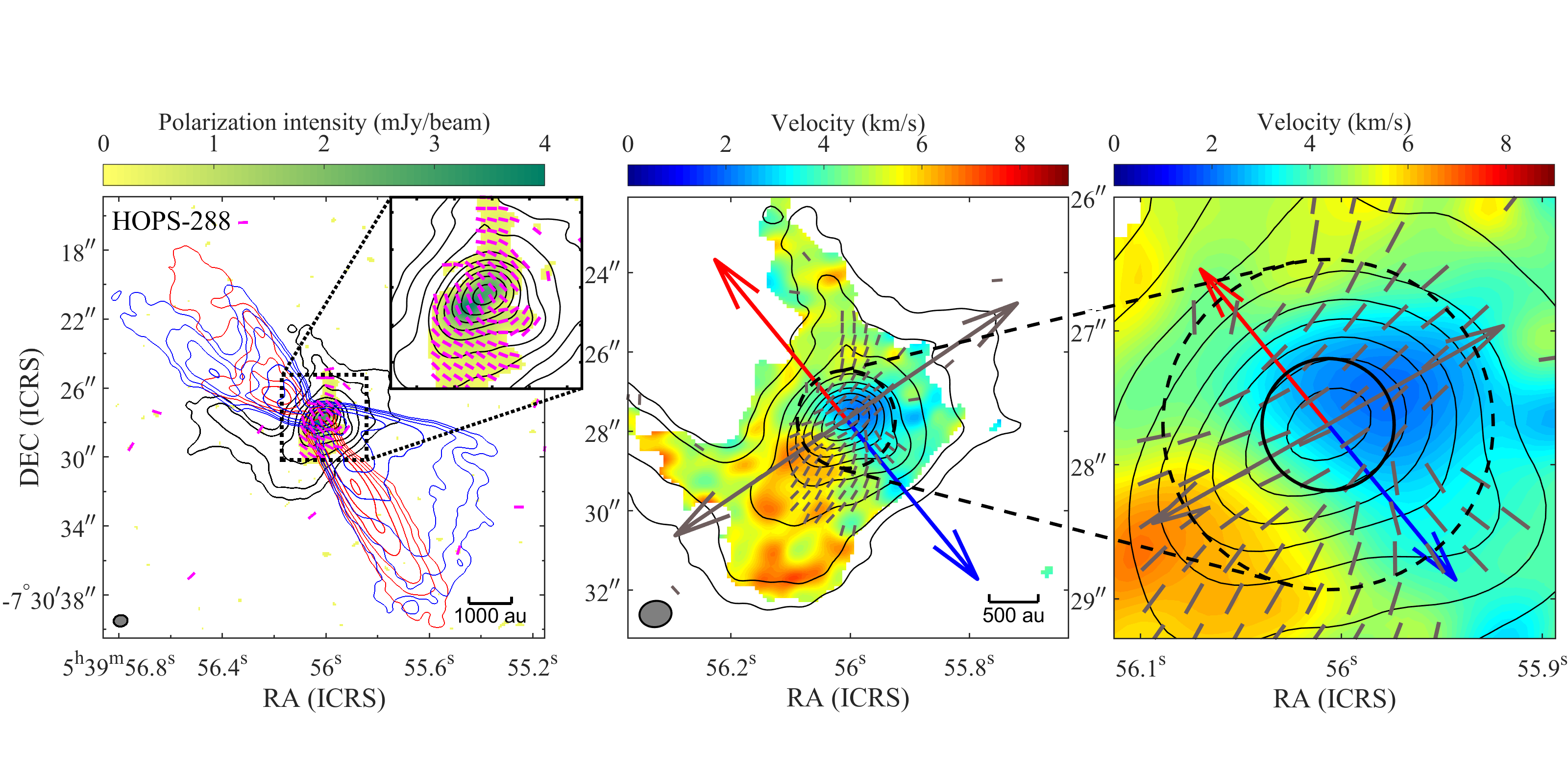}
\includegraphics[clip=true,trim=0cm 1cm 0cm 2cm,width=0.49 \textwidth]{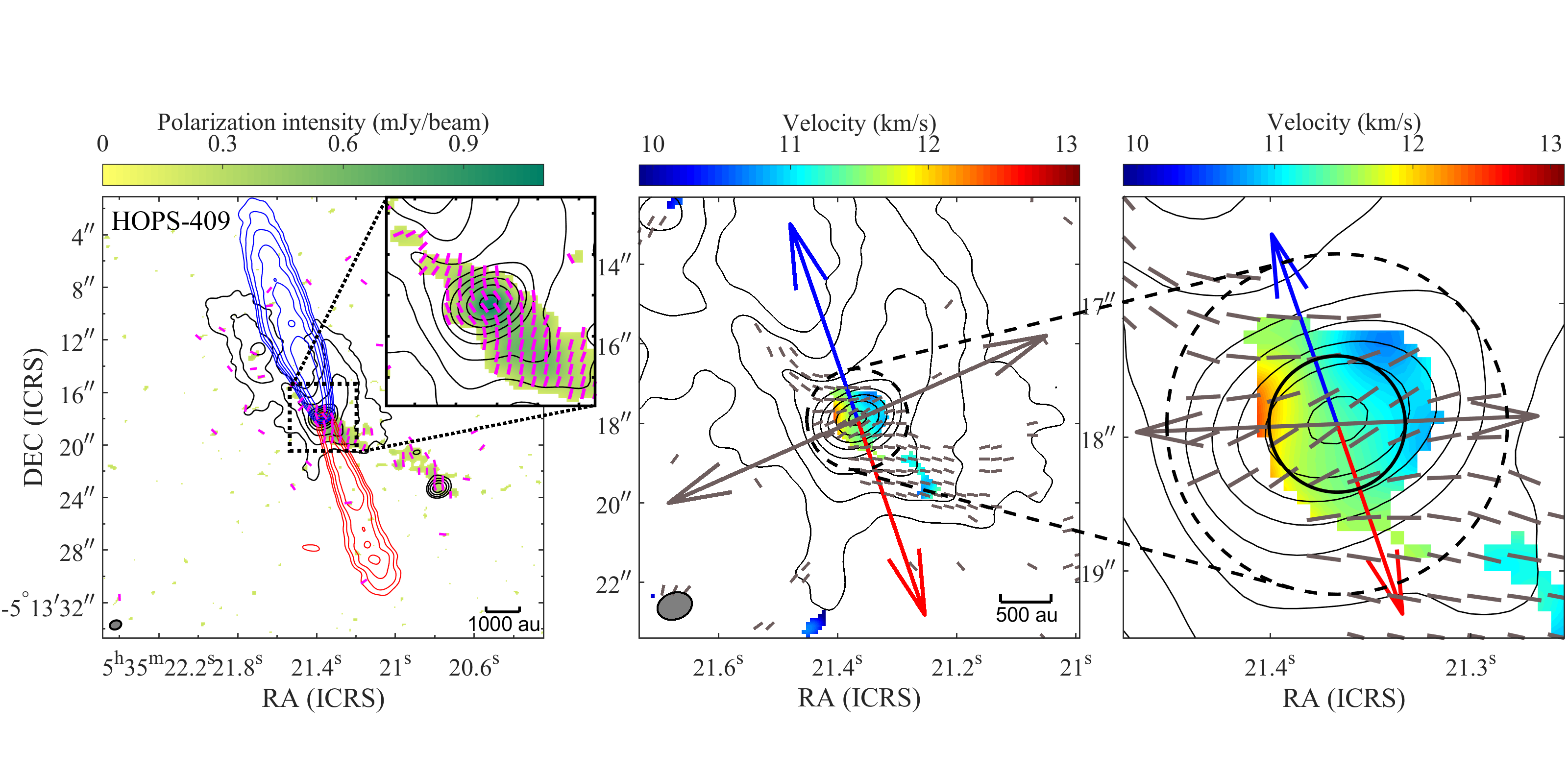}
\includegraphics[clip=true,trim=0cm 1cm 0cm 2cm,width=0.49 \textwidth]{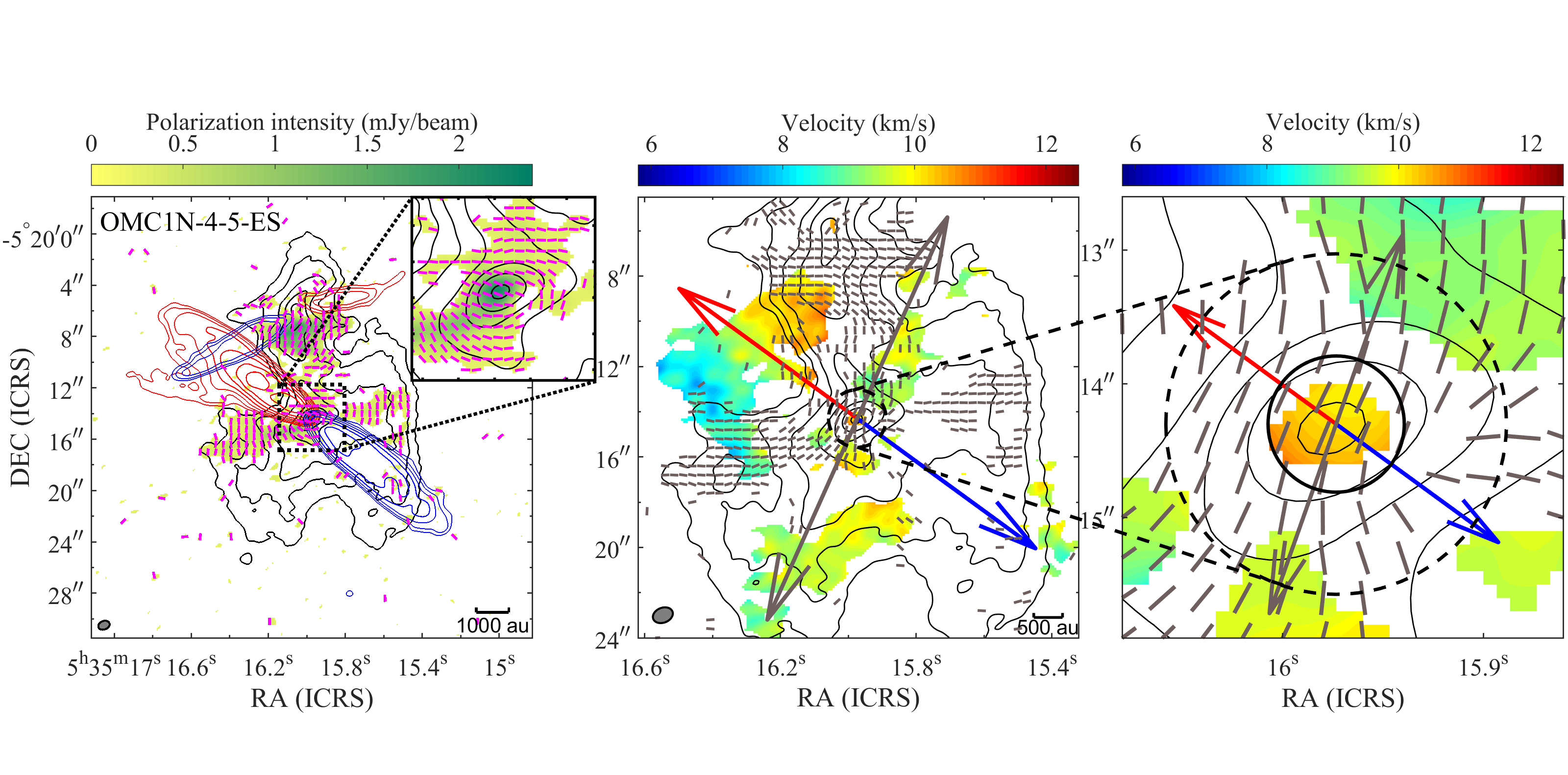}
\end{figure*}

\begin{figure*}
\centering
\caption{Nonperp-Type: velocity gradient direction is not perpendicular to outflow direction ($\vert \theta_{\rm Out}-\theta_{v} \vert < 67.5^{\circ}$). 
First column: 870~$\mu$m dust polarization intensity in color scale overlaid with the redshifted and blueshifted outflow lobes (obtained from the $^{12}$CO (3--2) line), polarization segments, and dust continuum emission (Stokes {\em I}) contours.
Blue contours indicate the blueshifted outflow, while red contours are redshifted outflow, with counter levels set at 5 times the outflow {\em rms} × (1, 2, 4, 8, 16, 32).
The magenta segments represent the polarization.
The regions of polarization intensity less than 3$\sigma$ have been masked.
Second column: the velocity field in color scale (obtained from the C$^{17}$O (3--2) line) overlaid with the {\textit B}-field segments (i.e., polarization rotated by 90\arcdeg) and Stokes {\em I} contours.
Third column: an enlarged perspective of 1000 au of the second column.
In the second and third panels,
the black segments represent the {\textit B}-fields.
The red and blue arrows indicate the mean direction of the red-shifted and blue-shifted outflows.
For the velocity field, regions with an S/N less than 4 have been flagged.
The grey arrows in the second and third columns indicate the mean \textit{B}-field directions weighted by the intensity including all the polarization segments, and weighted by the uncertainty within annular region of 400--1000 au, respectively.
In all panels, the black contour levels for the Stokes {\textit I} image are 10 times the {\em rms} × (1, 2, 4, 8, 16, 32, 64, 128, 256, 512).
The black dotted square in the first column corresponds to 2000 au scale, while the black dashed, and solid circles correspond to scales of 1000 au, and 400 au, respectively.
}
\includegraphics[clip=true,trim=0cm 1cm 0cm 2cm,width=0.49 \textwidth]{HOPS-182.png}
\includegraphics[clip=true,trim=0cm 1cm 0cm 2cm,width=0.49 \textwidth]{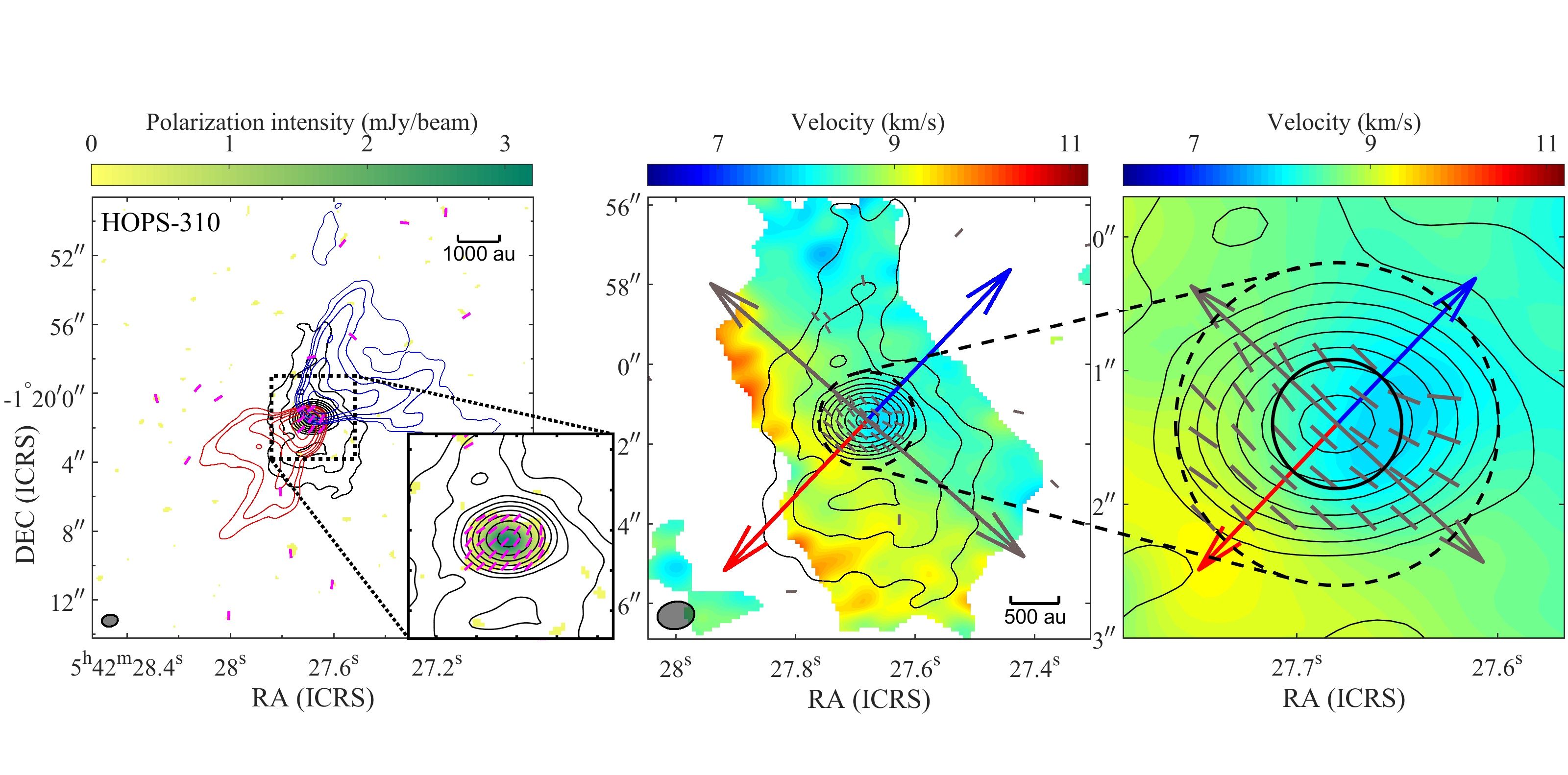}
\includegraphics[clip=true,trim=0cm 1cm 0cm 2cm,width=0.49 \textwidth]{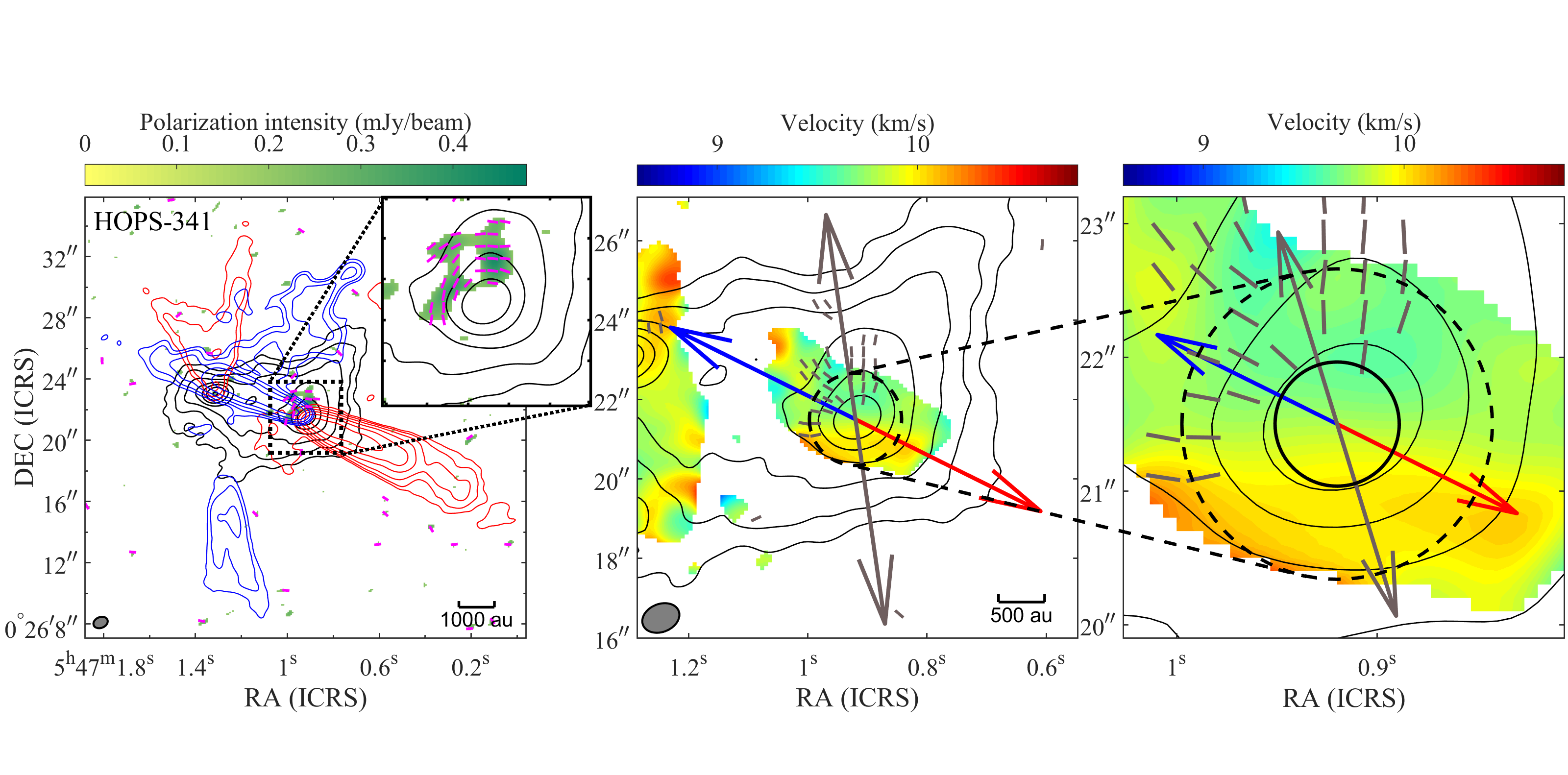}
\includegraphics[clip=true,trim=0cm 1cm 0cm 2cm,width=0.49 \textwidth]{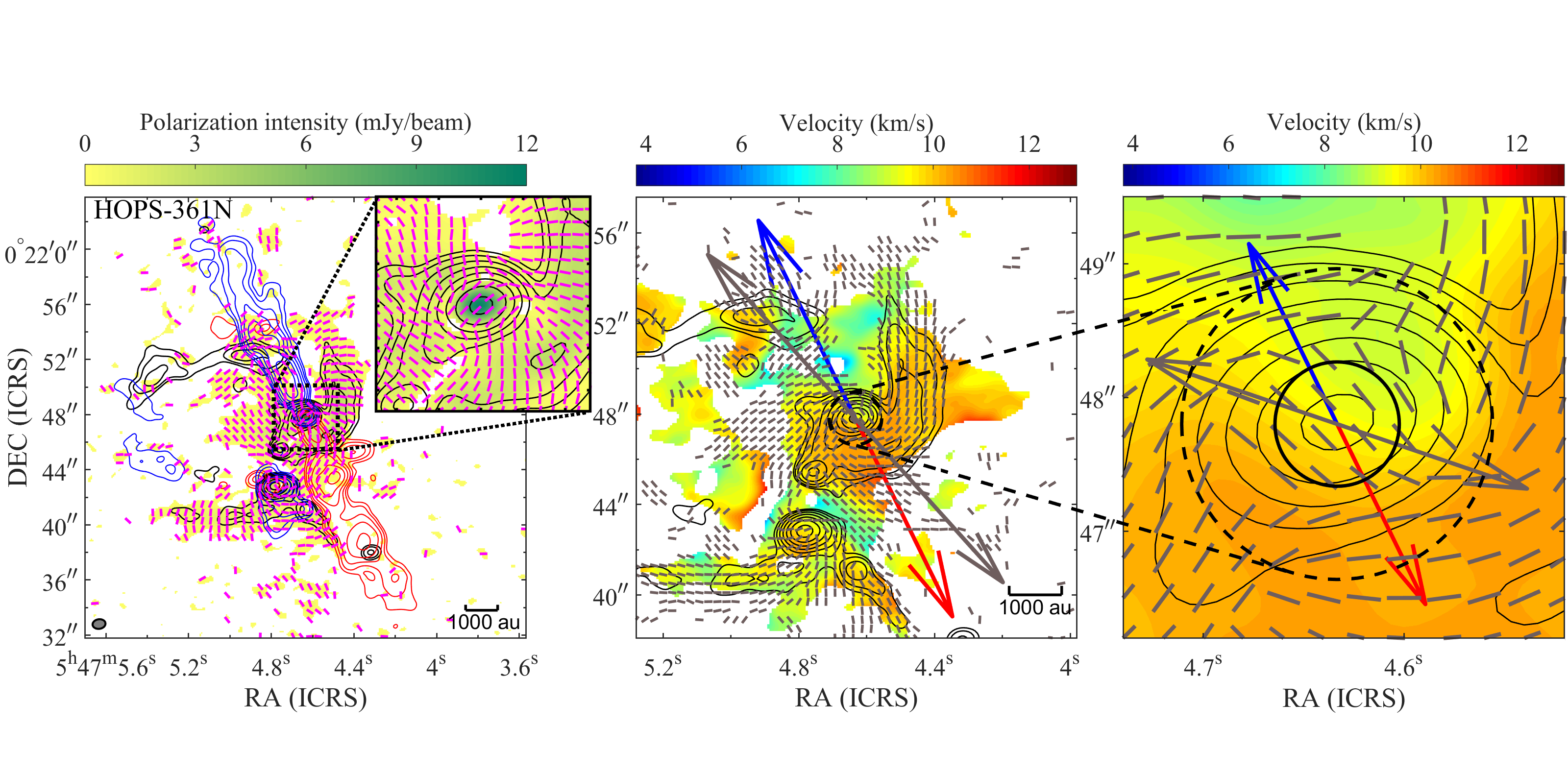}
\includegraphics[clip=true,trim=0cm 1cm 0cm 2cm,width=0.49 \textwidth]{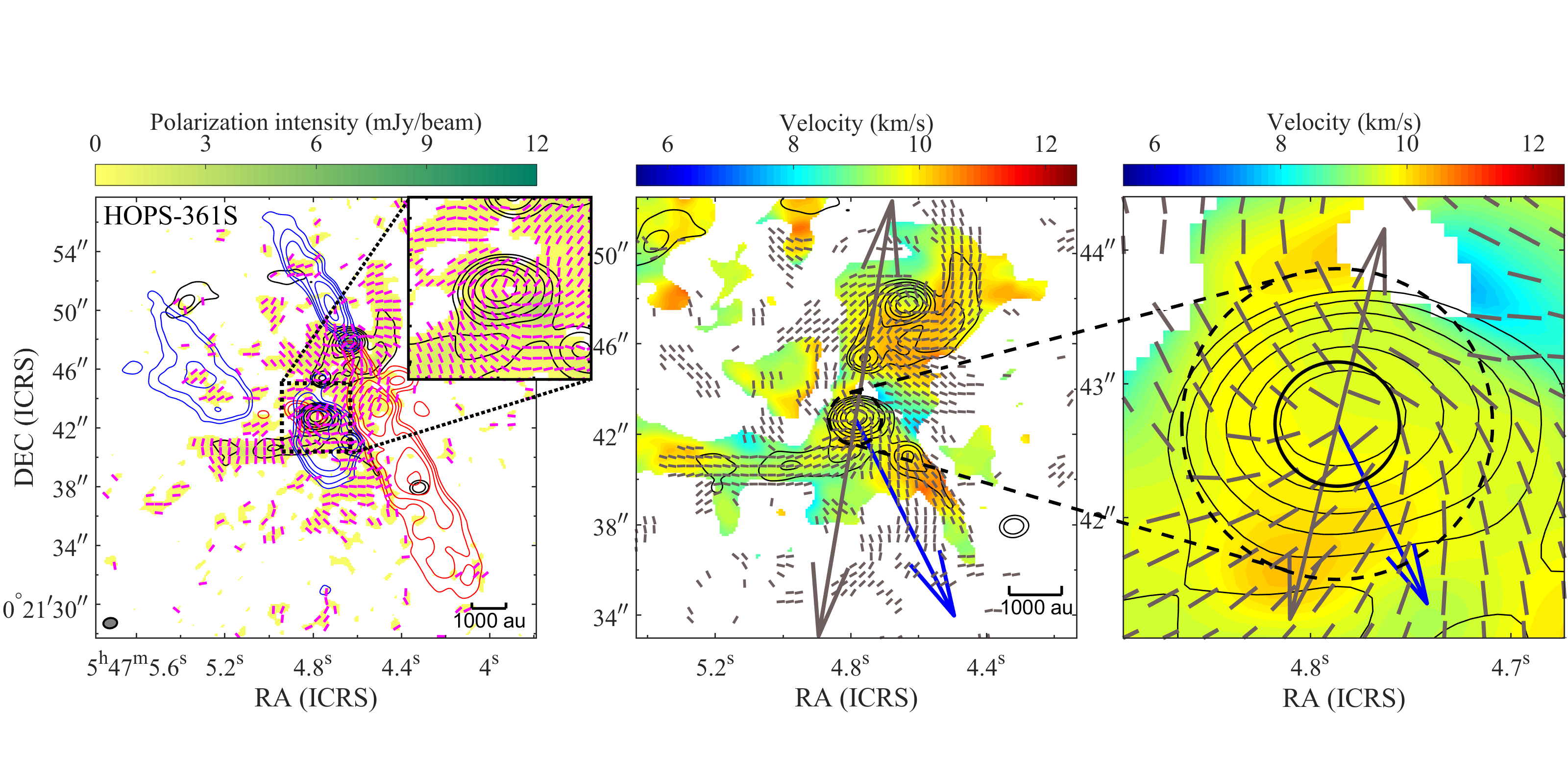}
\includegraphics[clip=true,trim=0cm 1cm 0cm 2cm,width=0.49 \textwidth]{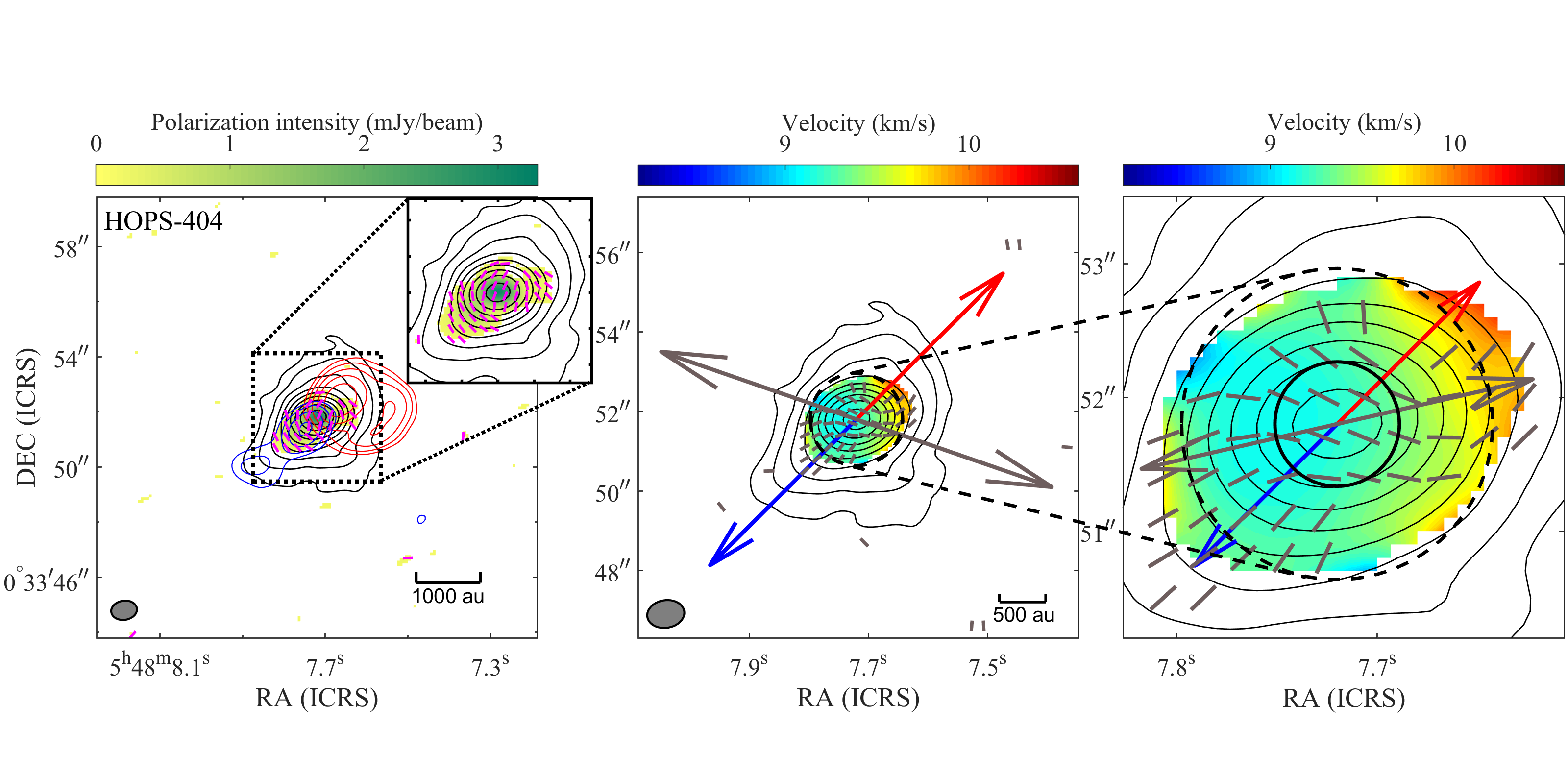}
\end{figure*}

\begin{figure*}
\centering
\caption{Unres-Type (26 including HOPS-250): velocity gradient is unresolved. 
First column: 870~$\mu$m dust polarization intensity in color scale overlaid with the redshifted and blueshifted outflow lobes (obtained from the $^{12}$CO (3--2) line), polarization segments, and dust continuum emission (Stokes {\em I}) contours.
Blue contours indicate the blueshifted outflow, while red contours are redshifted outflow, with counter levels set at 5 times the outflow {\em rms} × (1, 2, 4, 8, 16, 32).
The magenta segments represent the polarization.
The regions of polarization intensity less than 3$\sigma$ have been masked.
Second column: the velocity field in color scale (obtained from the C$^{17}$O (3--2) line) overlaid with the {\textit B}-field segments (i.e., polarization rotated by 90\arcdeg) and Stokes {\em I} contours.
Third column: an enlarged perspective of 1000 au of the second column.
In the second and third panels,
the black segments represent the {\textit B}-fields.
The red and blue arrows indicate the mean direction of the red-shifted and blue-shifted outflows.
For the velocity field, regions with an S/N less than 4 have been flagged.
The grey arrows in the second and third columns indicate the mean \textit{B}-field directions weighted by the intensity including all the polarization segments, and weighted by the uncertainty within annular region of 400--1000 au, respectively.
In all panels, the black contour levels for the Stokes {\textit I} image are 10 times the {\em rms} × (1, 2, 4, 8, 16, 32, 64, 128, 256, 512).
The black dotted square in the first column corresponds to 2000 au scale, while the black dashed, and solid circles correspond to scales of 1000 au, and 400 au, respectively.
}
\includegraphics[clip=true,trim=0cm 1cm 0cm 2cm,width=0.49 \textwidth]{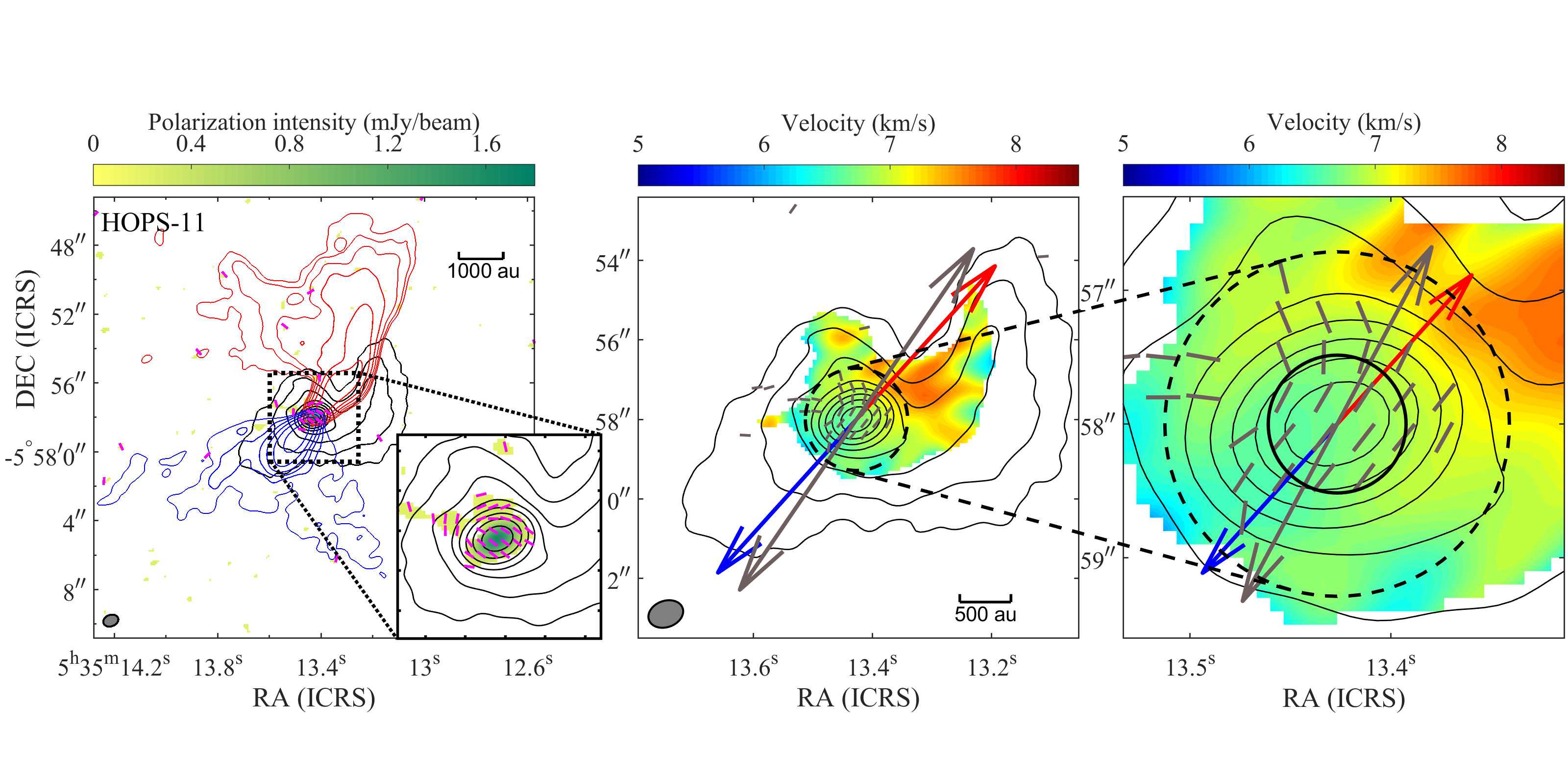}
\includegraphics[clip=true,trim=0cm 1cm 0cm 2cm,width=0.49 \textwidth]{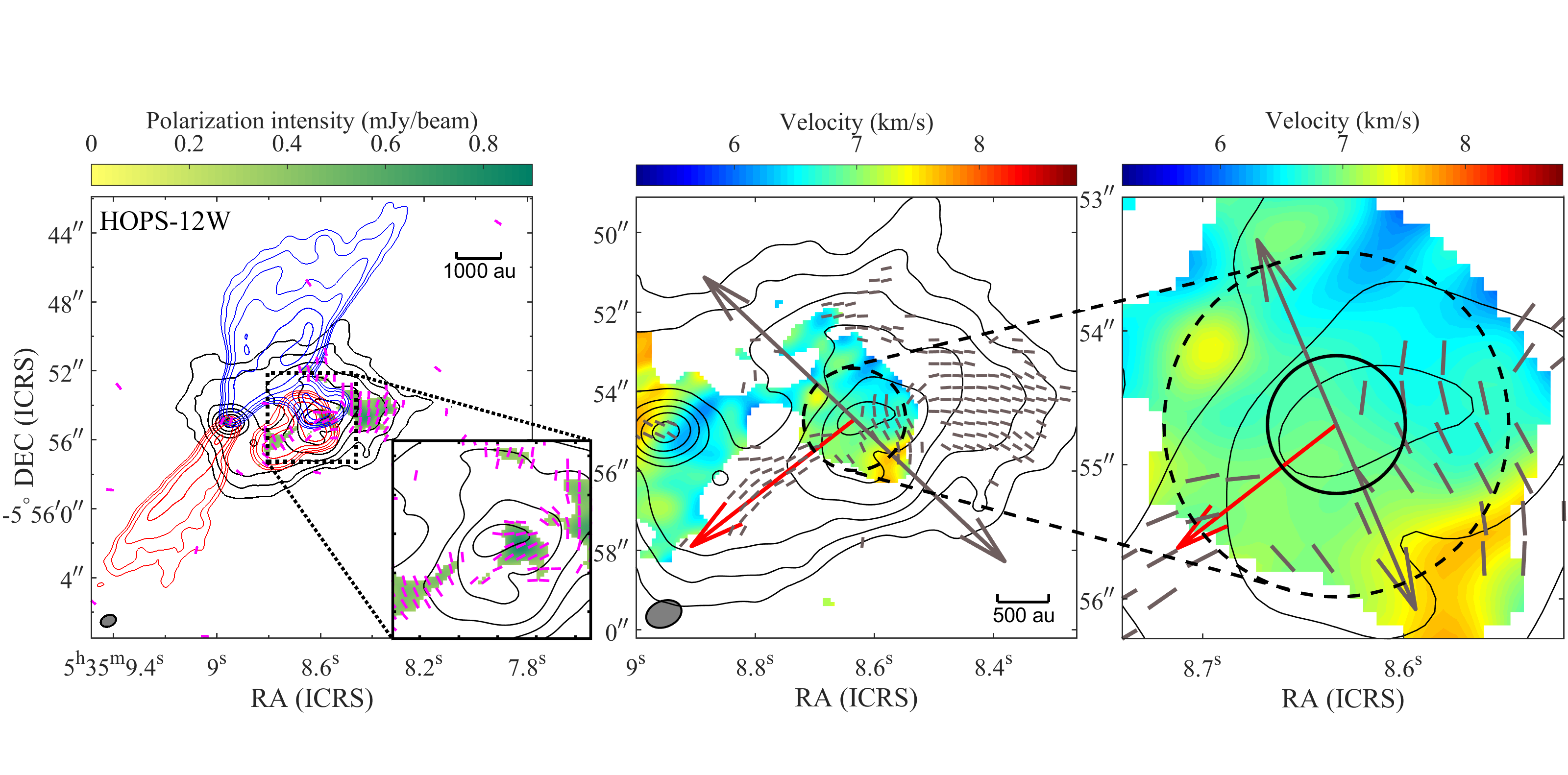}
\includegraphics[clip=true,trim=0cm 1cm 0cm 2cm,width=0.49 \textwidth]{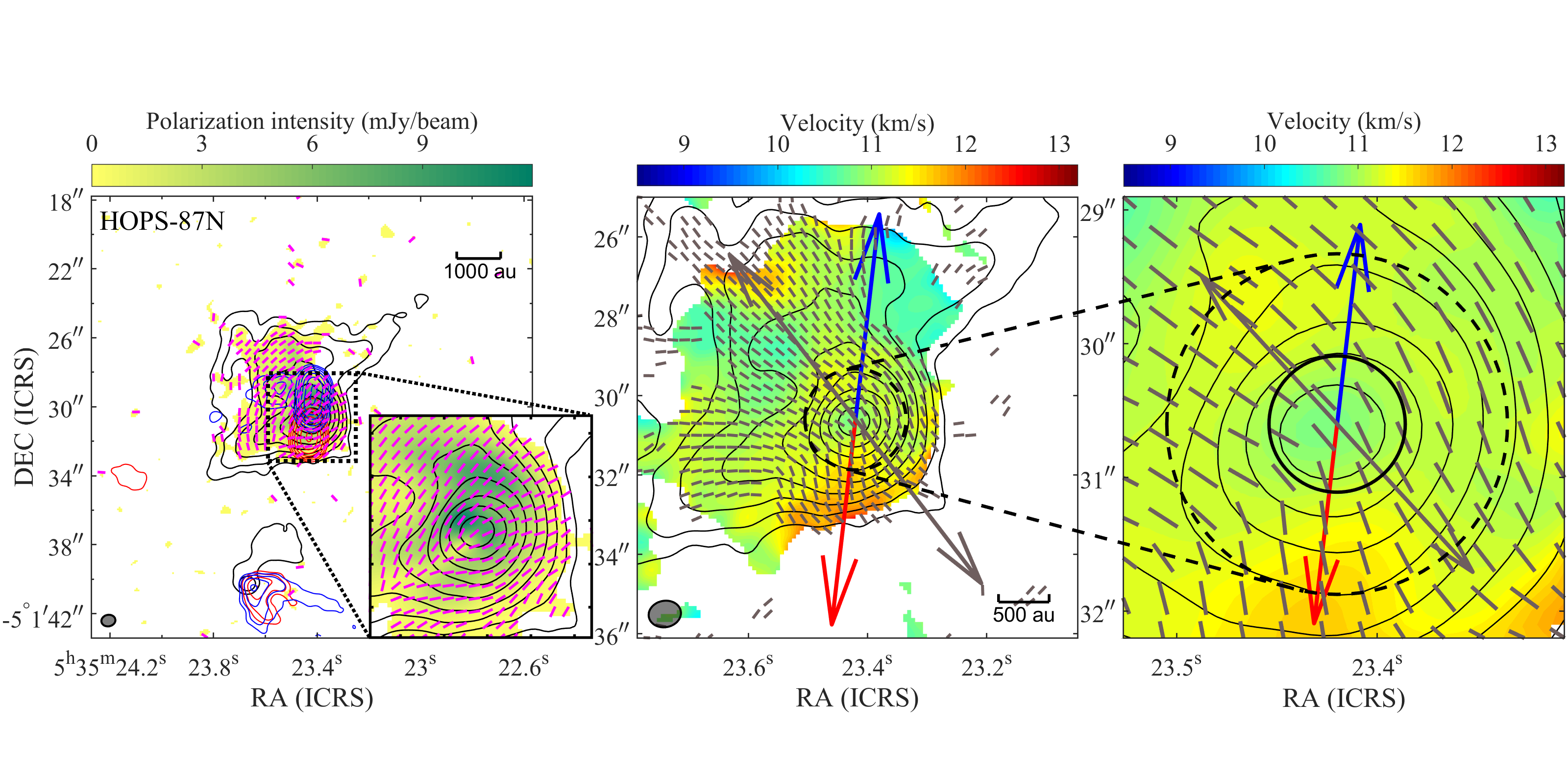}
\includegraphics[clip=true,trim=0cm 1cm 0cm 2cm,width=0.49 \textwidth]{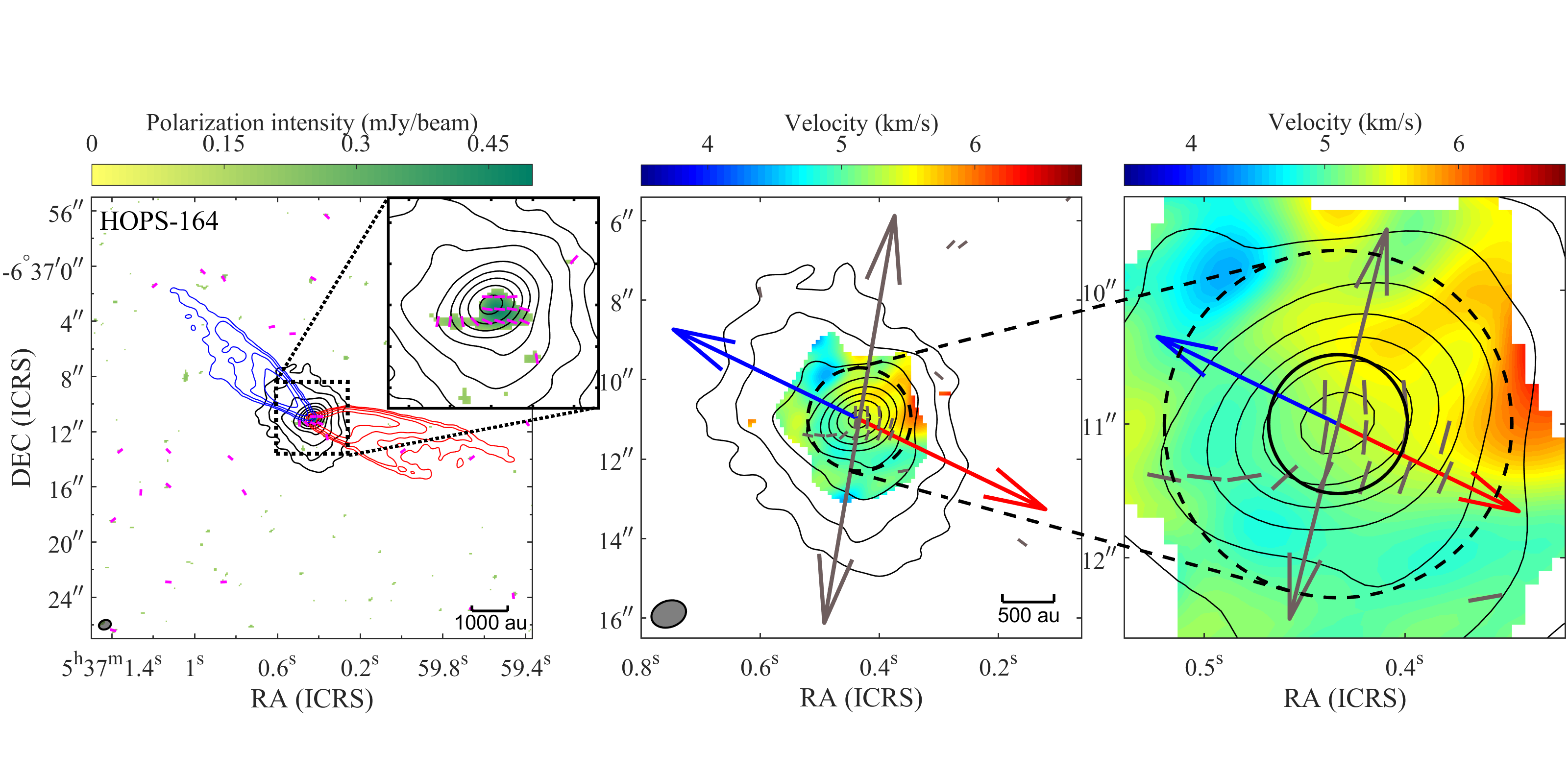}
\includegraphics[clip=true,trim=0cm 1cm 0cm 2cm,width=0.49 \textwidth]{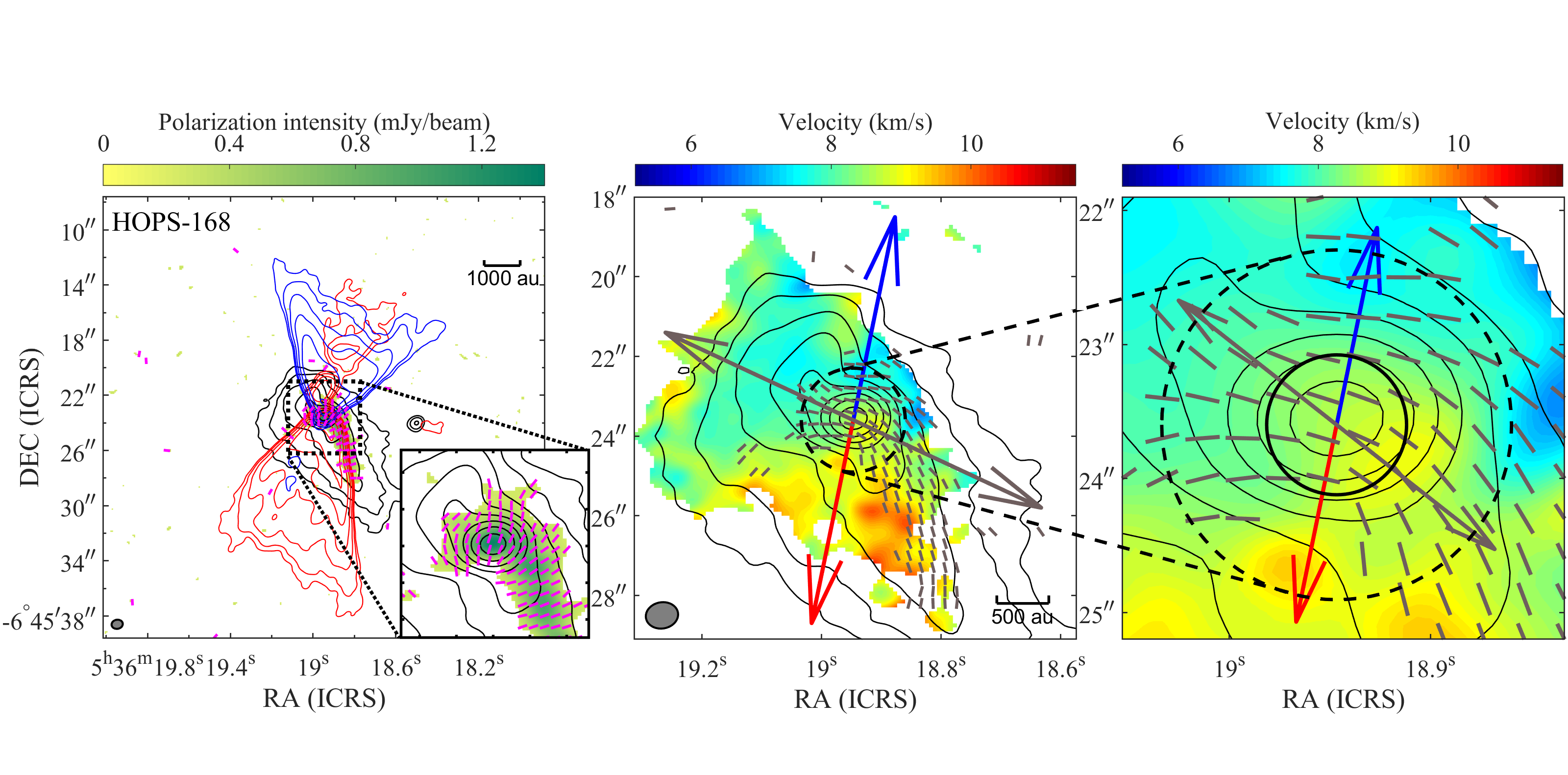}
\includegraphics[clip=true,trim=0cm 1cm 0cm 2cm,width=0.49 \textwidth]{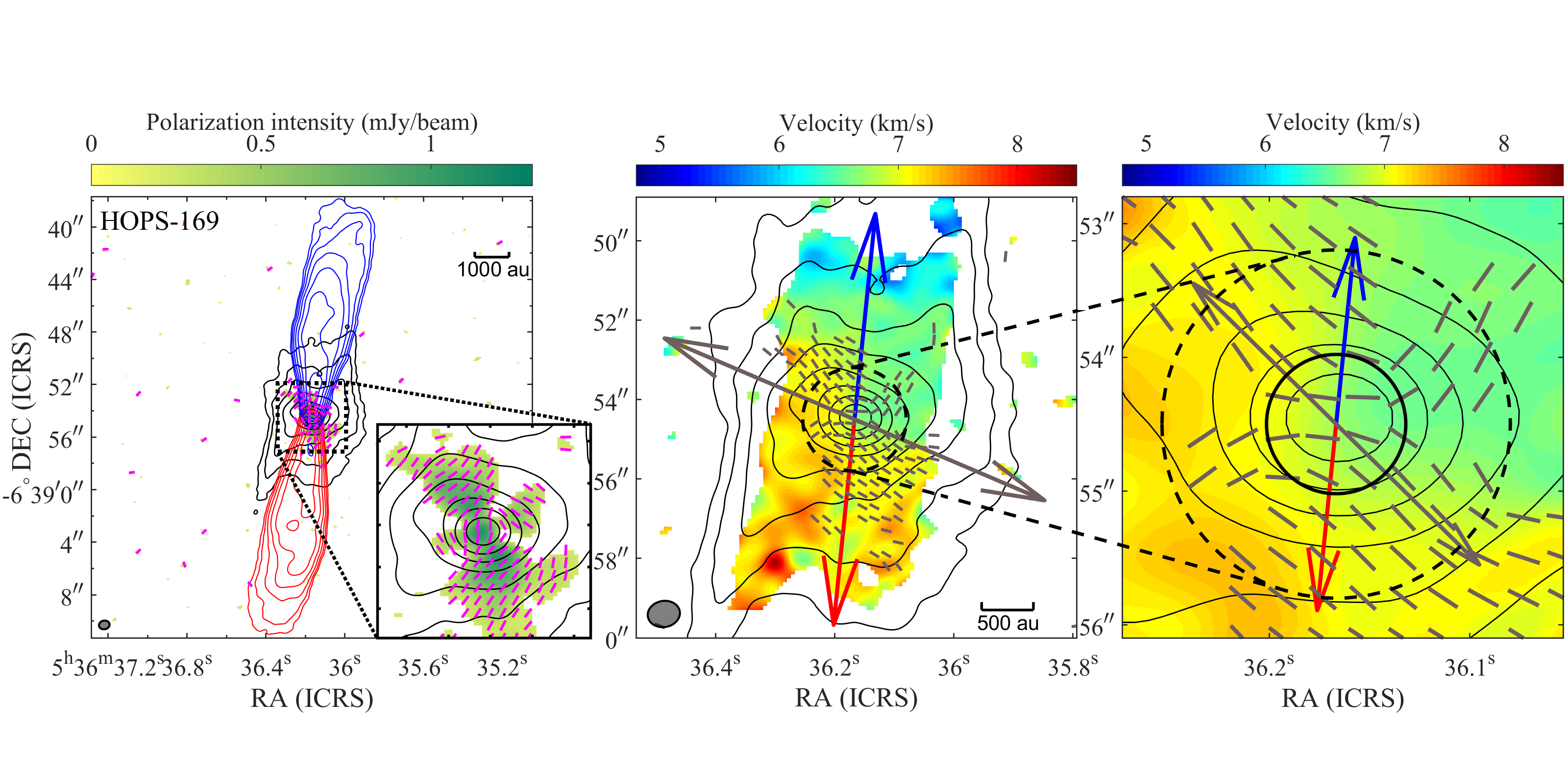}
\includegraphics[clip=true,trim=0cm 1cm 0cm 2cm,width=0.49 \textwidth]{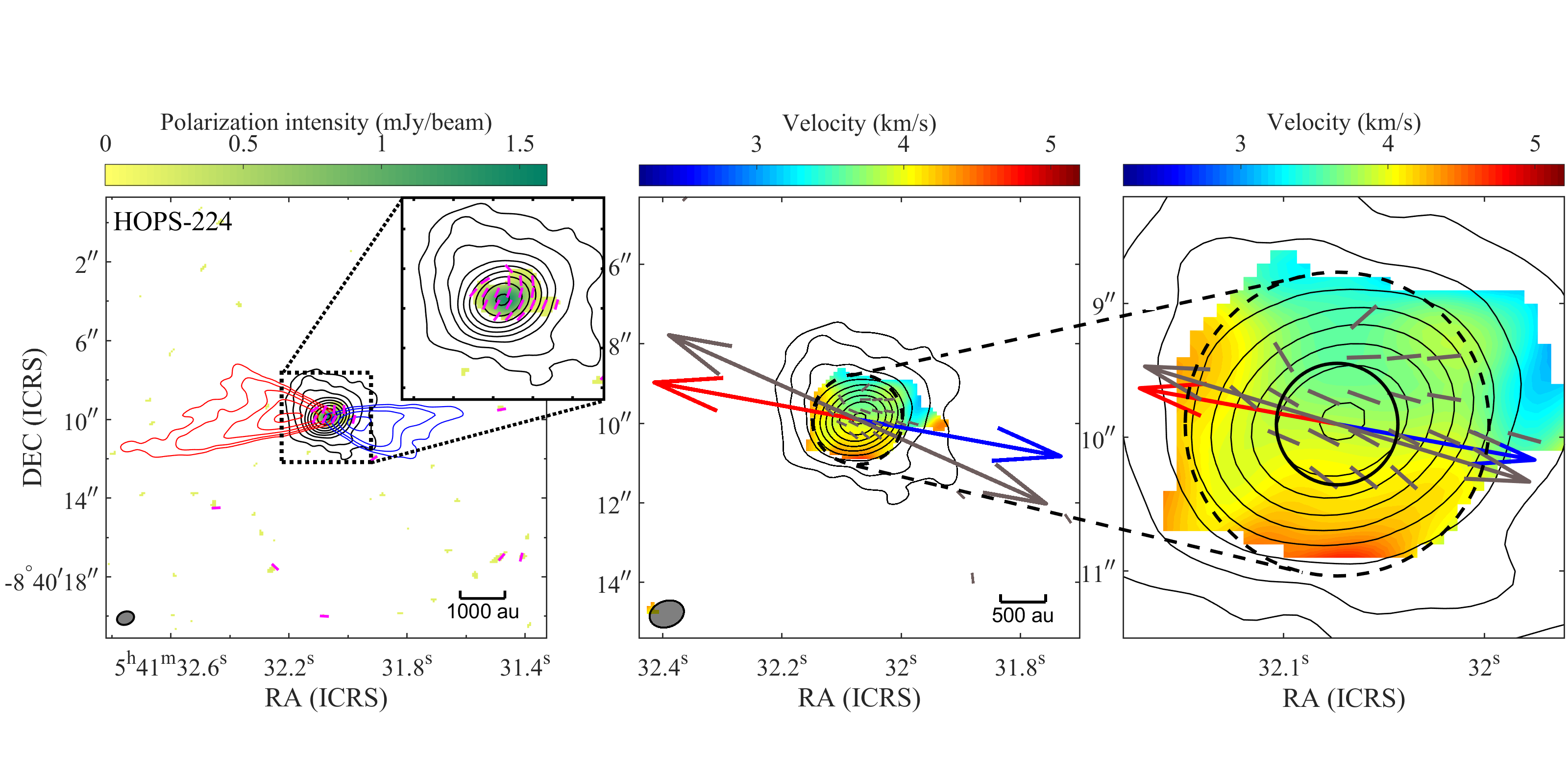}
\includegraphics[clip=true,trim=0cm 1cm 0cm 2cm,width=0.49 \textwidth]{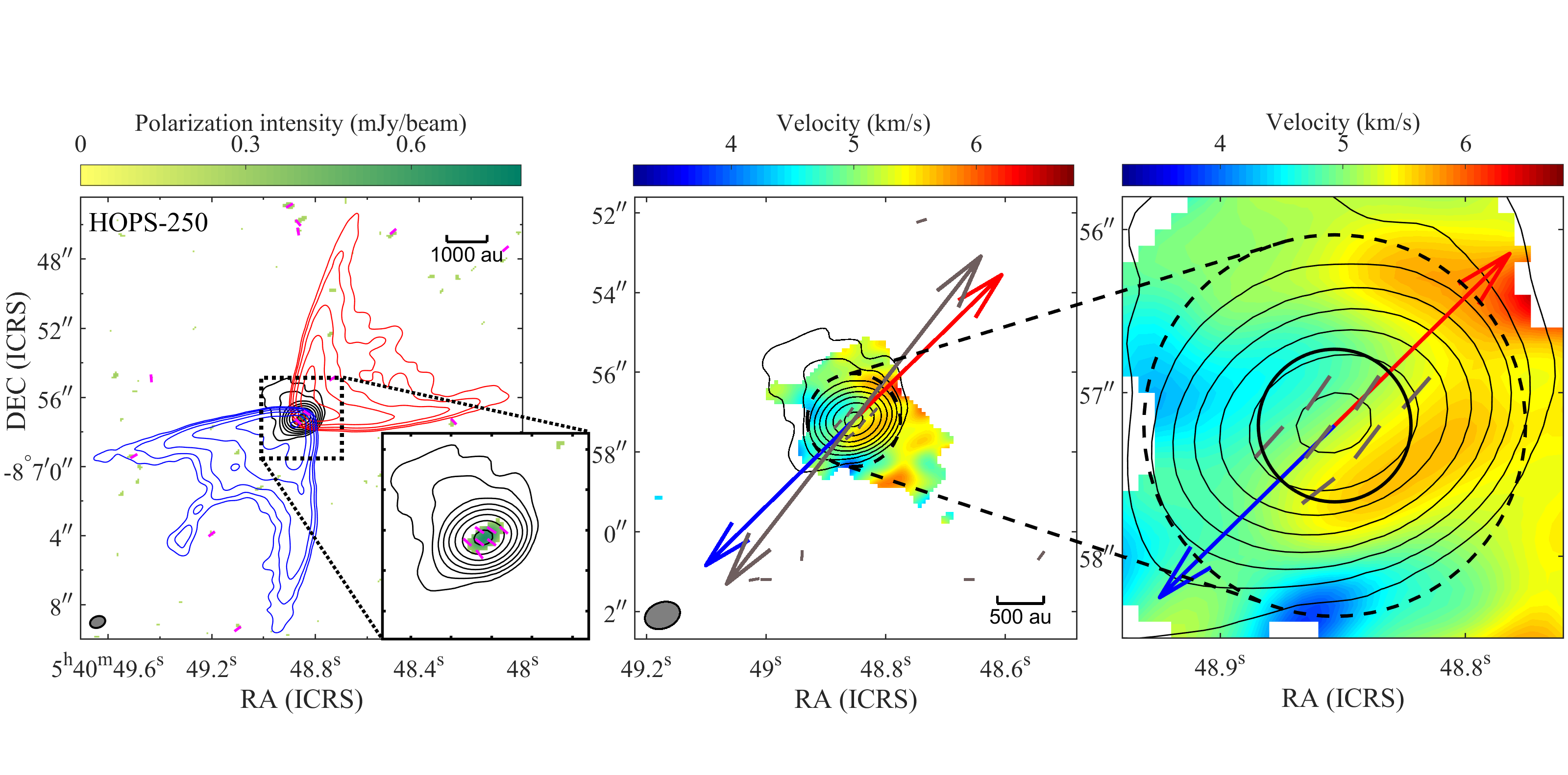}
\end{figure*}

\begin{figure*}
\centering
\text{\textbf{Unres-Type: velocity gradient is unresolved (Continued.)}} \\
\includegraphics[clip=true,trim=0cm 1cm 0cm 2cm,width=0.49 \textwidth]{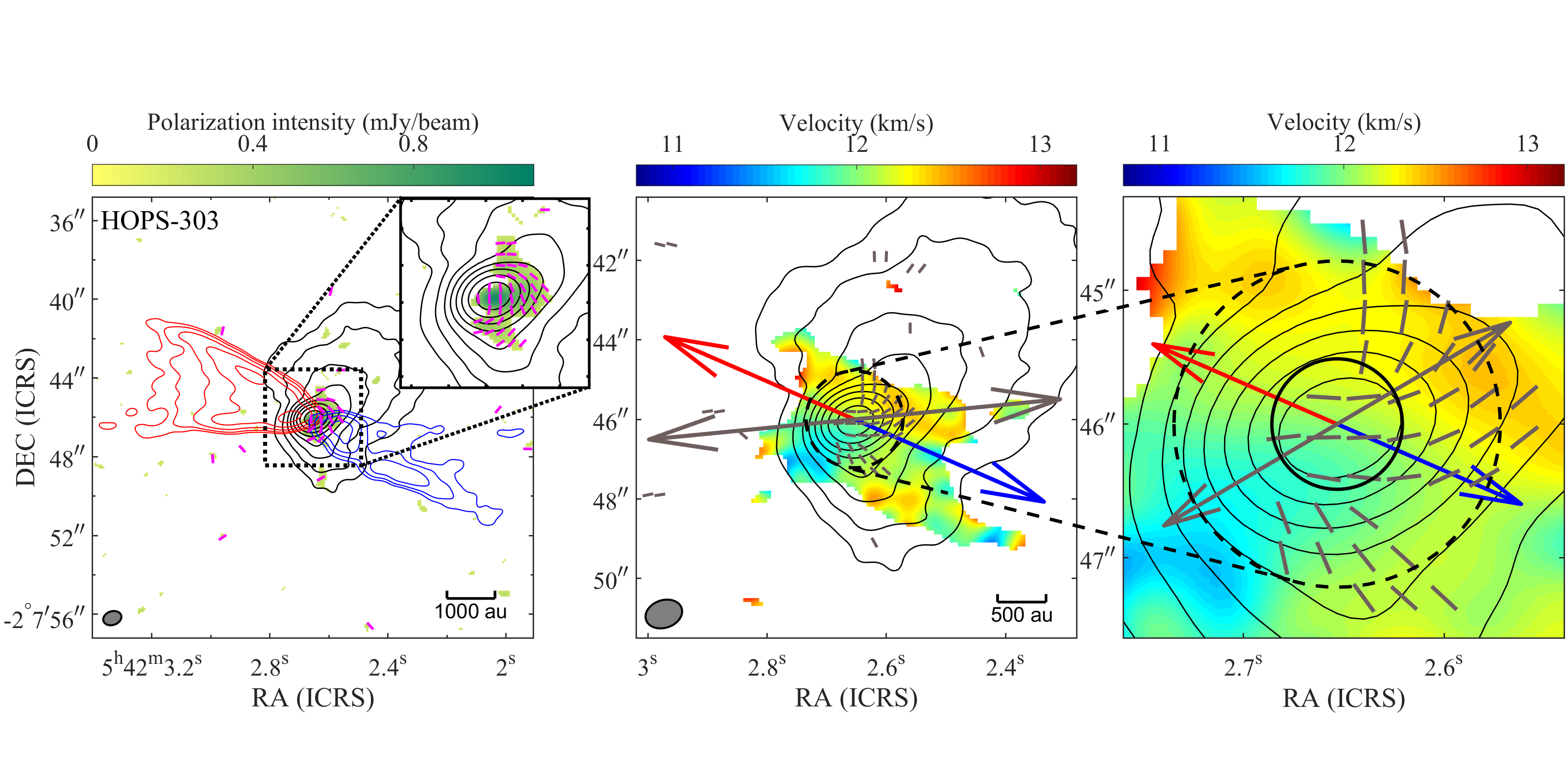}
\includegraphics[clip=true,trim=0cm 1cm 0cm 2cm,width=0.49 \textwidth]{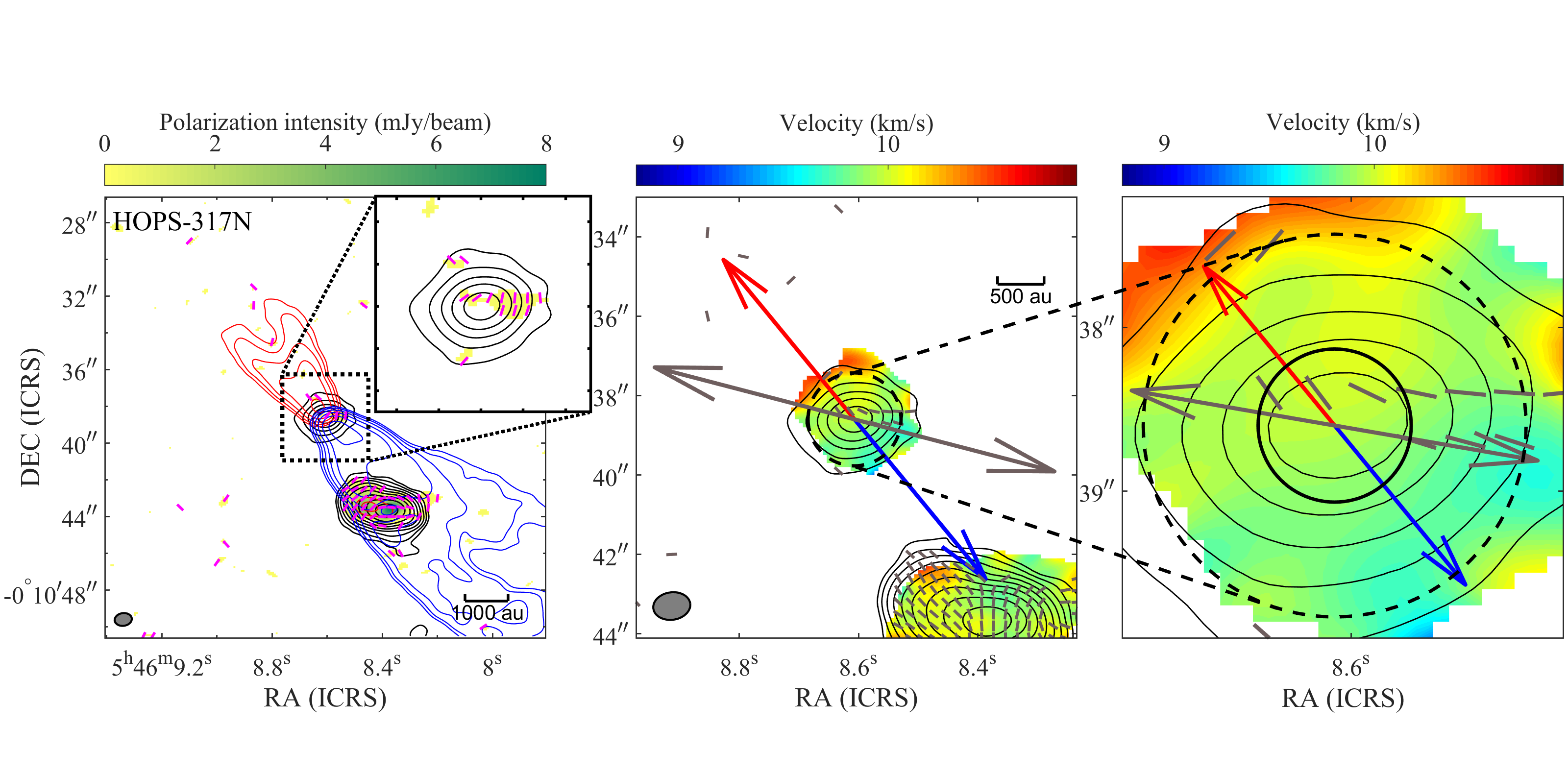}
\includegraphics[clip=true,trim=0cm 1cm 0cm 2cm,width=0.49 \textwidth]{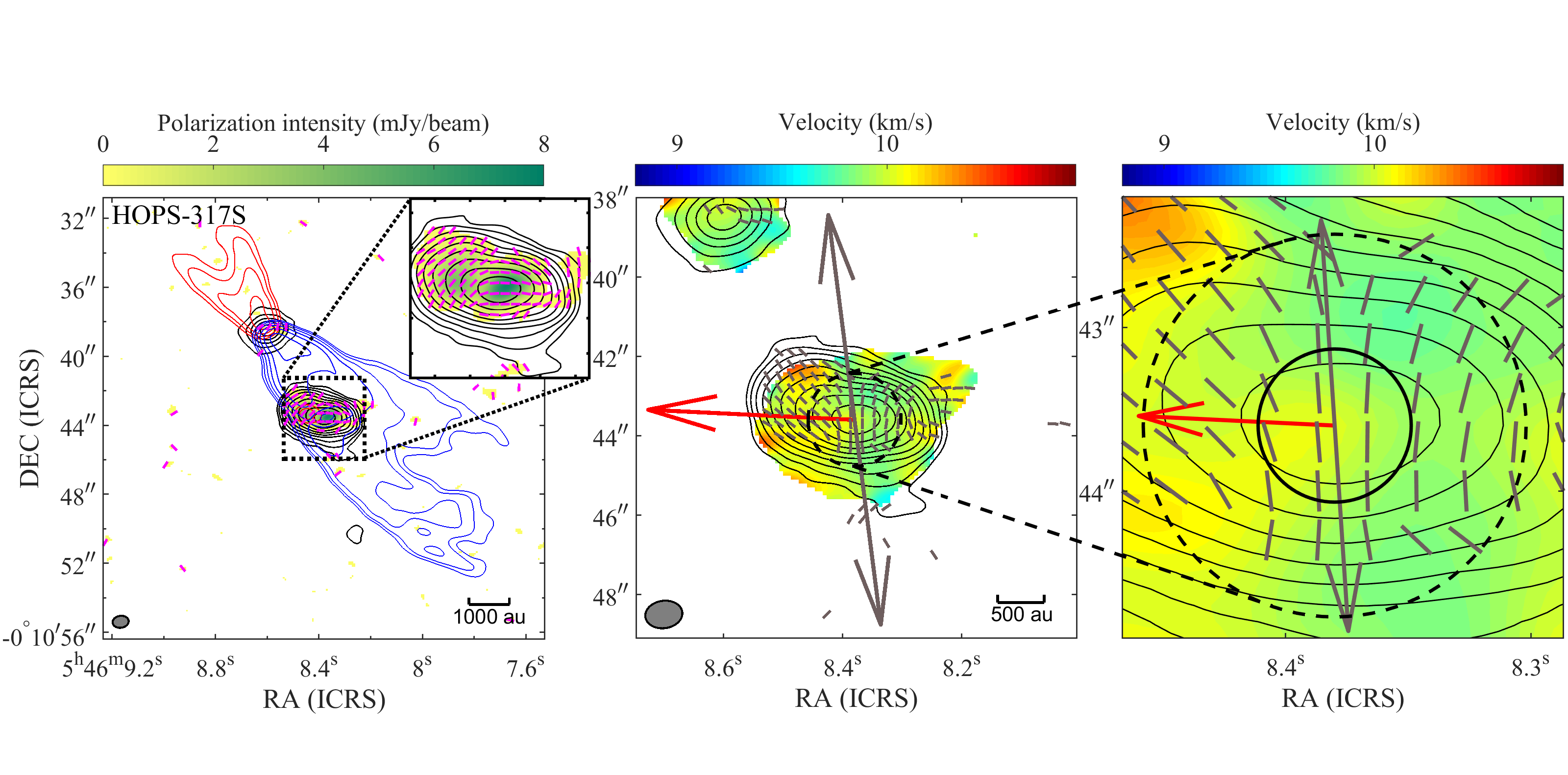}
\includegraphics[clip=true,trim=0cm 1cm 0cm 2cm,width=0.49 \textwidth]{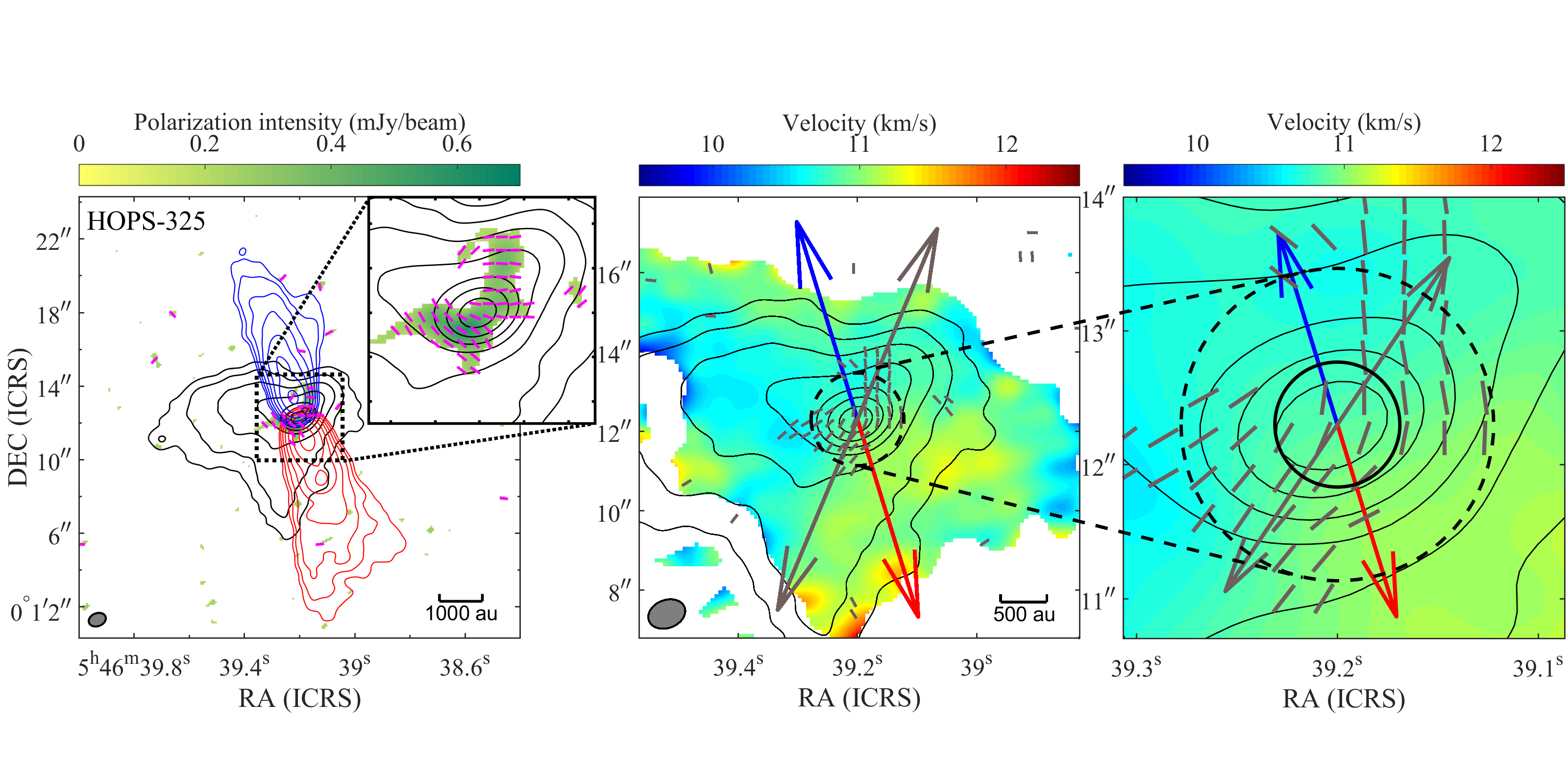}
\includegraphics[clip=true,trim=0cm 1cm 0cm 2cm,width=0.49 \textwidth]{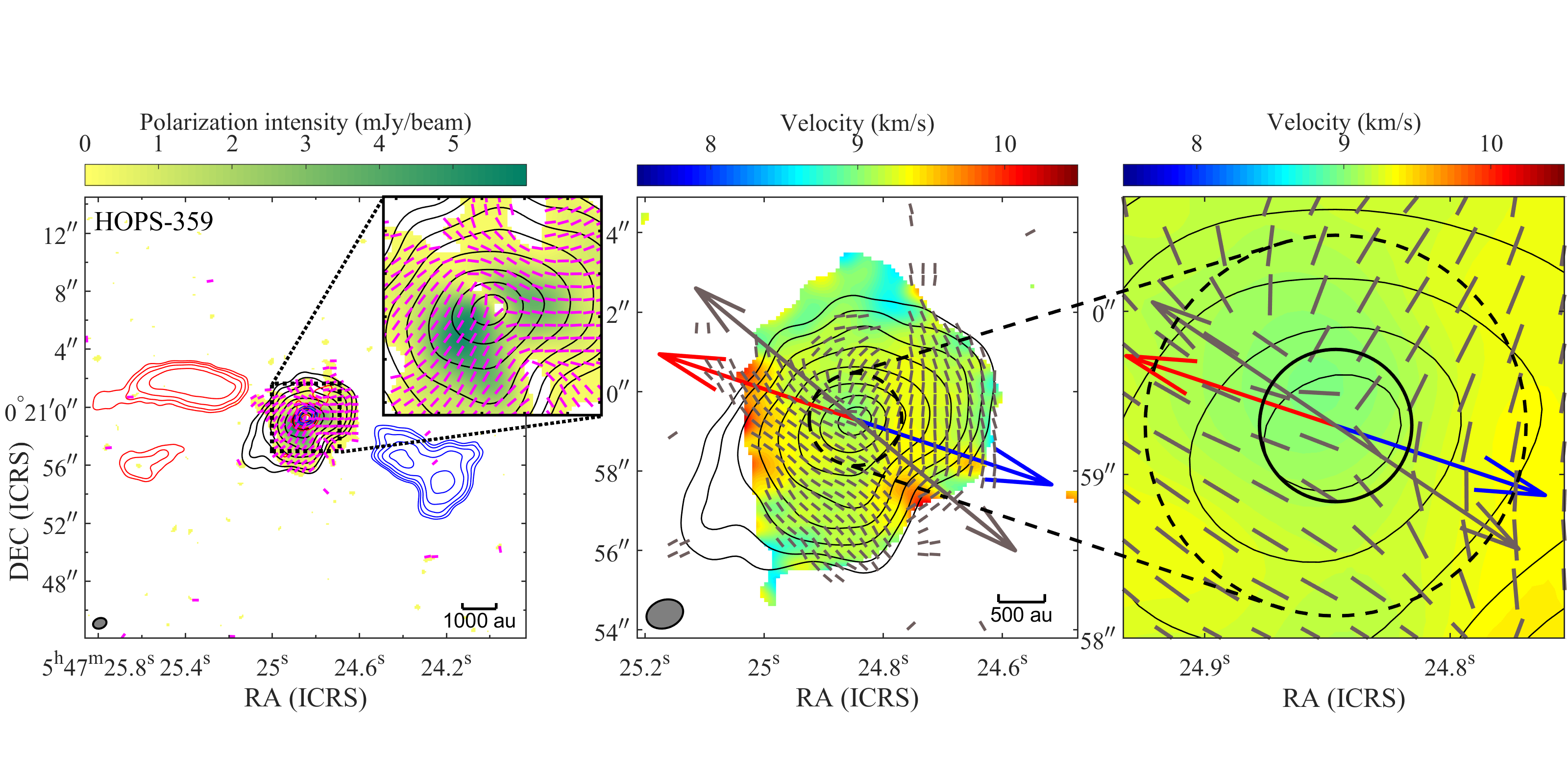}
\includegraphics[clip=true,trim=0cm 1cm 0cm 2cm,width=0.49 \textwidth]{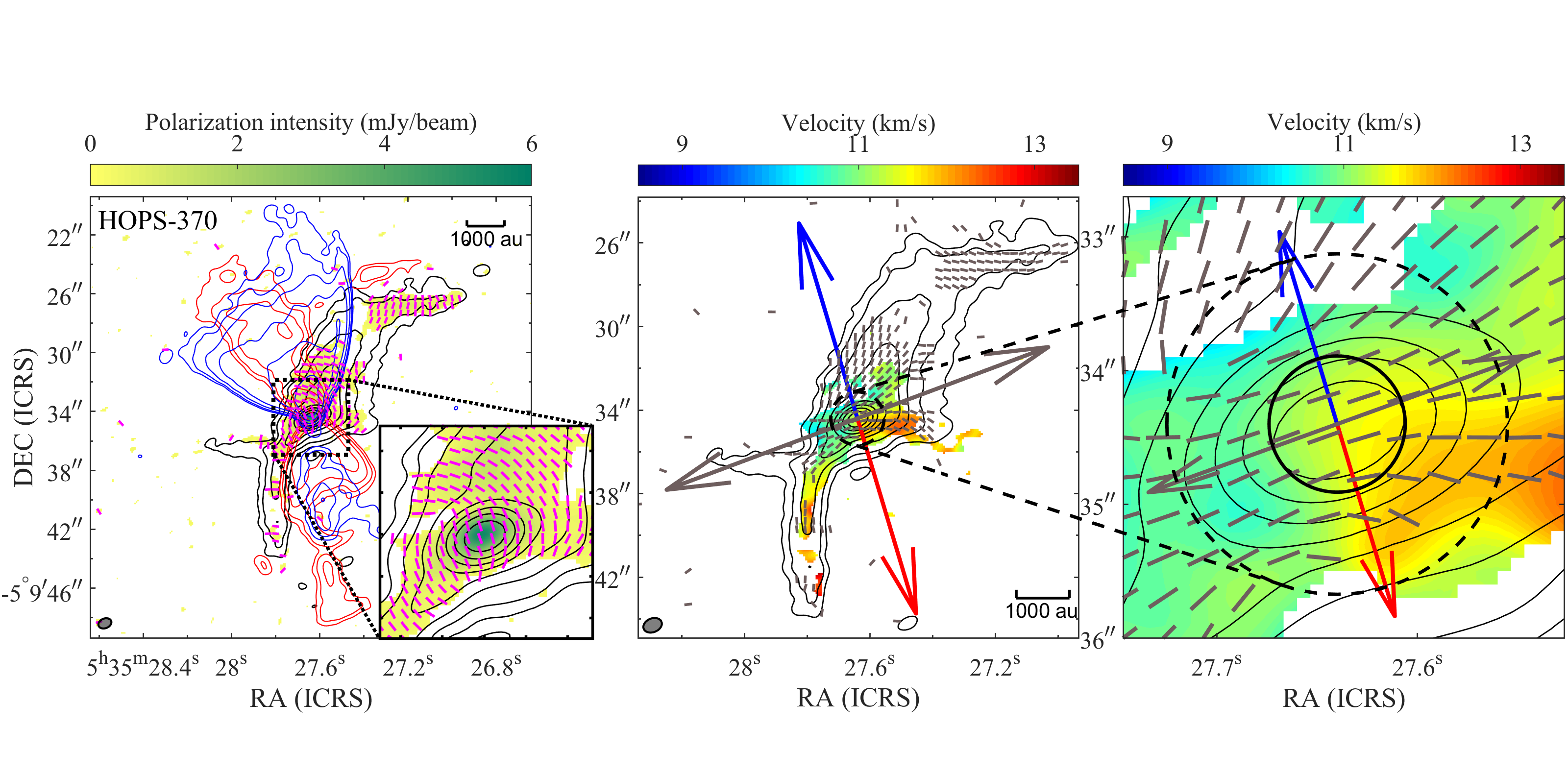}
\includegraphics[clip=true,trim=0cm 1cm 0cm 2cm,width=0.49 \textwidth]{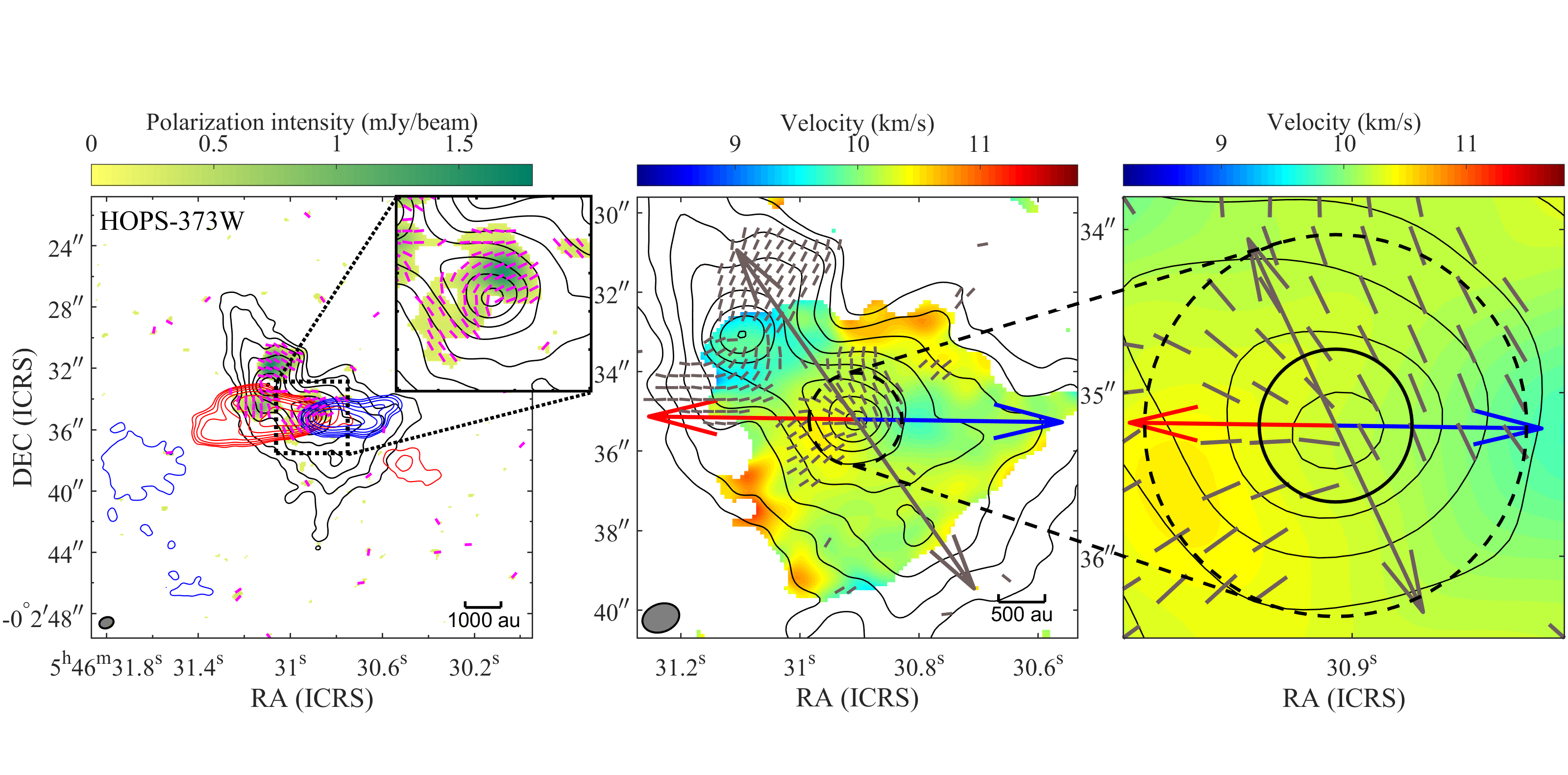}
\includegraphics[clip=true,trim=0cm 1cm 0cm 2cm,width=0.49 \textwidth]{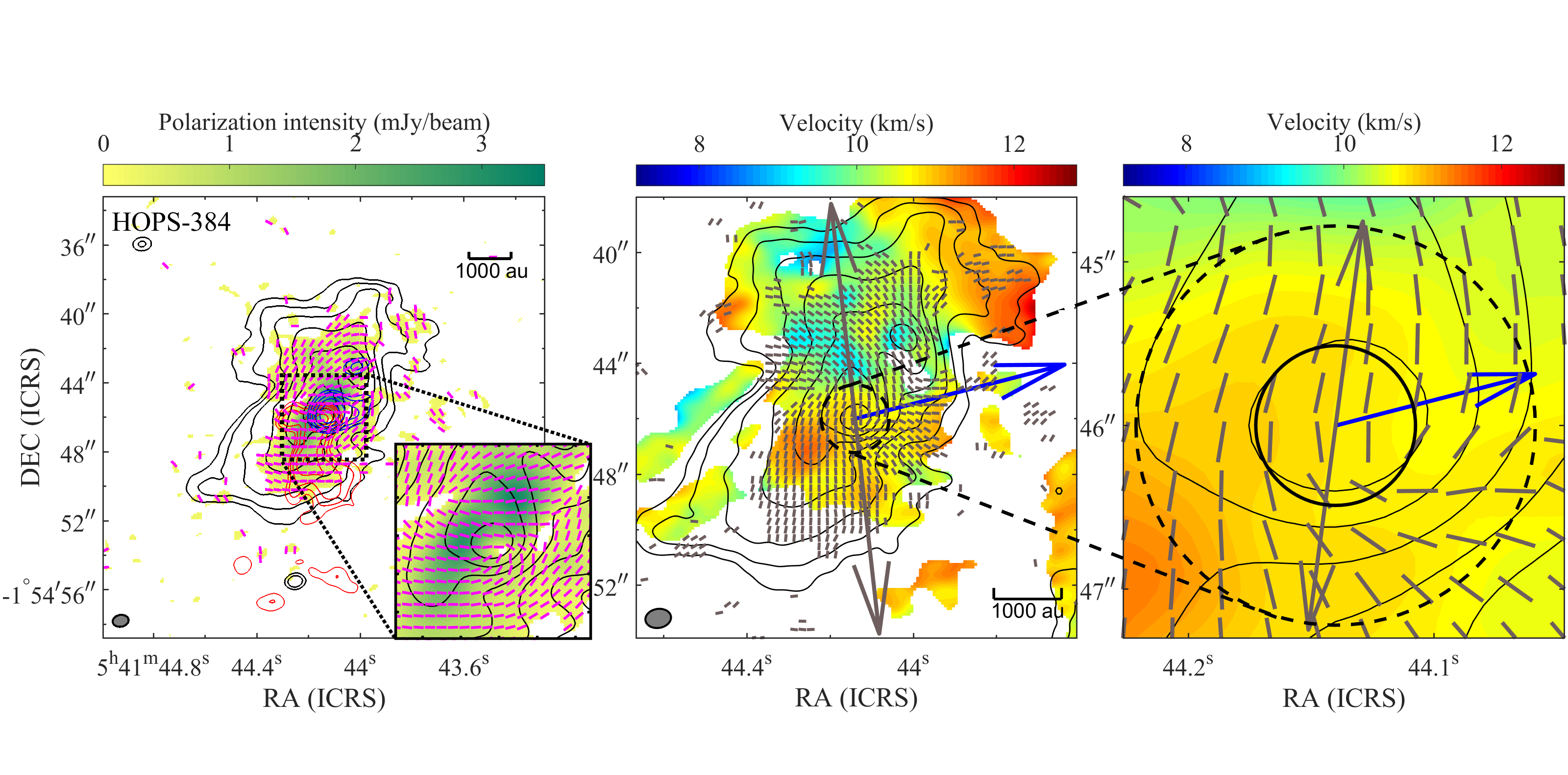}
\includegraphics[clip=true,trim=0cm 1cm 0cm 2cm,width=0.49 \textwidth]{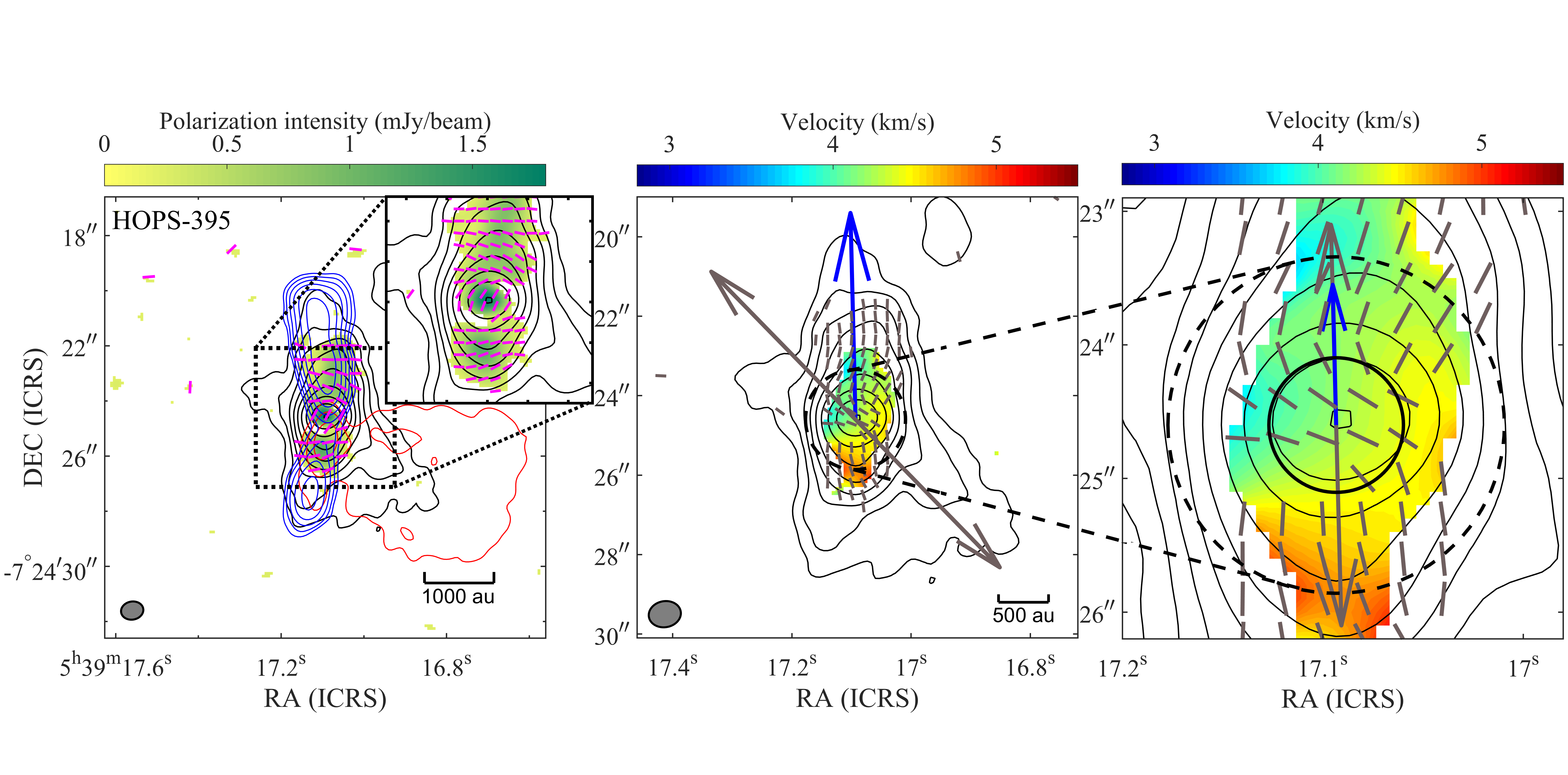}
\includegraphics[clip=true,trim=0cm 1cm 0cm 2cm,width=0.49 \textwidth]{HOPS-399.png}
\includegraphics[clip=true,trim=0cm 1cm 0cm 2cm,width=0.49 \textwidth]{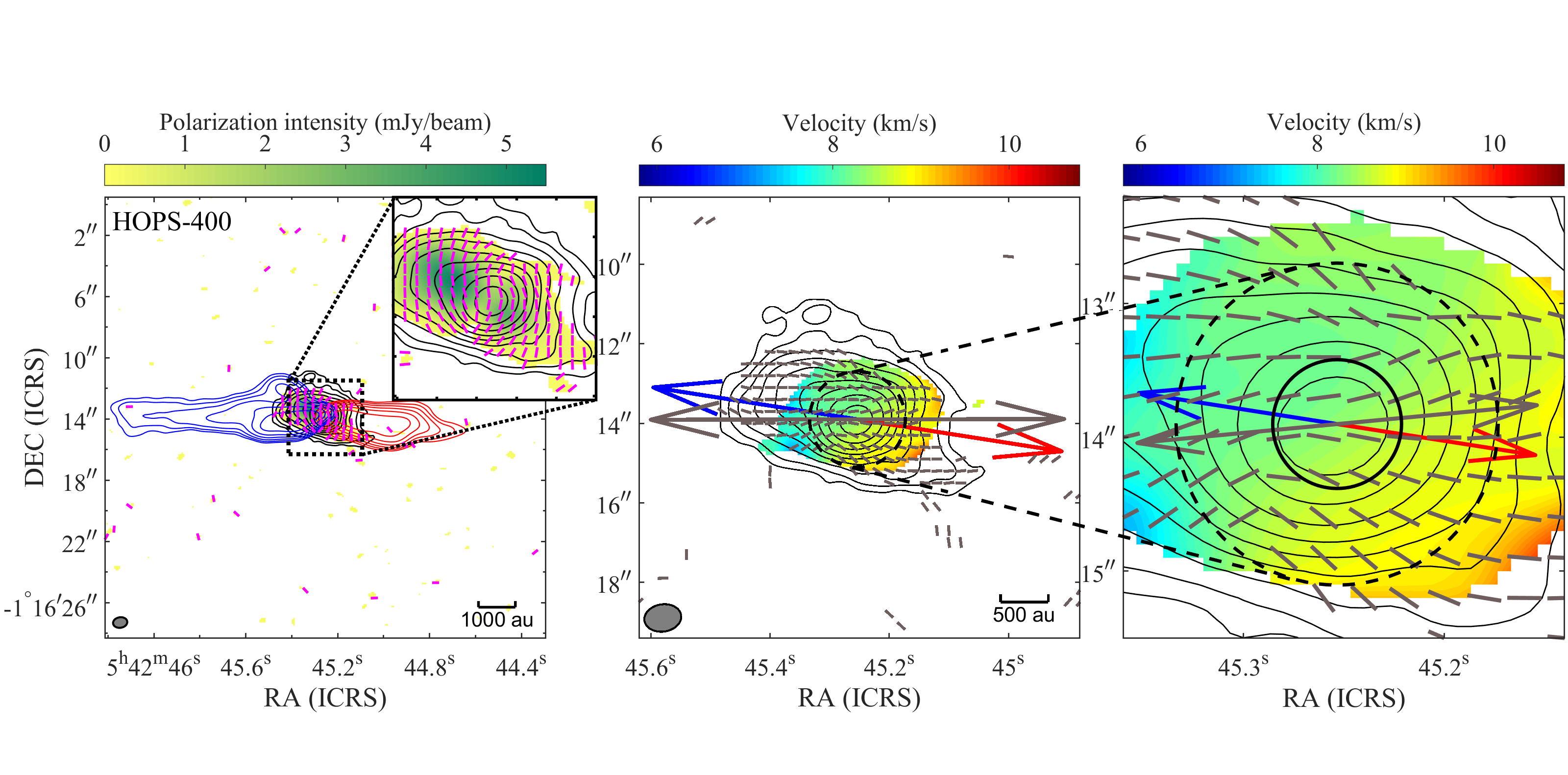}
\includegraphics[clip=true,trim=0cm 1cm 0cm 2cm,width=0.49 \textwidth]{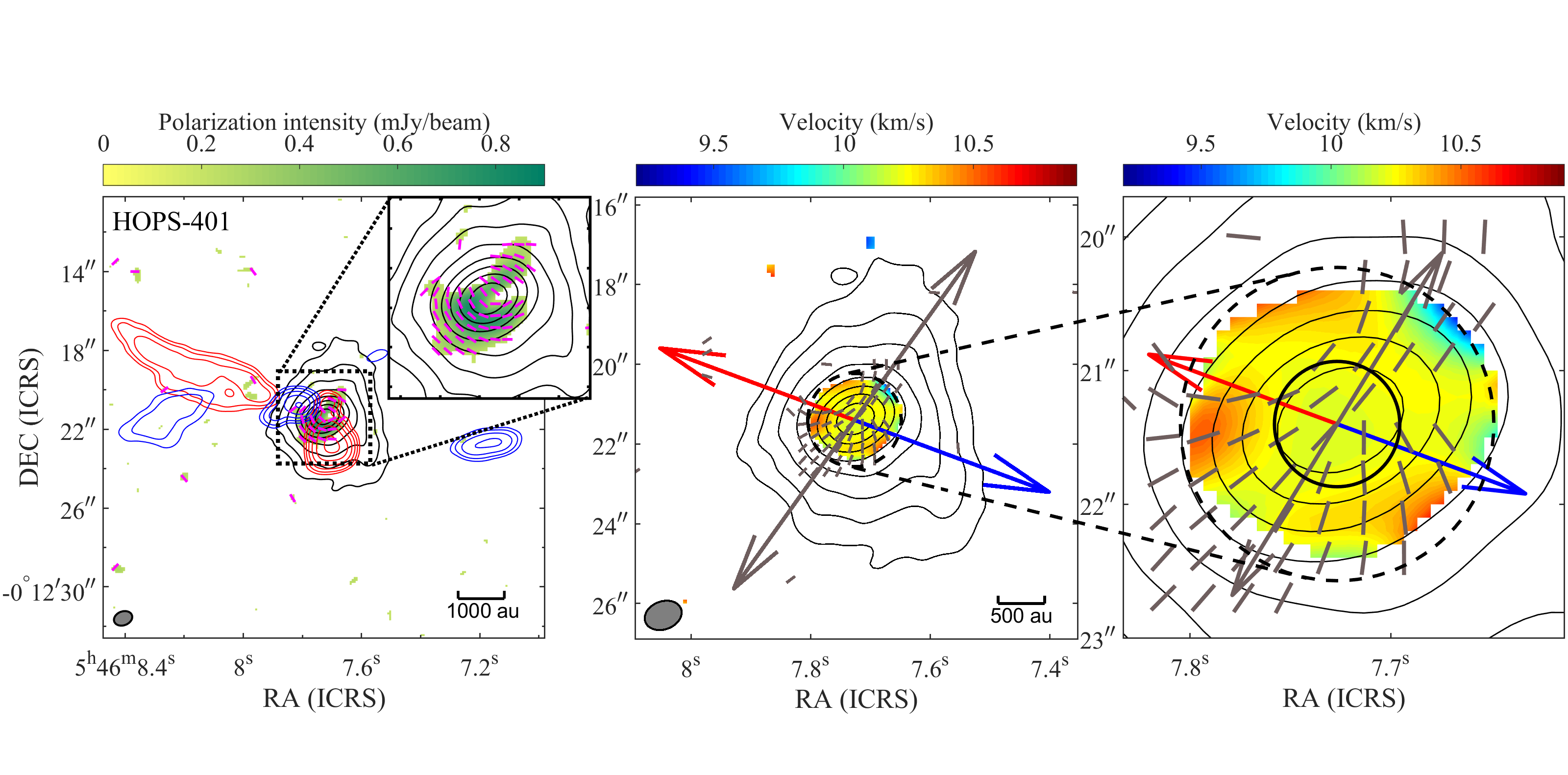}
\end{figure*}

\begin{figure*}
\centering
\text{\textbf{Unres-Type: velocity gradient is unresolved (Continued.)}}\\
\includegraphics[clip=true,trim=0cm 1cm 0cm 2cm,width=0.49 \textwidth]{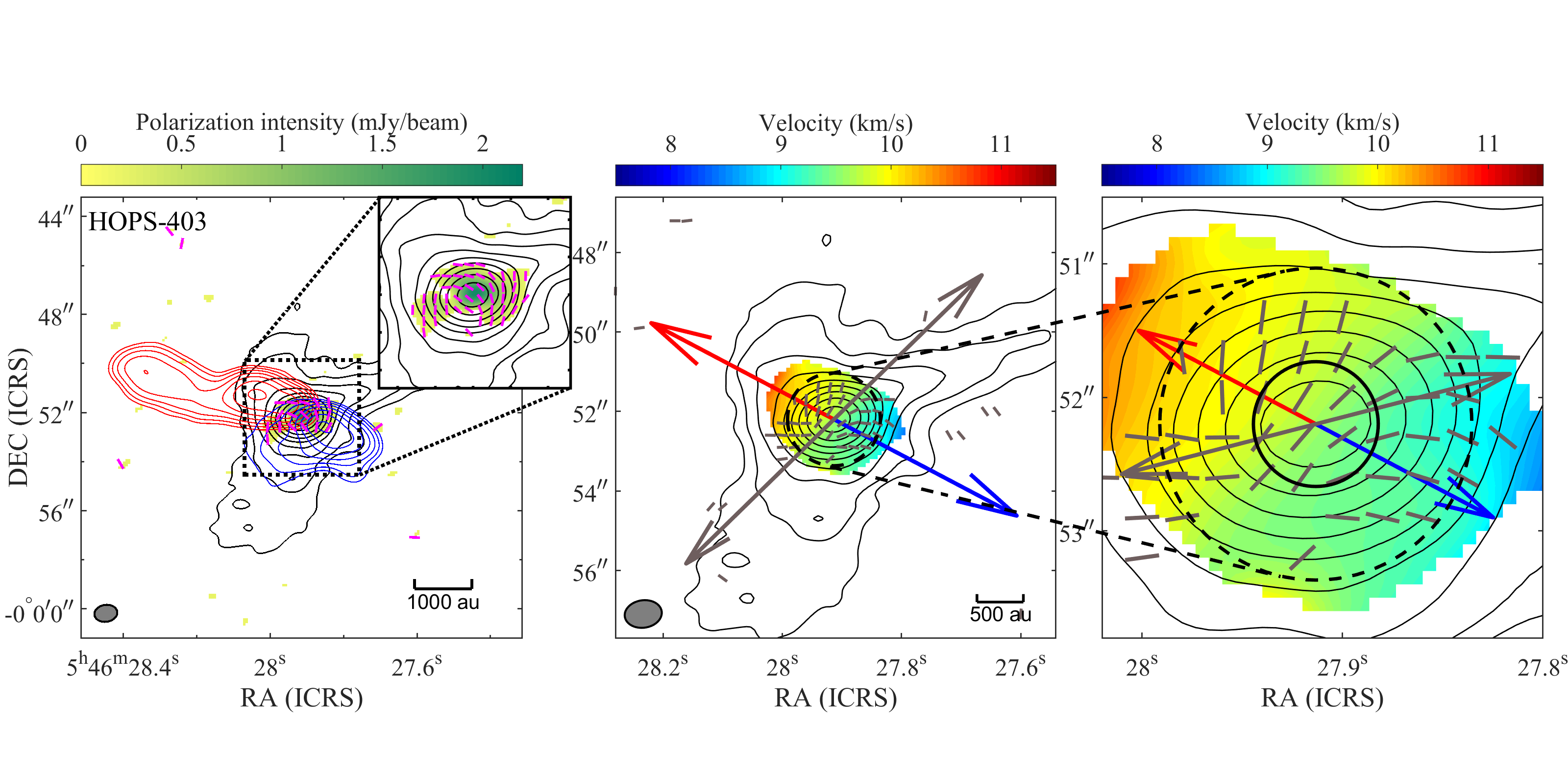}
\includegraphics[clip=true,trim=0cm 1cm 0cm 2cm,width=0.49 \textwidth]{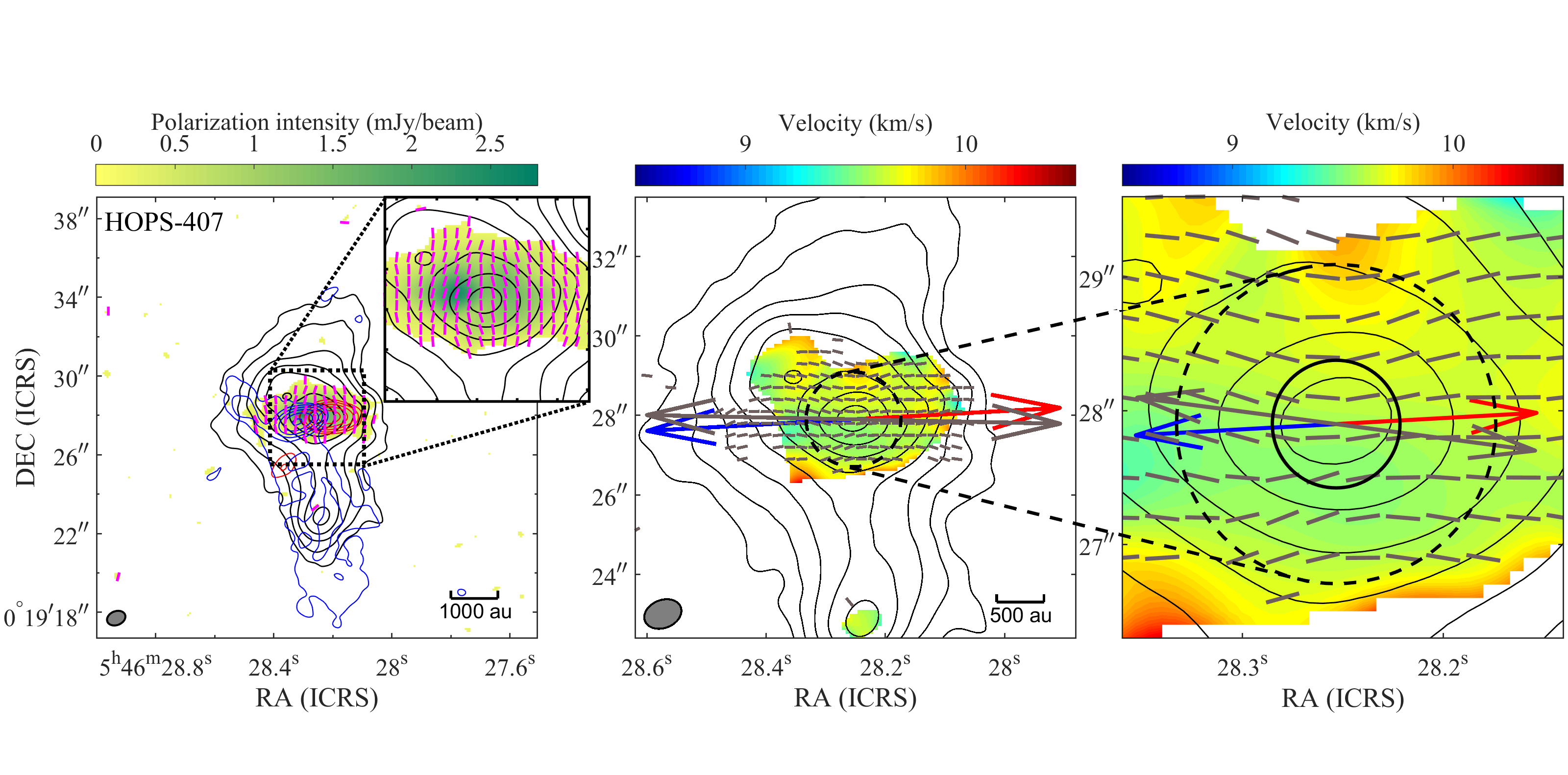}
\includegraphics[clip=true,trim=0cm 1cm 0cm 2cm,width=0.49 \textwidth]{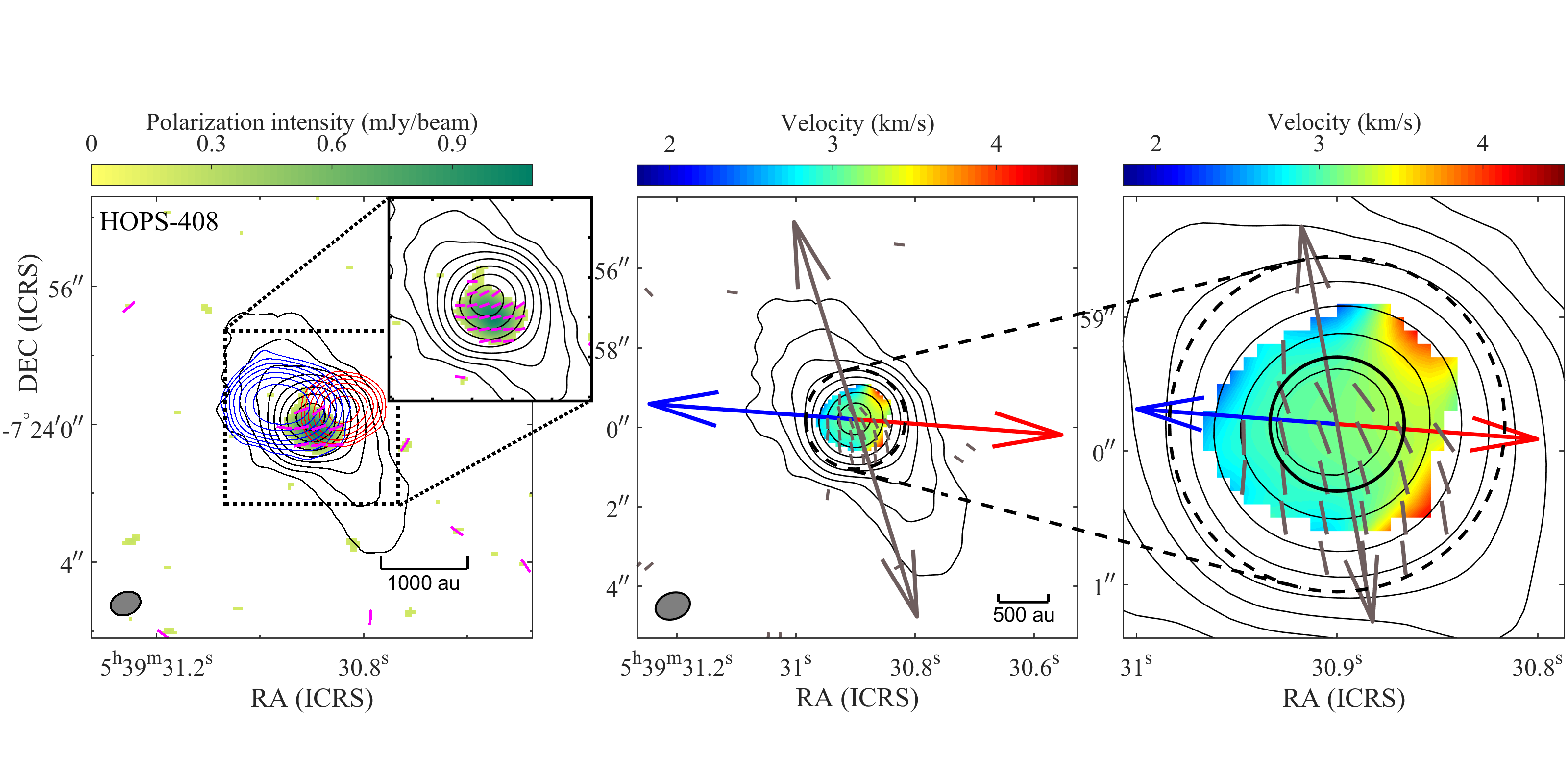}
\includegraphics[clip=true,trim=0cm 1cm 0cm 2cm,width=0.49 \textwidth]{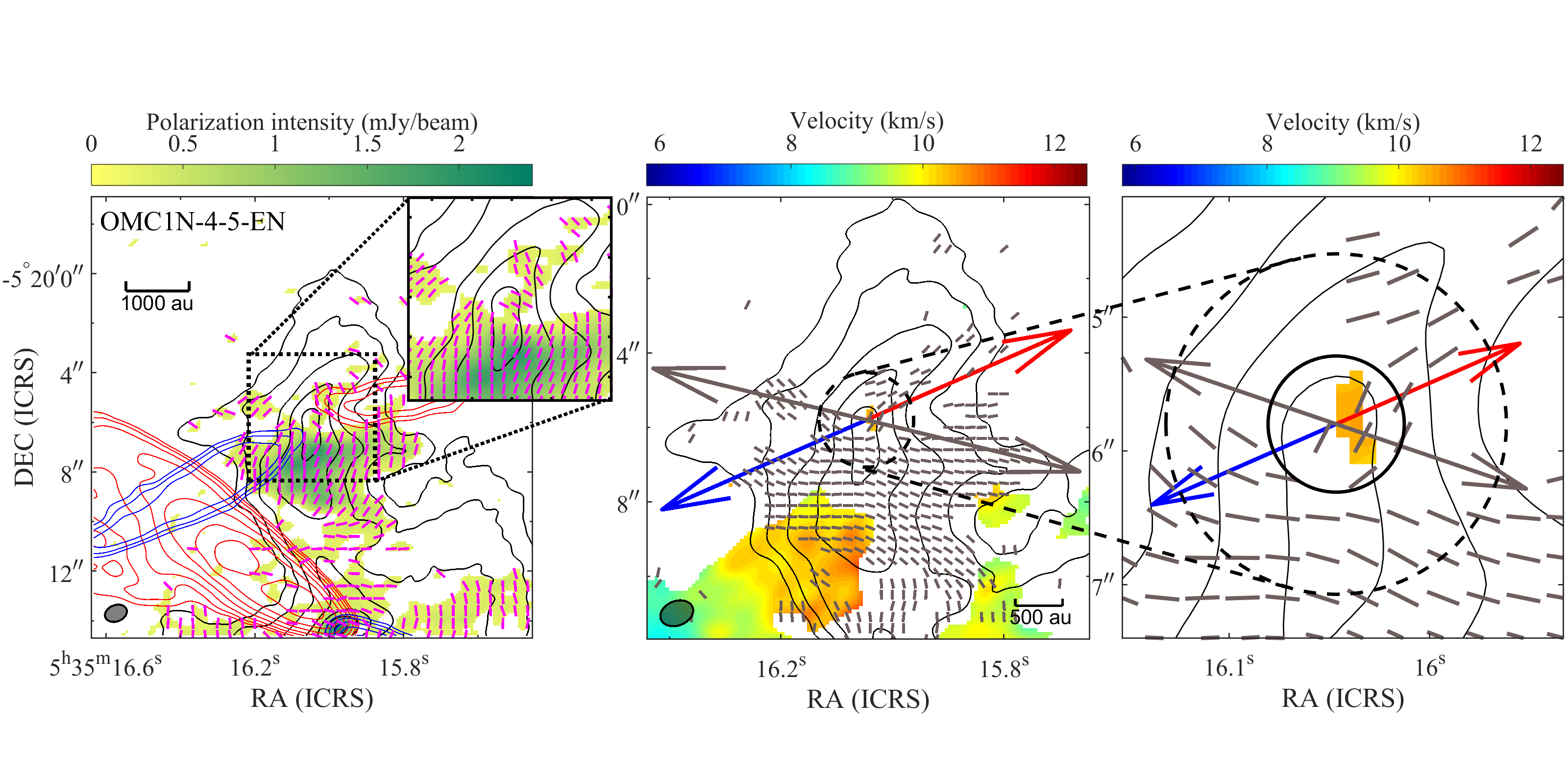}
\includegraphics[clip=true,trim=0cm 1cm 0cm 2cm,width=0.49 \textwidth]{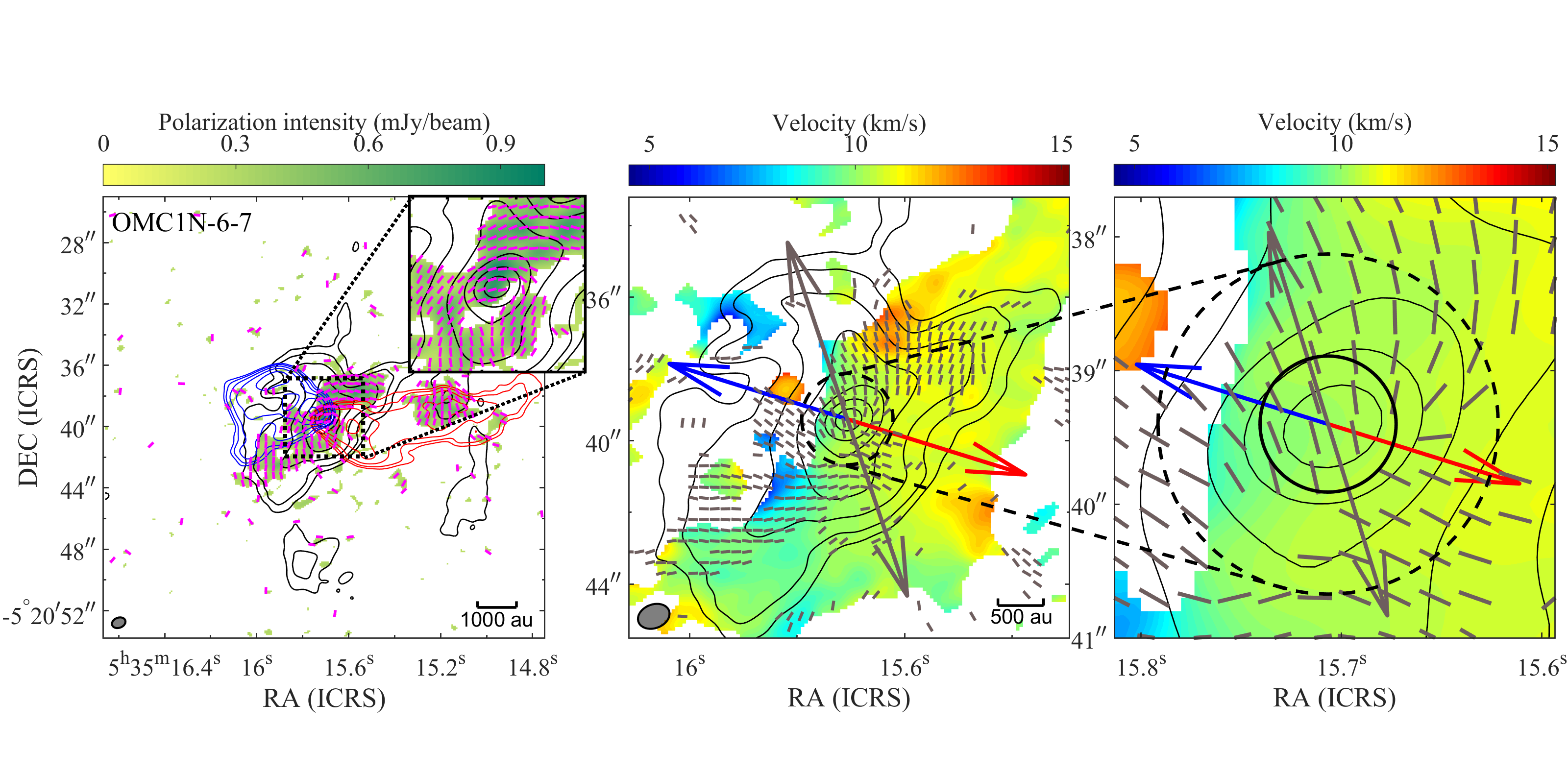}
\includegraphics[clip=true,trim=0cm 1cm 0cm 2cm,width=0.49 \textwidth]{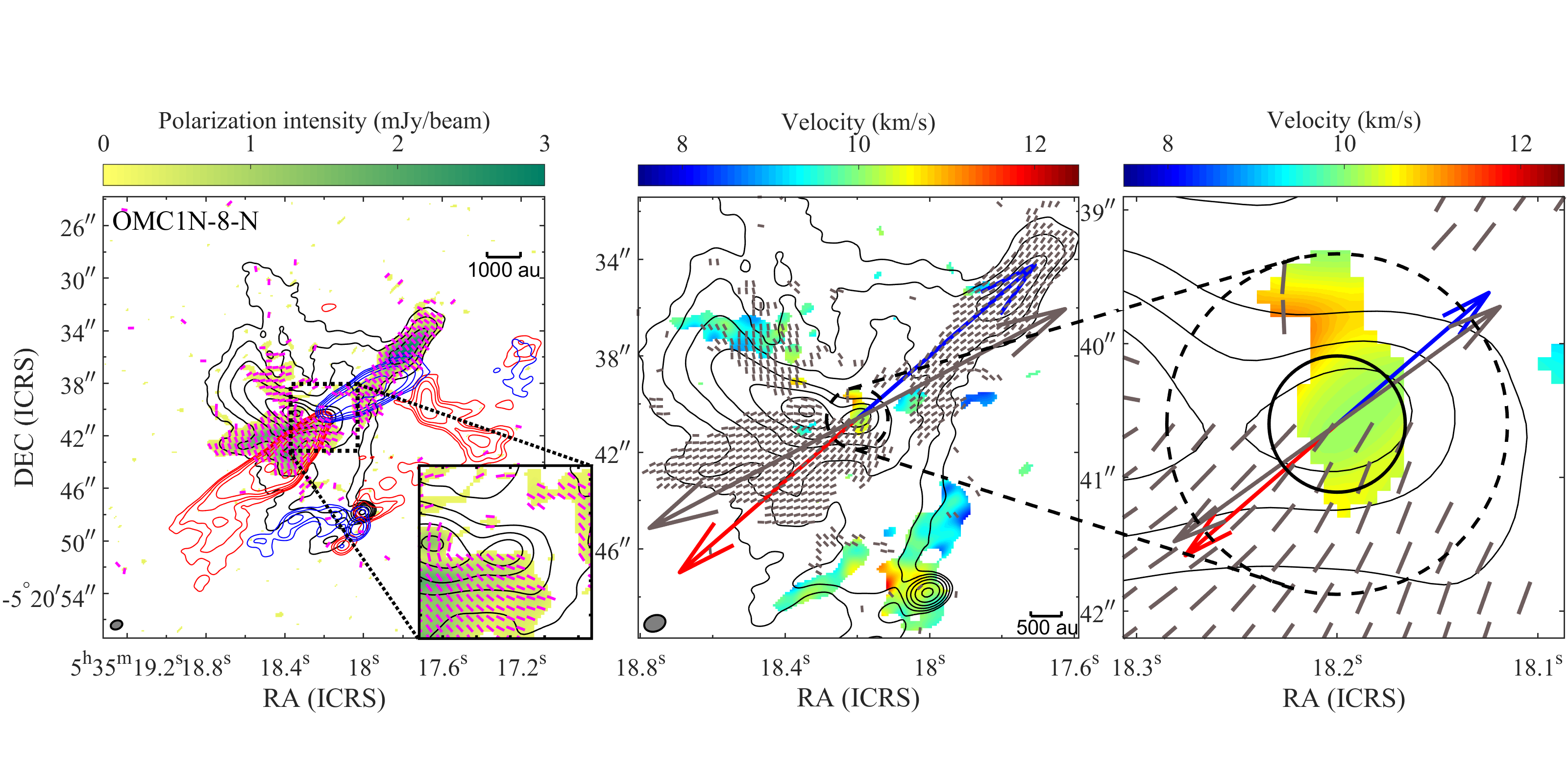} 
\end{figure*}

\begin{figure*}
\centering
\caption{Other: protostars without clear outflows and without 3$\sigma$ polarization detection.
First column: 870~$\mu$m dust polarization intensity in color scale overlaid with the redshifted and blueshifted outflow lobes (obtained from the $^{12}$CO (3--2) line), polarization segments, and dust continuum emission (Stokes {\em I}) contours.
Blue contours indicate the blueshifted outflow, while red contours are redshifted outflow, with counter levels set at 5 times the outflow {\em rms} × (1, 2, 4, 8, 16, 32).
The magenta segments represent the polarization.
The regions of polarization intensity less than 3$\sigma$ have been masked.
Second column: the velocity field in color scale (obtained from the C$^{17}$O (3--2) line) overlaid with the {\textit B}-field segments (i.e., polarization rotated by 90\arcdeg) and Stokes {\em I} contours.
Third column: an enlarged perspective of 1000 au of the second column.
In the second and third panels,
the black segments represent the {\textit B}-fields.
The red and blue arrows indicate the mean direction of the red-shifted and blue-shifted outflows.
For the velocity field, regions with an S/N less than 4 have been flagged.
The grey arrows in the second and third columns indicate the mean \textit{B}-field directions weighted by the intensity including all the polarization segments, and weighted by the uncertainty within annular region of 400--1000 au, respectively.
In all panels, the black contour levels for the Stokes {\textit I} image are 10 times the {\em rms} × (1, 2, 4, 8, 16, 32, 64, 128, 256, 512).
The black dotted square in the first column corresponds to 2000 au scale, while the black dashed, and solid circles correspond to scales of 1000 au, and 400 au, respectively.
}
\text{\textbf{Protostars without clear outflows}}\\
\includegraphics[clip=true,trim=0cm 1cm 0cm 2cm,width=0.49 \textwidth]{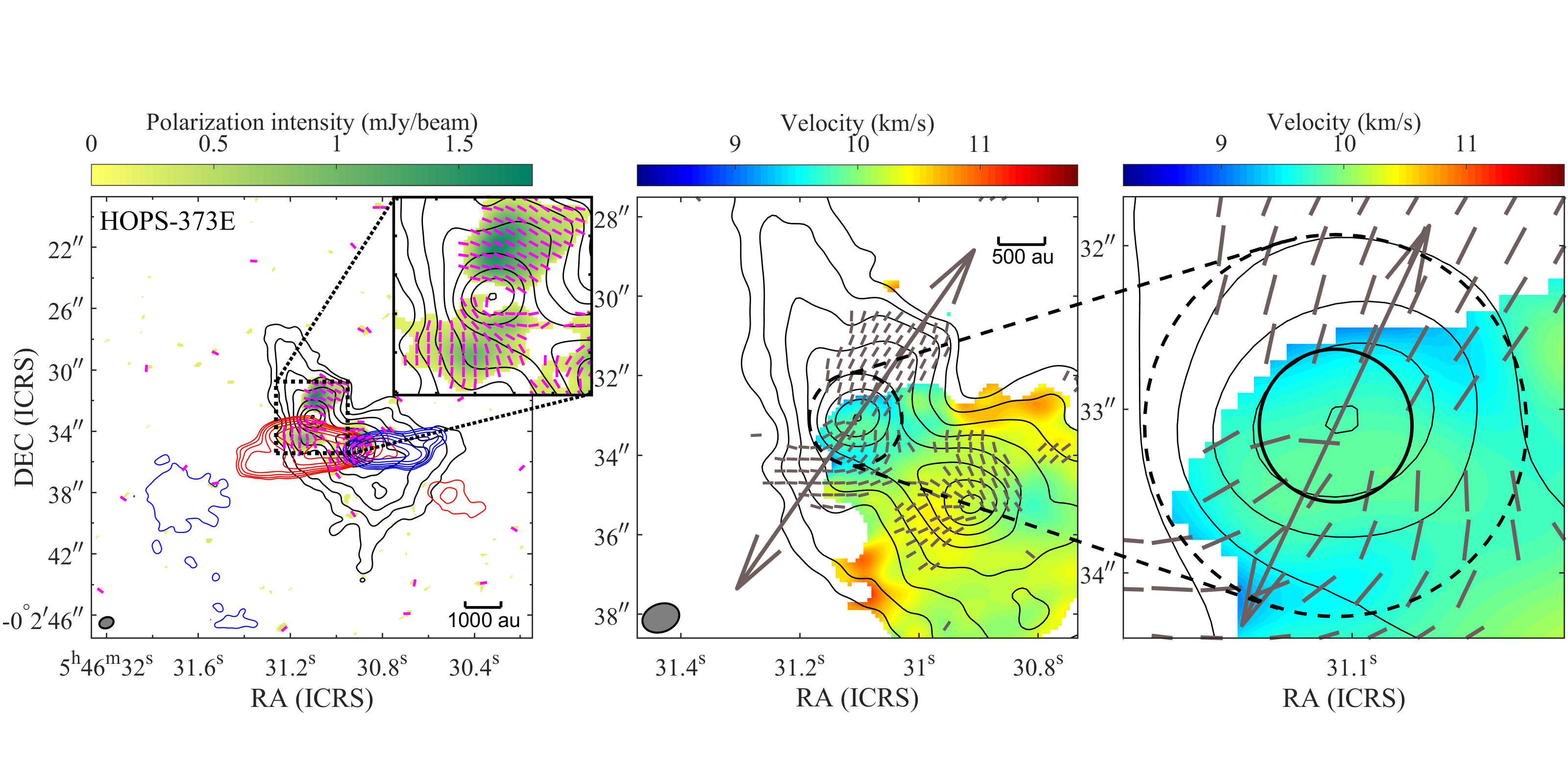}
\includegraphics[clip=true,trim=0cm 1cm 0cm 2cm,width=0.49 \textwidth]{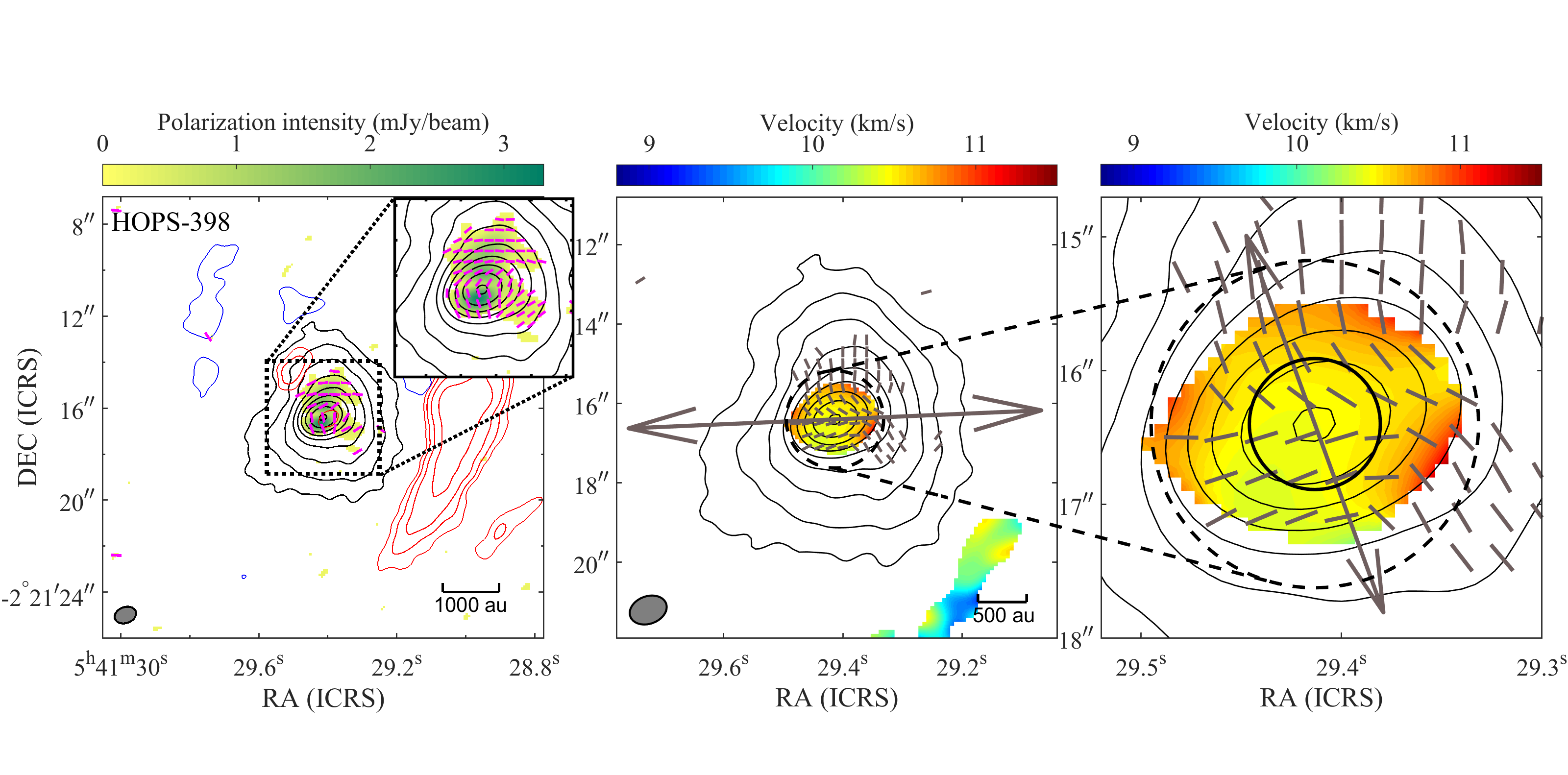}
\includegraphics[clip=true,trim=0cm 1cm 0cm 2cm,width=0.49 \textwidth]{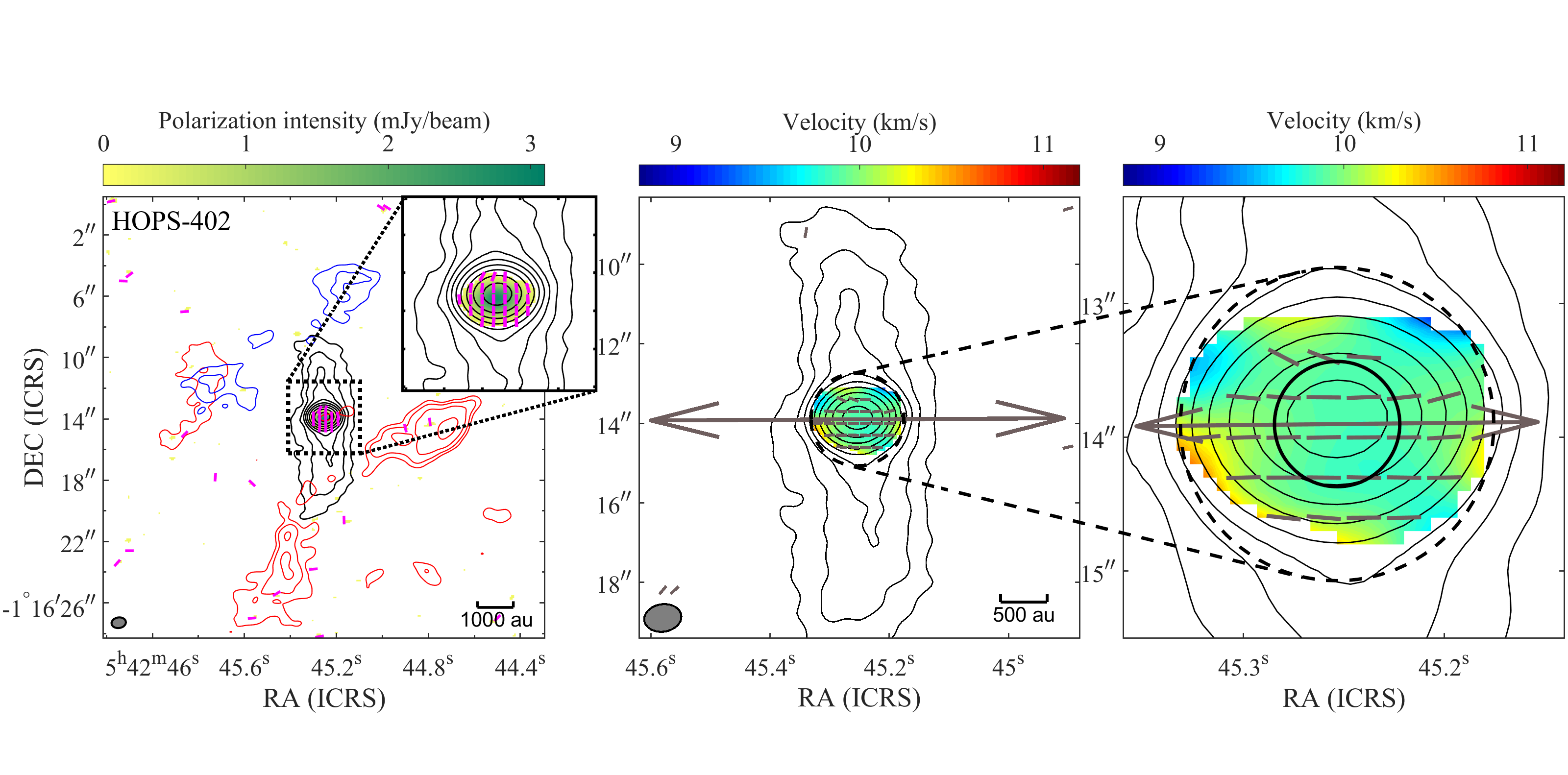} \\
\text{\textbf{Protostars without 3$\sigma$ polarization detection}}\\
\includegraphics[clip=true,trim=0cm 1cm 0cm 2cm,width=0.49 \textwidth]{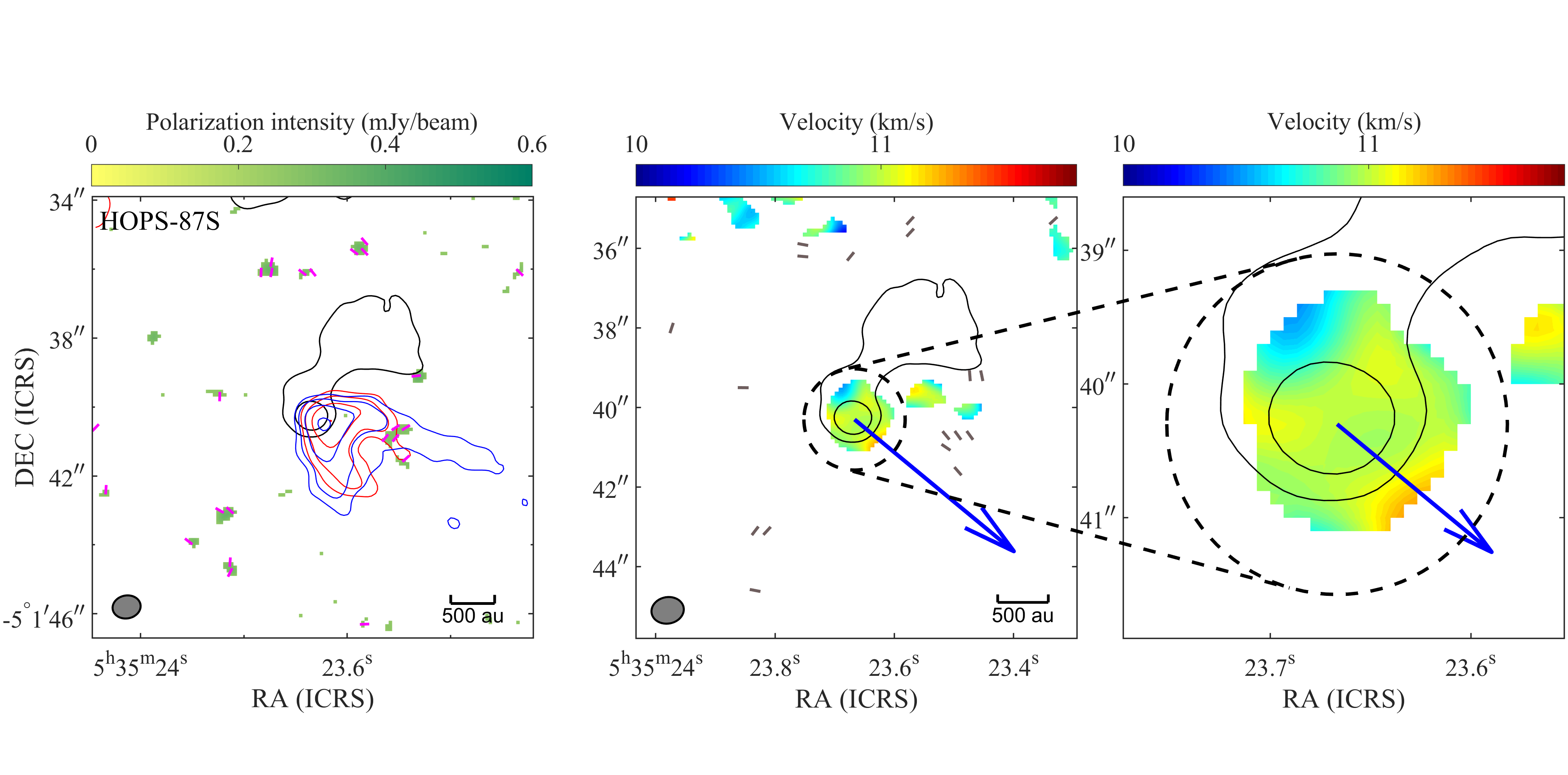}
\includegraphics[clip=true,trim=0cm 1cm 0cm 2cm,width=0.49 \textwidth]{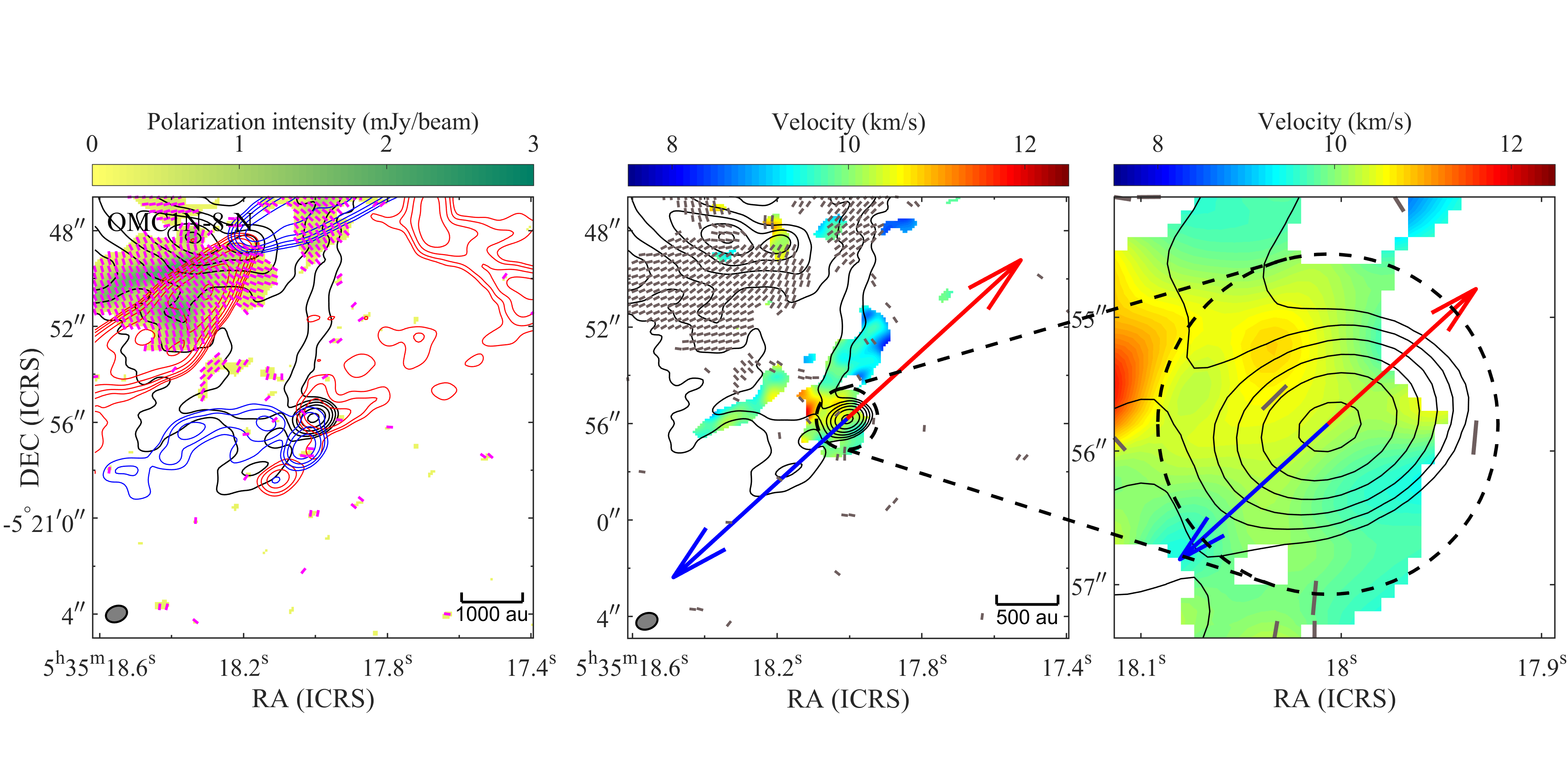}
\end{figure*}

\begin{figure*}[ht]
\centering
\caption{Starless core.
Color scale indicates the total intensity.}
\includegraphics[clip=true,trim=8cm 0cm 8cm 0.6cm,width=0.49 \textwidth]{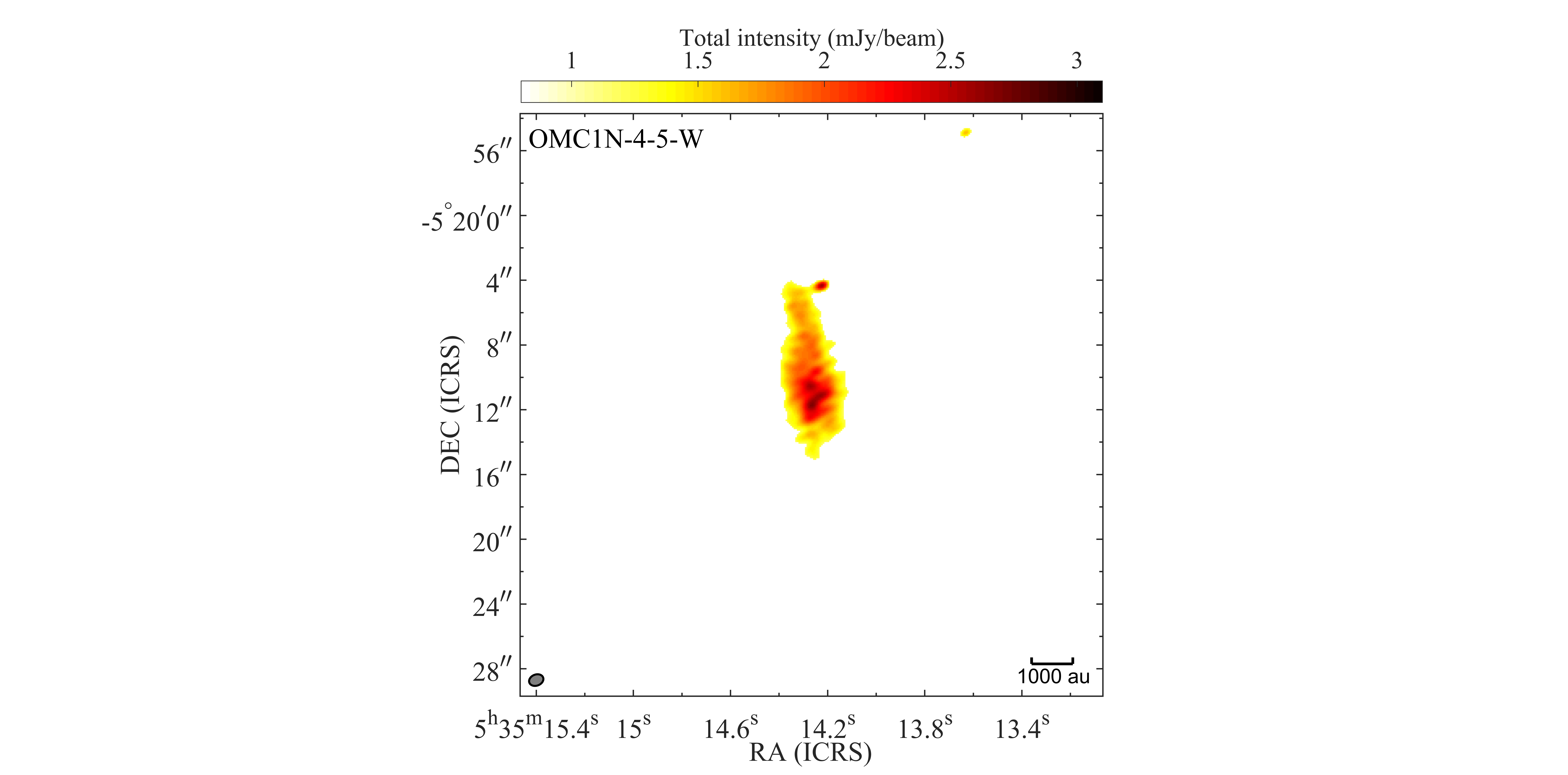}
\end{figure*}

\end{document}